%% file: author.tex
\documentclass[graybox]{svmult}

\usepackage{mathptmx}       
\usepackage{helvet}         
\usepackage{courier}        
\usepackage{type1cm}        
\usepackage{makeidx}         
\usepackage{graphicx}        
\usepackage{multicol}        
\usepackage[bottom]{footmisc}

\makeindex             

\usepackage{mathptmx}
\usepackage{amsfonts}
\usepackage[cmex10]{amsmath}
\usepackage{graphicx}
\usepackage{times}
\usepackage[tight,normalsize,sf,SF]{subfigure}

\usepackage{wrapfig,lipsum,booktabs}
\usepackage{amsfonts} 
\usepackage{amsmath} 
\usepackage{color}
\usepackage{relsize}
\usepackage{bbm}
\usepackage{algorithm}
\usepackage{algorithmic}
\usepackage[usenames,dvipsnames]{xcolor}
\usepackage{tikz}
\usetikzlibrary{positioning,shapes,shadows,arrows,trees,snakes}
\usepackage{listings}
\usepackage{mathptmx}
\usepackage{graphicx}
\usepackage{times}
\usepackage{amssymb}
\usepackage{subfigure}
\usepackage{txfonts}
\usepackage{graphicx}

\newcommand{\R}{\mathbb{R}}

\newcommand{\N}{\mathcal{N}}

\newcommand{\Span}{\operatorname{span}}

\newcommand{\neigh}{\operatorname{N}}

\newcommand{\proj}{\operatorname{proj}}

\newcommand{\diver}{\operatorname{div}}
\newcommand{\vv}{\mathbf{v}}
\newcommand{\id}{\mathbf{Id}}
\newcommand{\ww}{\mathbf{w}}
\newcommand{\xx}{\mathbf{x}}
\newcommand{\yy}{\mathbf{y}}
\newcommand{\kk}{\mathbf{k}}
\newcommand{\pp}{\mathbf{p}}

\newcommand{\nn}{\mathbf{n}}

\newcommand{\ee}{\mathbf{e}}
\newcommand{\ttb}{\mathbf{t}}
\newcommand{\uu}{\mathbf{u}}

\newcommand{\no}{$\textcolor{red}{\text{\rlap{$\times$}}\square}$}
\newcommand{\yes}{$\textcolor{Green}{\text{\rlap{$\checkmark$}}\square}$}

\usepackage[shadow,colorinlistoftodos]{todonotes}

\newcommand{\GRAMMAR}[1]{\todo[color=green!10]{Grammar!}}
\newcommand{\LOSENT}[1]{\todo[color=red!60]{Sentence too long!}}

\begin{document}

\title*{Feature Lines for Illustrating Medical Surface Models: Mathematical Background and Survey}
\titlerunning{Feature Lines for Illustrating Medical Surface Models}
\author{Kai Lawonn and Bernhard Preim}
\institute{Kai Lawonn \at Otto-von-Guericke University Magdeburg, Faculty of Computer Science, Department of Simulation and Graphics
\email{lawonn@isg.cs.uni-magdeburg.de }
\and Bernhard Preim \at Otto-von-Guericke University Magdeburg, Faculty of Computer Science, Department of Simulation and Graphics 
\email{preim@isg.cs.uni-magdeburg.de }}
%
%
\maketitle

\abstract{\input{abstract.tex}}

\input{introduction.tex}
\input{mathematical_background.tex}
\input{computer_graphic.tex}
\input{aspects.tex}
\input{feature_lines.tex}
\input{discussion.tex}

\input{conclusion.tex}

\bibliographystyle{styles/spmpsci}

\bibliography{bib}
\end{document}

%% file: abstract.tex
This paper provides a tutorial and survey for a specific kind of illustrative visualization technique: feature lines.
We examine different feature line methods.
For this, we provide the differential geometry behind these concepts and adapt this mathematical field to the discrete differential geometry.
All discrete differential geometry terms are explained for triangulated surface meshes.
These utilities serve as basis for the feature line methods.
We provide the reader with all knowledge to re-implement every feature line method.
Furthermore, we summarize the methods and suggest a guideline for which kind of surface which feature line algorithm is best suited.
Our work is motivated by, but not restricted to, medical and biological surface models.

%% file: introduction.tex
%
%
\section{Introduction}\label{introduction}
%
%

\begin{wrapfigure}{r}{0.5\textwidth}
 \centering
  \subfigure[]{
 \includegraphics[scale=1.5]{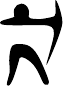}
  \label{fig:fl_s}
  }~
  \subfigure[]{
 \includegraphics[scale=1.5]{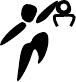}
 \label{fig:fl_rv}
 }~
 \subfigure[]{
 \includegraphics[scale=1.5]{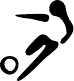}
 \label{fig:fl_sc}
 }~
  \subfigure[]{
 \includegraphics[scale=1.5]{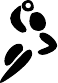}
 \label{fig:fl_ar}
 }
 \caption{Visual abstraction of the four Olympic disciplines: archery, basketball, football and handball in the style of the pictograms of the Olympic Games 2012 in London.}
 \label{fig:examples}
\end{wrapfigure}

The application of illustrative visualization has increased in recent years.
The principle goal behind the concept of illustrative visualization is a meaningful, expressive, and simplified depiction of a problem, a scene or a situation.
As an example, running people are represented running stickmans, which can be seen in the Olympic games, and other objects become simplified line drawings, see Figure~\ref{fig:examples}.
More complex examples can be found in medical atlases. 
Most anatomical structures are painted and illustrated with pencils and pens.
Gray's anatomy is one of the famous textbooks for medical teaching.
Most other textbooks in this area orient to depict anatomy with art drawing, too.

Other than simplified representation, illustrative visualization is not restricted to these fields.
Illustrative techniques are essential for \emph{focus-and-context} visualizations.
Consider a scene with anatomical structures and one specific (important) structure.
The specific structure may be strongly related to the surrounding objects.
Therefore, hiding the other objects is not a viable option.
In contrast, depicting all structures leads to visual clutter and optical distraction of the most important structures.
Focus-and-context visualization is characterized by a few local regions that are displayed in detail and with emphasis techniques, such as a saturated color.
Surrounding contextual objects are displayed in a less prominent manner to avoid distraction from focal regions.
Medical examples are vessels with interior blood flow, livers with inner structures including vascular trees and possible tumors, proteins with surface representation and interior ribbon visualization.
Focus-and-context visualization is not restricted to medical data.
An example is the vehicle body and the interior devices.
The user or engineer needs the opportunity to illustrate all devices in the same context. 

There are numerous methods for different illustration techniques.
This survey is focused on a specific illustrative visualization category: feature lines.
Feature lines are a special group of line drawing techniques.
Another class of line drawing methods is hatching.
Hatching tries to convey the shape by drawing a bunch of lines.
Here, the spatial impression of the surface is even more improved.
Several methods exist to hatch the surface mesh, see~\cite{Girshick2000,Hertzmann2000,Lawonn2013CGF,Praun2001,Webb2002,Zander2004}.
In contrast, feature lines try to generate lines at salient regions only.
Not only for illustrative visualization, feature lines can also be used for rigid registrations of anatomical surfaces~\cite{Stylianou2004} or for image and data analysis in medical applications~\cite{Eberly1996}.
The goal of this survey is to convey the reader to the different feature line methods and offer a tutorial with all the knowledge to be able to implement each of the methods.

\paragraph{Organization.} We first give an overview of the mathematical background.
In Section~\ref{differential_geometry_background}, we introduce the necessary fundamentals of differential geometry.
Afterwards, we adapt these fundamentals to triangulated surface meshes in Section~\ref{discrete_differential_geometry}.
Section~\ref{aspects} discusses general aspects and requirements for feature lines.
Next, we present different feature line methods in Section~\ref{feature_lines} and compare them in Section~\ref{discussion}.
Finally, Section~\ref{conclusion} holds the conclusion of this survey.
\begin{figure*}[tb]
 \centering
 \includegraphics[scale=1]{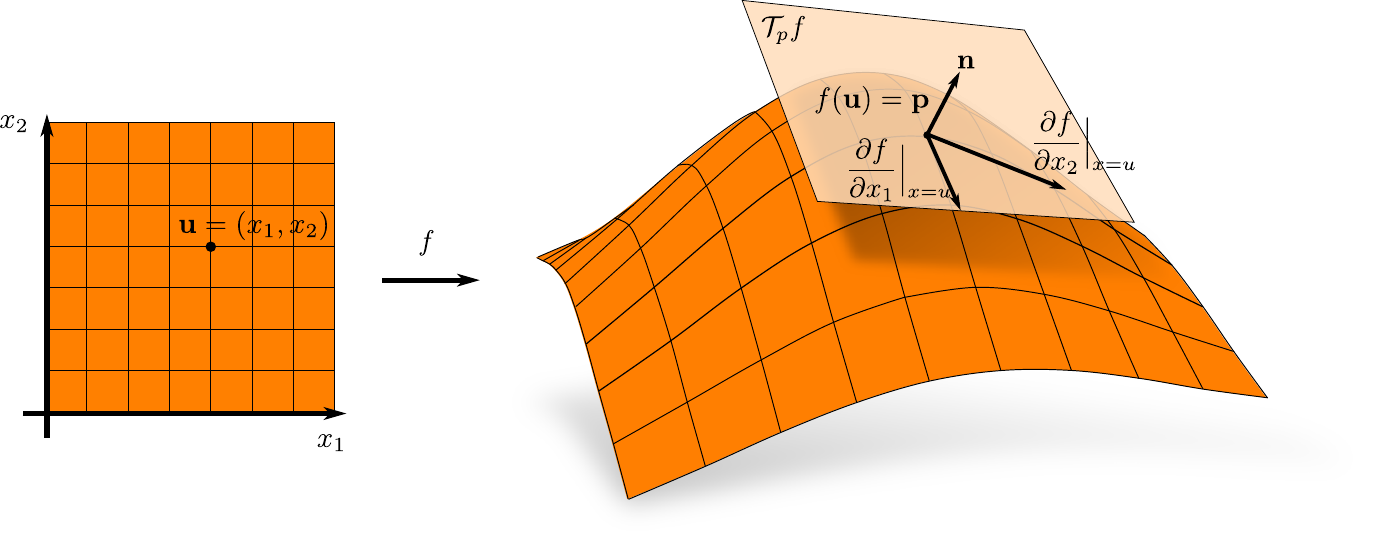}
 \caption{The basic elements for differential geometry. A parametric surface is given and the partial derivatives create the tangent space.}
 \label{fig:basic}
\end{figure*}

%% file: mathematical_background.tex
%
%
\section{Differential Geometry Background}\label{differential_geometry_background}
%
%
This section presents the fundamentals of differential geometry for feature line generation, which will be crucial for the further sections.
We present the basic terms and properties.
This section is inspired by differential geometry books~\cite{doCarmo1976,doCarmo1992,Kuehnel2006}.
%
%
\subsection{Basic Prerequisites}\label{subsec:basics}
%
%

A surface $f\colon I\subset\R^2\rightarrow\R^3$ is called a parametric surface if $f$ is an \textit{immersion}.
An \textit{immersion} means that all partial derivatives $\frac{\partial f}{\partial x_i}$ are injective at each point. 
The further calculations are mostly based on the tangent space of a surface.
The tangent space $\mathcal T_p f$ of $f$ is defined as the linear combination of the partial derivatives of $f$:
\[\mathcal{T}_p f\coloneqq\Span\Big\{{\frac{\partial f}{\partial x_1}\Big\vert_{\xx=\uu},\frac{\partial f}{\partial x_2}\Big\vert_{\xx=\uu} }\Big\}. \]
Here, $\Span$ is the space of all linear combinations. 
Formally: $\Span\{\vv_1,\vv_2\}\coloneqq\{\alpha\vv_1+\beta\vv_2\,|\,\alpha,\beta\in\R\}$.
With the tangent space, we can define a normalized normal vector $\nn$.
The (normalized) normal vector $\nn(\uu)$ at $\pp=f(\uu)$ is defined such that for all elements $\vv\in\mathcal T_p f$ the equation $\langle\vv,\nn(\uu)\rangle=0$ holds, where $\langle .,.\rangle$ denotes the canonical Euclidean dot product.
Therefore, $\nn(\uu)$ is defined as:
\[\nn(\uu)=\frac{\frac{\partial f}{\partial x_1}\Big\vert_{\xx=\uu}\times \frac{\partial f}{\partial x_2}\Big\vert_{\xx=\uu}}{\big\Vert\frac{\partial f}{\partial x_1}\Big\vert_{\xx=\uu}\times\frac{\partial f}{\partial x_2}\Big\vert_{\xx=\uu}\big\Vert}.\]
This map is also called the Gauss map.
Figure~\ref{fig:basic} depicts the domain of a parametric surface as well as the tangent space $\mathcal T_p f$ and the normal $\nn$.

%
%
\subsection{Curvature}\label{subsec:curvature}
%
%
The curvature is a fundamental property to identify salient regions of a surface that should be conveyed by feature lines.
Colloquially spoken, it is a measure of how far the surface bends at a certain point.
If we consider ourselves to stand on a sphere at a specific point, it does not matter in which direction we go, the bending will be the same.
If we imagine we stand on a plane at a specific point, we can go in any direction, there will be no bending.
Without knowing any measure of the curvature, we can state that a plane has zero curvature and that a sphere with a small radius has a higher curvature than a sphere with a higher radius.
This is due to the fact that a sphere with an increasing radius becomes locally more a plane.
Intuitively, the curvature depends also on the direction in which we decide to go.
On a cylinder, we have a bending in one direction but not in the other.
Painting the trace of a walk on the surface and view it in 3D space, we could treat this as a 3D curve.
The definition of the curvature of a curve may be adapted to the curvature of a surface.
The adaption of this concepts is explained in the following.
Let $c\colon I\subset\R\rightarrow\R^3$ be a 3D parametric curve with 
$\Vert \frac{d c}{d t}\Vert=1$ .
The property of constant length of the derivative is called arc length or natural parametrization.
One can show that such a parametrization exists for each continuous, differentiable curve that is an immersion.
So, if we want to measure the size of bending, we can use the norm of the second derivative of the curve.
Therefore, the (absolute) curvature $\kappa(t)$ at a time point $t$ is defined as:
\[\kappa(t)=\big\Vert c''(t) \big\Vert.\] 
\begin{wrapfigure}{r}{0.5\textwidth}
 \centering
   \includegraphics[scale=0.75 ]{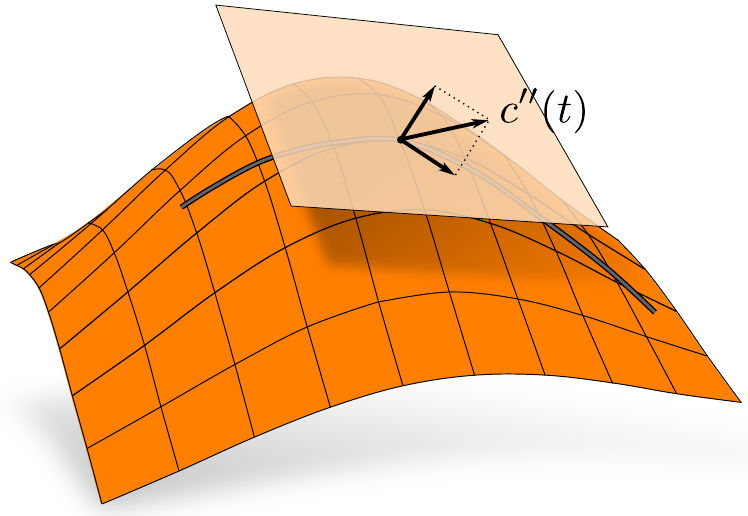}
 \caption{The curve's second derivative is decomposed into the tangential and normal part.}
 \label{fig:curve_decom}
\end{wrapfigure}
To determine the curvature on a certain point of the surface in a specific direction, we can employ a curve and calculate its curvature.
This approach is imperfect because curves that lie in a plane can have non-vanishing curvature, e.g., a circle, whereas we claimed to have zero curvature on a planar surface.
Therefore, we have to distinguish which part of the second derivative of the curve contributes to the tangent space and which contributes to the normal part of the surface.
Decomposing the second derivative of the curve into tangential and normal part of the surface yields: 
\[c''(t)=\underbrace{\proj_{\mathcal{T}_pf} c''(t)}_{\text{tangential part}} + \underbrace{\langle c''(t),\nn\rangle \nn}_{\text{normal part}},\]
where $c(t)=\pp$ and $\proj_E\xx$ means the projection of the point $\xx$ onto the space $E$, see Figure~\ref{fig:curve_decom}.
The curvature $\kappa_c(\pp)$ of the surface at $\pp$ along the curve $c$ is defined as the coefficient of the normal part:
\begin{align}\label{curv_surface}
\kappa_c(p)=\langle c''(t),\nn\rangle.
\end{align}

Hence, we know that $c'(t)\in\mathcal{T}_p f$ and $\langle c'(t),\nn\rangle=0$.
Deriving the last equation yields:
\begin{align*}
\frac{d}{dt}\langle c'(t),\nn\rangle &=0\\
\frac{d}{dt}\langle c'(t),\nn\rangle &= \langle c''(t),\nn\rangle + \langle c'(t),\frac{\partial \nn}{\partial t}\rangle.
\end{align*}
We obtain
\[\langle c''(t),\nn\rangle=- \langle c'(t),\frac{\partial \nn}{\partial t}\rangle.\]
Combining this equation with Equation~\ref{curv_surface} yields
\begin{align}\label{curv_surface_2}
\kappa_{c'(t)}(\pp)=-\langle c'(t),\frac{\partial \nn}{\partial t}\rangle.
\end{align}
Thus, the curvature of a surface at a specific point in a certain direction can be calculated by a theorem by Meusnier.
We call the vectors $\vv,\ww$ at $\pp$ the maximal/ minimal principle curvature directions of the maximal and minimal curvature, if $\kappa_{\vv}(\pp)\ge \kappa_{\vv'}(\pp)$, $\kappa_{\ww}(\pp)\le \kappa_{\vv'}(\pp)$ for all directions $\vv'\in\mathcal{T}_pf$.
If such a minimum and maximum exists, then $\vv$ and $\ww$ are perpendicular, see Section~\ref{subsec:weingarten} for a proof.
If we want to determine the curvature in direction $\uu$, we first need to normalize $\uu,\vv,\ww$ and can then determine $\kappa_{\uu}(\pp)$ by:
\begin{align}\label{curvature_direction}
\kappa_{\uu}(\pp)=\langle\uu,\vv\rangle^2 \kappa_{\vv}(\pp) + \langle\uu,\ww\rangle^2 \kappa_{\ww}(\pp).
\end{align}
The coefficients of the curvature are the decomposition of the principle curvature directions with the vector $\uu$.
%
%
\subsection{Covariant Derivative}\label{subsec:covariant_derivative}
\begin{figure*}[tb]
 \centering
   \includegraphics[scale=1]{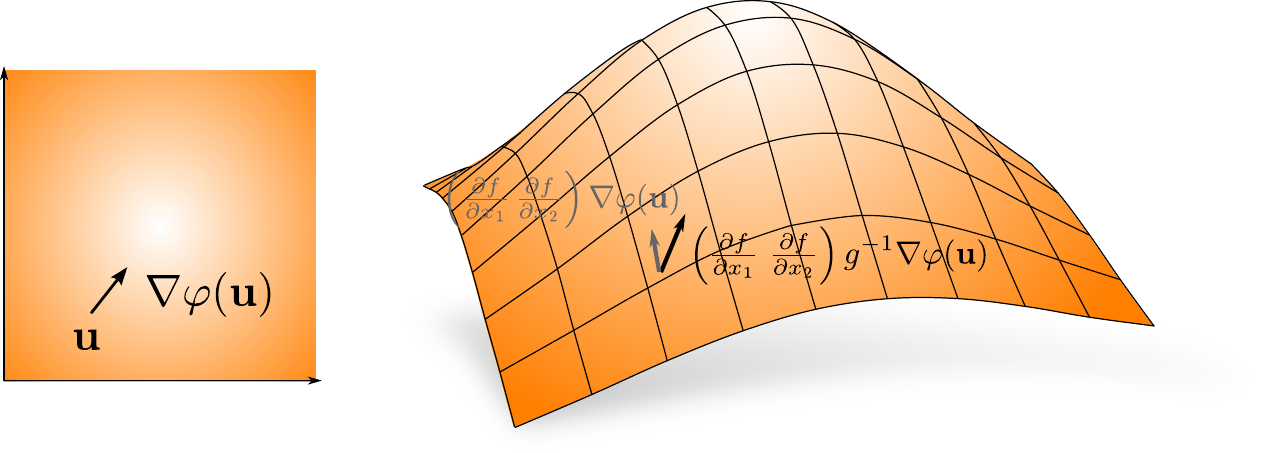}
 \caption{Given: a scalar field in the domain. Determining the gradient and using it as coefficient for the basis tangent vectors leads to the wrong result (grey). Balancing the distortion with the inverse of the metric tensor yields the correct gradient on the surface (black).}
 \label{fig:gradient}
\end{figure*}

%
%
The essence of the feature line generation is the analysis of local variations in a specific direction, i.e., the covariant derivative.
Therefore, the covariant derivative is a crucial concept for feature line methods.
We consider a scalar field on a parametric surface $\varphi\colon f(I)\rightarrow\R$.
One can imagine this scalar field as a heat or pressure (as well as a curvature) distribution.
The directional derivative of $\varphi$ in direction $\vv$ can be written as $D_\vv\varphi$ and is defined by:
\[
D_\vv\varphi(\xx)=\lim_{h\to 0}\frac{\varphi(\xx+h\vv)-\varphi(\xx)}{h}.
\] 
If $\varphi$ is differentiable at $\xx$, the directional derivative can be simplified:
\[
D_\vv\varphi(\xx)=\langle\nabla \varphi(\xx),\vv\rangle,
\]
where $\nabla$ denotes the gradient.
The gradient is an operator applied to a scalar field and results in a vector field.
When we want to extend the definition of the derivative to an arbitrary surface, we first need to define the gradient of surfaces.
In the following, we make use of the \emph{covariant derivative}.
The standard directional derivative results in a vector which lies somewhere in the 3D space, whereas the covariant derivative is restricted to stay in the tangent space of the surface.
The gradient is a two-dimensional vector.
Actually, we need a three-dimensional vector in the tangent space of the surface.
Here, we employ the gradient and use it as coefficients of the tangential basis.
Unfortunately, this leads to wrong results because of the distortions of the basis of the tangent space, see Figure~\ref{fig:gradient}.
The basis is not necessarily an orthogonal normalized basis as in the domain space and can therefore lead to distortions of the gradient on the surface.
One way to calculate this vector is to use the plain scalar field $\varphi\colon \R^3\rightarrow\R$.
Afterwards, we are able to attain the gradient in three-dimensional space and project it on the tangent space.
However, we want to use the gradient of $\varphi\colon \R^2\rightarrow\R$ in the domain of a parametric surface and compensate the length distortion such that we can use it as coordinates with the basis in the tangent space.
One important fact is when multiplying the gradient with the $i$-th basis vector, one obtains the partial derivative of $\varphi$ with $x_i$.
Hence, we know that the three-dimensional gradient $\nabla\varphi$ lies in the tangent space.
Therefore, it can be represented as a linear combination of $\frac{\partial f}{\partial x_1},\frac{\partial f}{\partial x_2}$ with coefficients $\alpha,\beta$:
\[
\nabla\varphi=\alpha\cdot\frac{\partial f}{\partial x_1}+\beta\cdot\frac{\partial f}{\partial x_2}.
\]
Multiplying both sides with the basis vectors and using the relation $\frac{\partial \varphi}{\partial x_i}=\langle\nabla\varphi,\frac{\partial f}{\partial x_i}\rangle$, we obtain an equation system:
\begin{align*}
\begin{pmatrix}
\frac{\partial \varphi}{\partial x_1}\\ \frac{\partial \varphi}{\partial x_2}
\end{pmatrix} ~=~\begin{pmatrix}
\alpha\cdot \langle \frac{\partial f}{\partial x_1},\frac{\partial f}{\partial x_1}\rangle + \beta\cdot\langle\frac{\partial f}{\partial x_1},\frac{\partial f}{\partial x_2}\rangle\\
\alpha\cdot \langle \frac{\partial f}{\partial x_1},\frac{\partial f}{\partial x_2}\rangle + \beta\cdot\langle\frac{\partial f}{\partial x_2},\frac{\partial f}{\partial x_2}\rangle
\end{pmatrix} 
~=~\underbrace{\begin{pmatrix}
\langle \frac{\partial f}{\partial x_1},\frac{\partial f}{\partial x_1}\rangle & \langle\frac{\partial f}{\partial x_1},\frac{\partial f}{\partial x_2}\rangle\\
\langle \frac{\partial f}{\partial x_1},\frac{\partial f}{\partial x_2}\rangle & \langle\frac{\partial f}{\partial x_2},\frac{\partial f}{\partial x_2}\rangle
\end{pmatrix}}_{g\coloneqq}\begin{pmatrix}
\alpha\\ \beta
\end{pmatrix}.
\end{align*}
The matrix $g$ is called the \textit{metric tensor}.
This tensor describes the length and area distortion from $\R^2$ to the surface.
The last equation yields the coefficients $\alpha,\beta$ when multiplied with the inverse of $g$:
\[
\begin{pmatrix}
\alpha\\ \beta
\end{pmatrix}=g^{-1} \begin{pmatrix}
\frac{\partial \varphi}{\partial x_1}\\ \frac{\partial \varphi}{\partial x_2}
\end{pmatrix}.
\]
This leads to a general expression of the gradient for a scalar field $\varphi\colon\R^n\rightarrow\R$:
\begin{align}\label{gradient}
\nabla\varphi=\sum_{i,j=1}^n \Bigg(g^{ij}\frac{\partial\varphi}{\partial x_j}\Bigg)\frac{\partial}{\partial x_i},
\end{align}
where $g^{ij}$ is the $i,j$-th matrix entry from the inverse of $g$ and $\frac{\partial}{\partial x_i}$ means the basis.
Now, we are able to determine the covariant derivative of a scalar field by first determining its gradient and afterwards using the dot product:
\[
D_w\varphi=\langle\nabla\varphi,w\rangle.
\]
%
%
\subsection{Laplace-Beltrami Operator}\label{subsec:laplace_operator}
%
%
The Laplace-Beltrami operator is needed for a specific feature line method and will therefore be introduced.
The Laplace operator is defined as a composition of the \textit{gradient} and the \textit{divergence}.
When interpreting the vector field as a flow field, the divergence is a measure of how much more flow leaves a specific region than flow enters.
In the Euclidean space, the divergence $\diver \Phi$ of a vector field $\Phi\colon\R^n\rightarrow\R^n$ is the sum of the partial derivatives of the components $\Phi_i$:
\begin{align*}
\diver \Phi=\sum_{i=1}^n\frac{\partial}{\partial x_i} \Phi_i.
\end{align*}
The computation of the divergence for a vector field $\Phi\colon\R^n\rightarrow\R^n$ in Euclidean space is straightforward. 
However, for computing the divergence to an arbitrary surface we have to be aware of the length and area distortions.
Without giving a derivation of the divergence, the components $\Phi_i$ of the vector field have to be weighted by the square root of the determinant $\sqrt{\vert g\vert}$ of the metric tensor $g$ before taking the derivative.
The square root of the determinant of $g$ describes the distortion change from the Euclidean space to the surface.
Formally, the divergence of a vector field $\Phi\colon\R^n\rightarrow\R^n$ with a given metric tensor $g$ is given by:
\begin{align}
\diver\Phi=\frac{1}{\sqrt{\vert g\vert}}\sum_{i=1}^n \frac{\partial}{\partial x_i}\Bigg(\sqrt{\vert g\vert}~ \Phi_i \Bigg).
\end{align}
Given the definition of the gradient and the divergence, we can compose both operators to obtain the Laplace-Beltrami operator $\Delta \varphi$ of a scalar field $\varphi\colon\R^n\rightarrow\R$ on surfaces:
\begin{align}\label{laplace}
\Delta \varphi=\diver\nabla\varphi=\frac{1}{\sqrt{\vert g\vert}}\sum_{i,j=1}^n \frac{\partial}{\partial x_i}\Bigg(\sqrt{\vert g\vert}\, g^{ij}~ \frac{\partial \varphi}{\partial x_j} \Bigg).
\end{align} 
%
%
\subsection{Shape Operator}\label{subsec:weingarten}
%
%
In Section~\ref{subsec:curvature}, we noticed that the curvature of a parametric surface at a specific point $\pp$ in a certain direction can be determined by Equation~\ref{curv_surface_2}:
\[
\kappa_{c'(t)}(\pp)=-\langle c'(t),\frac{\partial \nn}{\partial t}\rangle.
\]
Actually, this means that the curvature in the direction $c'(t)$ is a measure of how much the normal changes in this direction, too.
Given is $\vv\in\mathcal{T}_\pp f$ with $\pp=f(\uu)$ and $\vert \vv\vert=1$.
Then, we determine the coefficients $\alpha,\beta$ of $\vv$ with the basis $\frac{\partial f}{\partial x_1},\frac{\partial f}{\partial x_2}$:
\[
\begin{pmatrix}
\alpha\\ \beta
\end{pmatrix}=g^{-1} \begin{pmatrix} \langle \vv,\frac{\partial f}{\partial x_1}\rangle\\ \langle\vv,\frac{\partial f}{\partial x_2}\rangle
\end{pmatrix}.
\]
We use $(\alpha,\beta)$ to determine the derivative of $\nn$ along $\vv$ by using the two-dimensional curve $\tilde c(t)=\uu+t\binom{\alpha}{\beta}$ and calculate:
\[
D_\vv\nn\coloneqq\frac{d}{dt}\nn(\tilde{c}(t)).
\]
We define $S(\vv)\coloneqq - D_\vv\nn$. 
This linear operator is called \textit{Shape Operator} (also \textit{Weingarten Map or Second Fundamental Tensor}).
One can see that $S(\frac{\partial f}{\partial x_i})=\frac{\partial \nn}{\partial x_i}$ holds.
Note that this operator can directly operate on the 3D space with a three-dimensional vector in the tangent space, as well as the 2D space with the coefficients of the basis.
Therefore, it can be represented by a matrix $S$.
Recall Equation~\ref{curv_surface_2}, we substitute $c'$ with $\vv$ and $\frac{\partial \nn}{\partial t}$ by $S\vv$:
\[
\kappa_\vv(\pp)=\langle\vv, S\vv\rangle.
\]
We want to show that the principle curvature directions are the eigenvectors of $S$.
Assuming $\vv_1,\vv_2\in\R^2$ are the normalized eigenvectors with the eigenvalues $\lambda_1\ge\lambda_2$.
Every normalized vector $\ww$ can be written as a linear combination of $\vv_1,\vv_2$: $\ww=\alpha\vv_1+\beta\vv_2$ with $\Vert w\Vert=\Vert \alpha\vv_1+\beta\vv_2\Vert=\alpha^2+\beta^2+2\alpha\beta\langle\vv_1,\vv_2\rangle=1$.
Therefore, we obtain:
\begin{align}
\kappa_\ww(\pp)&=\langle\ww, S\ww\rangle 
=\frac{1}{2}[(\alpha^2-\beta^2)(\lambda_1-\lambda_2)+\lambda_1+\lambda_2].\label{Eq:max_curv}
\end{align}
One can see from Equation~\ref{Eq:max_curv} that $\kappa_\ww(\pp)$ reaches a maximum if $\beta=0$, $\alpha=1$, and a minimum is reached if $\alpha=0$, $\beta=1$.
If the eigenvalues (curvatures) are not equal, we can show that the principle curvature directions are perpendicular.
For this, we need to show that $S$ is a self-adjoint operator. 
Thus, the equation $\langle S\vv,\ww\rangle=\langle\vv,S\ww\rangle$ holds.
We show this by using the property $\langle\nn,\frac{\partial f}{\partial x_i}\rangle=0$ and derive this with $x_j$:
\[
\langle\frac{\partial \nn}{\partial x_j}, \frac{\partial f}{\partial x_i}\rangle+\langle \nn, \frac{\partial^2 f}{\partial x_i\partial x_j}\rangle=0.
\]
We demonstrate that $S$ is a self-adjoint operator with the basis $\frac{\partial f}{\partial x_i}$:
\begin{align*}
\langle S(\frac{\partial f}{\partial x_i}),\frac{\partial f}{\partial x_j}\rangle&=\langle -\frac{\partial \nn}{\partial x_i},\frac{\partial f}{\partial x_j}\rangle =\langle \nn,\frac{\partial^2 f}{\partial x_i\partial x_j}\rangle =\langle -\frac{\partial \nn}{\partial x_j},\frac{\partial f}{\partial x_i}\rangle =\langle S(\frac{\partial f}{\partial x_j}),\frac{\partial f}{\partial x_i}\rangle.
\end{align*}
Now, we show that the eigenvectors (principle curvature directions) are perpendicular if the eigenvalues (curvatures) are different:
\[
\lambda_1\langle\vv_1,\vv_2\rangle=\langle S\vv_1,\vv_2\rangle=\langle \vv_1,S\vv_2\rangle=\lambda_2\langle \vv_1,\vv_2\rangle.
\]
The equation is only true if $\vv_1,\vv_2$ are perpendicular (and $\lambda_1\neq\lambda_2$ holds).

%% file: computer_graphic.tex
%
%
\section{Discrete Differential Geometry}\label{discrete_differential_geometry}
%
%

This section adapts the continuous differential geometry to discrete differential geometry, the area of polygonal meshes that approximate continuous geometries.
The following notation is used in the remainder of this paper.
Let $M\subset\R^3$ be a triangulated surface mesh.
The mesh consists of vertices $i\in V$ with associated positions
$\pp_i\in\R^3$, edges $E=\{(i,j)\,|\,i,j\in{}V\}$, and triangles
$T=\{(i,j,k)\,|\, (i,j),(j,k),(k,i)\in E\}$.
We write $\nn_i$ as the normalized normal vector at vertex $i$.
If nothing else is mentioned, we refer to normal vectors at vertices.
Furthermore, $\N(i)$ denotes the neighbors of $i$.
So, for every $j\in\N(i)$, $(i,j)\in E$ holds.
Furthermore, if we use a triangle for calculation, we always use this notation: given a triangle $\triangle=(i,j,k)$ with vertices $\pp_i,\pp_j,\pp_k$, and the edges are defined as $\ee_1=\pp_i-\pp_j,\ee_2=\pp_j-\pp_k,\ee_3=\pp_k-\pp_i$.
%
%
\subsection{Voronoi Area}\label{subsec:voronoi_area}
%
%
\begin{figure}[tb]
 \centering
  \subfigure[Points in 2D]{
 \includegraphics[scale=0.75]{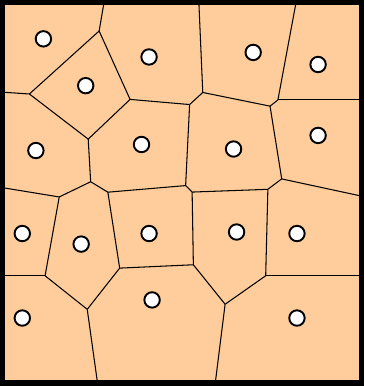}
  \label{fig:voroI}
  }\qquad
  \subfigure[Points on a surface mesh]{
 \includegraphics[scale=0.75]{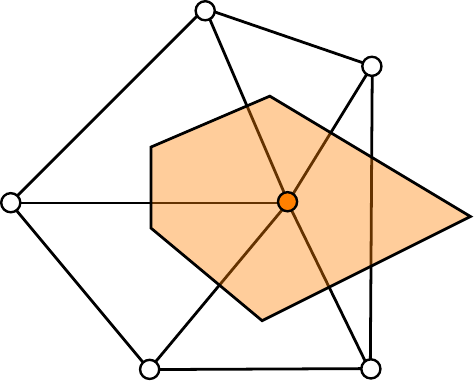}
 \label{fig:voroII}
 }
 \caption{The Voronoi diagram of different settings. In \subref{fig:voroI} a Voronoi diagram of a set of points is determined. In \subref{fig:voroII} the Voronoi area is calculated. If one of the triangles is obtuse, the area leaves the triangle.}
 \label{fig:voronoi}
\end{figure}
We need to introduce the term Voronoi area, as it is important for the determination of the curvature.
So, given are points in a 2D space.
Every point is spread out in equal speed.
If two fronts collide, they stop to spread out further at this region.
After all fronts stopped, every point lies in a region that is surrounded by a front.
This region is called a \textit{Voronoi region}.
Formally, given distinct points $\xx_i\in\R^2$ in the plane, the Voronoi region for the point $\xx_k$ is defined as the set of points $V(\xx_k)$ with
\begin{align*}
V(\xx_k)=\{\xx\in\R^2\,:~\Vert \xx-\xx_k\Vert\le\Vert \xx-\xx_j\Vert,\,j\ne k\}.
\end{align*}
See Figure~\ref{fig:voroI} for an example of a Voronoi diagram.
To obtain the Voronoi area of a vertex on a surface mesh, the Voronoi area of each incident triangle is accumulated.
The Voronoi area calculation is based on the method by Meyer et al.~\cite{Meyer2002}.
In case of a non-obtuse triangle, the Voronoi area at $\pp_i$ is determined by the perpendicular bisector of the edges incident to $\pp_i$.
The point of intersection, the midpoint of the incident edges and the point itself define the endpoints of the Voronoi area.
The triangle area of the Voronoi region equals:
\begin{align*}
\mathcal A_\triangle(\pp_i)=\frac{1}{8} \Big( \Vert \ee_1 \Vert^2\cdot\cot(\ee_2,\ee_3) + \Vert \ee_3 \Vert^2\cdot\cot(\ee_1,\ee_2) \Big).
\end{align*}
In case of an obtuse triangle, the Voronoi area is equal half of the triangle area if the angle at $\pp_i$ is obtuse.
Otherwise it is a quarter of the triangle area, see Figure~\ref{fig:voroII}.
%
\subsection{Discrete Curvature}\label{subsec:discrete_curvature}
%
%
The calculation of the curvatures as well as the principle curvature directions are important for a number of feature line techniques.
Several approaches exist to approximate the curvatures.
Some methods try to fit an analytic surface (higher order polynomials) to the mesh and determine the curvatures analytically \cite{Cazals2003,Goldfeather2004}.
Another approach estimates the normal curvature along edges first and then estimates the shape operator~\cite{Chen1992,Hameiri2003,Meyer2002,Page2001,Taubin1995_curvature}.
Other approaches are based on the calculation of the shape operator $S$~\cite{Alliez2003,Cohen-Steiner2003,Rusinkiewicz2004}.
We use the curvature estimation according to~\cite{Rusinkiewicz2004}.
After $S$ is determined on a triangle basis, it is adapted to vertices.
We already defined that $S\vv$ yields the change of the normal in the direction of $\vv$:
\[S\vv=D_\vv\nn.\]
This property is used to assess $S$ for each triangle.
When applying $S$ to the edge $\ee_1$, it should result in $\nn_i-\nn_j$ because of the change of the normals along the edge.
We need a basis of the tangent space of the triangle:
\[\tilde\ee_1=\frac{\ee_1}{\vert \ee_1\vert},~~~~~\tilde\ee_2=\frac{\ee_2}{\vert \ee_2\vert}.\]
Afterwards, we build the orthogonal normalized basis vectors $\xx_\triangle,\yy_\triangle$ by:
\begin{equation}
\begin{aligned}
\xx_\triangle\coloneqq\tilde\ee_1,~~~~~
\yy_\triangle\coloneqq \frac{\xx_\triangle\times(\tilde\ee_2\times\xx_\triangle)}{\Vert \xx_\triangle\times(\tilde\ee_2\times\xx_\triangle)\Vert}.
\end{aligned}
\label{triangle_basis}
\end{equation}

Applying the aforementioned property of the shape operator to all edges according to the basis leads to the following equation system:
\begin{align}\label{Eg:shape}
\begin{split}
S\begin{pmatrix} \langle \ee_1,\xx_\triangle \rangle\\ \langle \ee_1,\yy_\triangle \rangle \end{pmatrix}&=\begin{pmatrix} \langle \nn_i-\nn_j,\xx_\triangle \rangle\\ \langle \nn_i-\nn_j,\yy_\triangle \rangle \end{pmatrix}\\
S\begin{pmatrix} \langle \ee_2,\xx_\triangle \rangle\\ \langle \ee_2,\yy_\triangle \rangle \end{pmatrix}&=\begin{pmatrix} \langle \nn_j-\nn_k,\xx_\triangle \rangle\\ \langle \nn_j-\nn_k,\yy_\triangle \rangle \end{pmatrix}\\
S\begin{pmatrix} \langle \ee_3,\xx_\triangle \rangle\\ \langle \ee_3,\yy_\triangle \rangle \end{pmatrix}&=\begin{pmatrix} \langle \nn_k-\nn_i,\xx_\triangle \rangle\\ \langle \nn_k-\nn_i,\yy_\triangle \rangle \end{pmatrix},
\end{split}
\end{align}
\begin{wrapfigure}{r}{0.5\textwidth}
 \centering
 \includegraphics[scale=1]{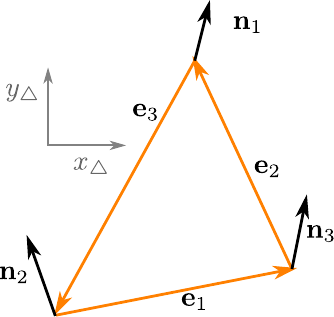}
 \caption{The shape operator estimation is based on a local coordinate system, the edges and the normals.}
 \label{fig:discrete_curvature}
\end{wrapfigure}
see Figure~\ref{fig:discrete_curvature} for an illustration.
Here, we have three unknowns (the matrix entries of the symmetric matrix $S=\binom{e~f}{f~g}$) and six linear equations.
Thus, a least square method can be applied to fit the shape operator to approximate curvature for each triangle.
Next, we need to calculate $S$ for each vertex of the mesh.
As the triangle basis normally differs from each vertex tangent space basis, we need to transform the shape operator according to the new coordinate system.
First, we assume that the normal $\nn_\triangle$ of the face is equal to the incident vertex normal $\nn_i$.
Hence, the basis $(\xx_\triangle,\yy_\triangle)$ of the triangle is coplanar to the basis $(\xx_i,\yy_i)$ of the incident vertex $i$.
Assuming we have the shape operator given in the vertex basis, then the entries can be determined by:
\begin{align*}
e_p&=\begin{pmatrix} 1 & 0 \end{pmatrix}\begin{pmatrix} e_p & f_p \\ f_p & g_p \end{pmatrix}\begin{pmatrix} 1\\0 \end{pmatrix} = \xx_i^T S\xx_i \\
f_p&=\begin{pmatrix} 1 & 0 \end{pmatrix}\begin{pmatrix} e_p & f_p \\ f_p & g_p \end{pmatrix}\begin{pmatrix} 0\\1 \end{pmatrix} = \xx_i^T S\yy_i \\
g_p&=\begin{pmatrix} 0 & 1 \end{pmatrix}\begin{pmatrix} e_p & f_p \\ f_p & g_p \end{pmatrix}\begin{pmatrix} 0\\1 \end{pmatrix} = \yy_i^T S\yy_i. \\
\end{align*}
As we have determined the shape operator in the basis $(\xx_\triangle,\yy_\triangle)$, we can express the basis of the vertex by expressing the new coordinate system with the old basis $\xx_i=\alpha\xx_\triangle+\beta\yy_\triangle$:
\begin{align*}
\alpha&=\langle\xx_i,\xx_\triangle\rangle\\
\beta&=\langle\xx_i,\yy_\triangle\rangle.\\
\end{align*}
The entry $e_p$ can be determined by:
\begin{align}\label{Eq:shape_newBasis}
e_p=\begin{pmatrix} \alpha & \beta \end{pmatrix}S\begin{pmatrix} \alpha\\\beta \end{pmatrix}.
\end{align} 
The other entries can be calculated by analogous calculations.
For the second case, we rotate the coordinate system of the triangle around the cross product of the normal such that the basis of the vertex and the triangle are coplanar.
Finally, we use this to determine the shape operator of the vertices.
We determine the shape operators for all incident triangles of a vertex.
Afterwards, we rotate the coordinate systems of the triangles to be coplanar with the basis of the vertex.
Next, we re-express the shape operator in terms of the basis of the vertex.
Then, we weight the shape operator according to the Voronoi area of the triangle and accumulate this tensor.
Finally, we divide the accumulated shape operator by the sum of the weights.
The eigenvalues provide the principle curvatures, and the eigenvectors give the principle curvature directions according to the basis.
The pseudo-code~\ref{algo:shape_operator} summarizes the algorithm.
\begin{algorithm}
\caption{Pseudo-code for the curvature estimation.}
\label{algo:shape_operator}
\begin{lstlisting}[boxpos=b,mathescape=true] 
for each triangle:
   $\text{Build basis accord. to Eq.~\ref{triangle_basis}}$
   $\text{Determine S accord. to Eq.~\ref{Eg:shape}}$
   for each vertex incident to the triangle:
      $\text{Rotate the triangle basis to the vertex basis}$
      $\text{Determine S in the new basis accord. to Eq.~\ref{Eq:shape_newBasis}}$
      $\text{Add this tensor weighted by the voronoi area}$
   end
end	
for each vertex:
   $\text{Divide S by the sum of the weights}$
   $\text{Determine the eigenvalues and eigenvectors}$
end   

\end{lstlisting}
\end{algorithm}

Please note that this algorithm can be generalized to obtain higher-order derivatives.
It can be used to determine the derivative of the curvature as it is important for a specific feature line method.
Formally, the derivative of the shape operator has the form:
\begin{align}
C=\begin{pmatrix} D_\vv S & D_\ww S \end{pmatrix}=\begin{pmatrix}\label{Eq:C}
\begin{pmatrix}
a & b \\ b & c
\end{pmatrix} ~
\begin{pmatrix}
b & c \\ c & d
\end{pmatrix}
\end{pmatrix}.
\end{align}
For the determination of the change of the curvature in direction $\uu$, the tensor $C$ has to be multiplied multiple times:
\begin{align}
D_\uu\kappa=\langle\uu,\begin{pmatrix}\label{Eq:D_ukappa}
D_\vv S\cdot\uu & D_\ww S\cdot\uu
\end{pmatrix}\cdot\uu\rangle.
\end{align}
%
%
\subsection{Discrete Covariant Derivative}\label{subsec:discrete_covariant_derivative}
%
%
First, we consider a linear 2D scalar field $\varphi(\xx)=\alpha\cdot x_1+\beta\cdot x_2+\gamma$ and its gradient:
\begin{align}\label{linear_gradient}
\nabla \varphi=\begin{pmatrix} \frac{\partial}{\partial x_1}\varphi\\\frac{\partial}{\partial x_2}\varphi \end{pmatrix}=\begin{pmatrix} \alpha\\\beta \end{pmatrix}.
\end{align}
To determine the gradient of a triangle $\triangle=(i,j,k)$ with scalar values $\varphi_i\coloneqq\varphi(\pp_i)$,  $\varphi_j\coloneqq\varphi(\pp_j)$, and  $\varphi_k\coloneqq\varphi(\pp_k)$, we build a basis according to Equation~\ref{triangle_basis} and transform the points $\pp_i,\pp_j,\pp_k\in\R^3$ to $\pp'_i,\pp'_j,\pp'_k\in\R^2$ by:
\begin{align*}
\pp'_i =\begin{pmatrix} 0\\0  \end{pmatrix}\qquad \pp'_j =\begin{pmatrix} \langle\pp_j-\pp_i,\xx_\triangle\rangle\\\langle\pp_j-\pp_i,\yy_\triangle\rangle  \end{pmatrix} \qquad \pp'_k =\begin{pmatrix} \langle\pp_k-\pp_i,\xx_\triangle\rangle\\\langle\pp_k-\pp_i,\yy_\triangle\rangle  \end{pmatrix}.
\end{align*}
This transformation describes an isometric and conformal map.
The next step is a linearization of the scalar values $\varphi_i,\varphi_j,\varphi_k$.
We want to determine a scalar field $\varphi'(\xx')=\alpha\cdot x_1' + \beta\cdot x_2' + \gamma$ such that 
\begin{align*}
\varphi'(\pp'_i) = \varphi_i\qquad
\varphi'(\pp'_j) = \varphi_j\qquad
\varphi'(\pp'_k) = \varphi_k
\end{align*}
holds.
These conditions yield the following equation system:
\begin{align*}
\begin{pmatrix} \alpha & \beta \end{pmatrix} \begin{pmatrix} \pp'_i & \pp'_j & \pp'_k \end{pmatrix} + \begin{pmatrix} \gamma & \gamma & \gamma \end{pmatrix} =
\begin{pmatrix} \varphi_i & \varphi_j &\varphi_k \end{pmatrix}.
\end{align*}
With $\pp'_i=\tbinom{0}{0}$ we obtain the following solution:
\begin{align*}
\gamma&=\varphi_i, \\
\begin{pmatrix} \alpha & \beta \end{pmatrix} &= \begin{pmatrix} \varphi_j-\varphi_i & \varphi_k-\varphi_j \end{pmatrix} \begin{pmatrix} \pp'_j & \pp'_k \end{pmatrix} ^{-1}.
\end{align*}
According to Equation~\ref{linear_gradient}, the gradient of $\varphi'$ is determined by $\tbinom{\alpha}{\beta}$.
\begin{wrapfigure}{r}{0.5\textwidth}\label{fig:discrete_gradient}
 \centering
 \includegraphics[scale=0.75]{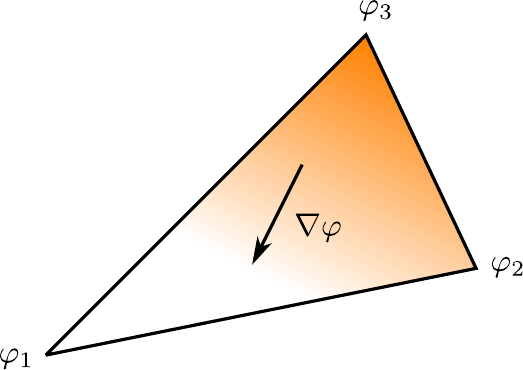}

\caption{A triangle with different scalar values.}
 \label{fig:discrete_gradient}
\end{wrapfigure}
The basis $\xx_\triangle,\yy_\triangle$ yields the gradient in 3D:
\begin{align*}
\nabla\varphi=\alpha\cdot\xx_\triangle+\beta\cdot\yy_\triangle.
\end{align*}
Figure~\ref{fig:discrete_gradient} illustrates the gradient of a triangle.
To determine the gradient per vertex, we use the same procedure as for the shape operator estimation.
We transform the basis and weight the triangle gradient according to its proportion of the Voronoi area.
%
%
\subsection{Discrete Laplace-Beltrami Operator}\label{subsec:discrete_laplace_operator}
%
%
Several methods exist to discretize the Laplace-Beltrami operator on surface meshes.
For an overview, we recommend the state of the art report by Sorkine~\cite{Sorkine2005}.
The operator can be presented by the generalized formula:
\begin{align*}
\Delta \varphi (\pp_i) = \sum_j w_{ij}~ \Big(\varphi(\pp_j)-\varphi(\pp_i)\Big).
\end{align*}
Different weights $w_j$ give different discrete Laplace-Beltrami operators.
For presenting different versions of this operator it is preferable that it fulfills some properties motivated by the smooth Laplace-Beltrami operator:
\begin{itemize}
\item[]\textbf{(Sym)} The weights should be symmetric $w_{ij}=w_{ji}$.
\item[]\textbf{(Loc)} If $(i,j)\not\in E$ then $w_{ij}=0$.
\item[]\textbf{(Pos)} All weights should be non-negative.
\item[]\textbf{(Lin)} If $\pp_i$ is contained in a plane and $\varphi$ is linear, then $\Delta\varphi(\pp_i)=0$ should hold.
\end{itemize}
In the following, we introduce different discrete Laplace-Beltrami operators.\\ \\
\textbf{Combinatorial:} For the combinatorial Laplace-Beltrami operator we have:
\begin{align*}
w_{ij}=\begin{cases}  1,&\text{if}~(i,j)\in E \\
 0,&\text{otherwise}.
\end{cases}
\end{align*}
This version may result in non-zero values on planar surfaces for linear scalar fields.
Therefore, it violates (Lin).\\ \\
\textbf{Uniform:} Taubin~\cite{Taubin1995} suggested the uniform Laplace-Beltrami operator.
The weights are determined by the number of neighbors of $\pp_i$:
\begin{align*}
w_{ij}=\begin{cases}  \frac{1}{\neigh(i)},&\text{if}~(i,j)\in E \\
 0,&\text{otherwise}.
\end{cases}
\end{align*}
These weights also violate (Lin).\\ \\
\begin{figure}[tb]
 \centering
 \includegraphics[scale=1]{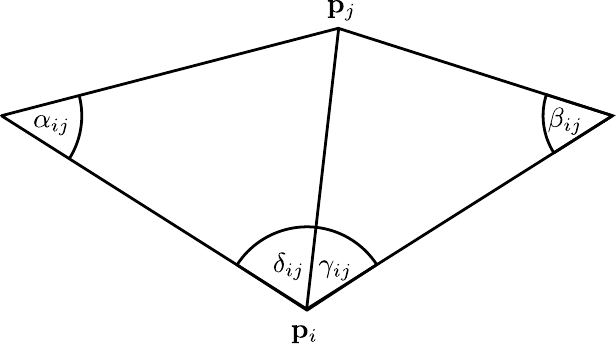}
 \caption{This figure illustrates the triangles with the angles for the weight calculation.}
 \label{fig:laplace_weight}
\end{figure}
\textbf{Floater's mean value:} Floater~\cite{Floater2003} proposed the mean value weights by the tangent of the corresponding angles:
\begin{align*}
w_{ij}=\begin{cases}  \frac{\tan(\delta_{ij}/2)+\tan(\gamma_{ij}/2)}{\Vert \pp_i-\pp_j\Vert},&\text{if}~(i,j)\in E \\
 0,&\text{otherwise}.
\end{cases}
\end{align*}
See Figure~\ref{fig:laplace_weight} for the angles.
These weights violate (Sym). \\ \\
\textbf{Cotangent weights:} MacNeal~\cite{Macneal1949} suggested the cotangent weights:
\begin{align*}
w_{ij}=\begin{cases} \cot(\alpha_{ij})+\cot(\beta_{ji}),&\text{if}~(i,j)\in E \\
 0,&\text{otherwise}.
\end{cases}
\end{align*}
See Figure~\ref{fig:laplace_weight} for the angles.
On general meshes the weights can violate (Pos).\\ \\
\textbf{Belkin weights:} Belkin~\cite{Belkin2008} suggested to determine weights over the whole surface:
\begin{align*}
\Delta\varphi(\pp_i)=\frac{1}{4\pi h^2(\pp_i)}\sum_{\triangle_k}\frac{A(\triangle_k)}{3}\sum_{j\in\triangle_k} e^{-\frac{\Vert \pp_i-\pp_j\Vert^2}{4h(\pp_i)}}\Big(\varphi(\pp_j)-\varphi(\pp_i) \Big),
\end{align*}
where $A(\triangle_k)$ denotes the area of the triangle $\triangle_k$ and $h$ corresponds intuitively to the size of the neighborhood.
This violates the (Loc) property.\\ \\
\textbf{Results:} The discussion leads to the question if there is any discrete Laplace-Beltrami operator which fulfills all required properties for an arbitrary surface mesh.
Wardetzky et al.~\cite{Wardetzky2007} showed that there is no such operator.
The proof is based on a Sch\"onhardt polytope which demonstrate that there is no Laplace-Beltrami operator, which does not violate any condition.
%
%
\subsection{Isolines on Discrete Surfaces}\label{subsec:isolines}
%
%

\begin{figure}[tb]
 \centering
  \subfigure[]{
 \includegraphics[width=0.55\columnwidth]{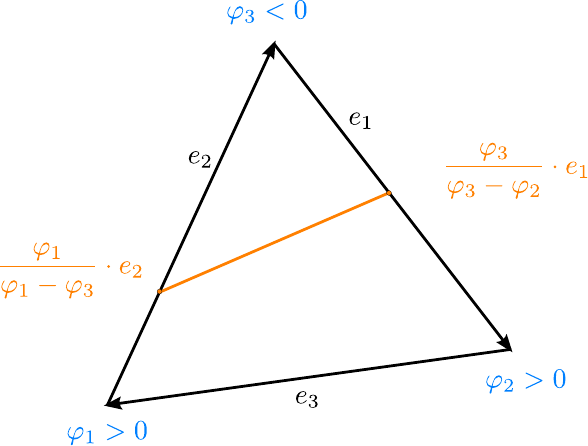}
  \label{fig:isoI}
  }~
  \subfigure[]{
 \includegraphics[width=0.425\columnwidth]{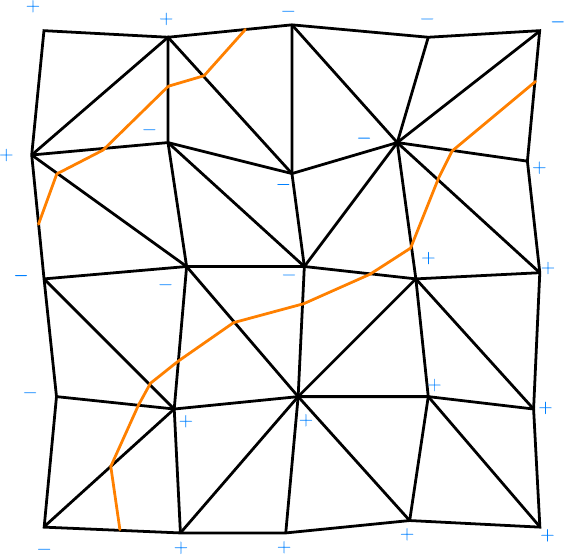}
 \label{fig:isoII}
 }
 \caption{In \subref{fig:isoI} the position of the zero crossing is determined and the points are connected. In \subref{fig:isoII} the isoline through a mesh is depicted. }
 \label{fig:isolines}
\end{figure}

For feature line methods, it is essential not to restrict the lines to the edges, as it is not desirable to perceive the mesh edges.
Given is a surface mesh and a scalar field, we want to depict the zero crossing of the scalar field.
Therefore, we linearize the scalar values for each triangle according to the values of the incident points.
Afterwards, we look for points on an edge such that the linearized values of the scalar values of the connecting points are equal to zero.
Having two points on two edges of a triangle, we connect them.
Suppose we have a triangle with scalar values $\varphi_i>0$, $\varphi_j>0$ and $\varphi_k<0$.
Thus, we know that somewhere on edge $\ee_2$ and $\ee_3$ there is a zero crossing.
We determine $t=\frac{\varphi_k}{\varphi_k-\varphi_j}$ and multiply $t$ with edge $\ee_2$.
This yields the position of the zero crossing on the edge.
The position on the edge $\ee_3$ is determined as well.
Afterwards, both points will be connected, see Figure~\ref{fig:isolines}.

%% file: aspects.tex
%
%
\section{General Requirements of Feature Lines}\label{aspects}
%
%
The generation of feature lines leads to several requirements, which have to be considered for acquiring appropriate results.\\ \\
\textbf{Smoothing:}
Most of the feature line methods use higher order derivatives.
Therefore, the methods assume sufficiently smooth input data.
For data acquired with laser scanners or industrial measurement process, smoothness cannot be expected.
Discontinuities represent high frequencies in the surface mesh and lead to the generation of distracting (and erroneous) lines.
Several algorithms exist, which smooth the surface by keeping relevant features.
Depending on the feature line method, different smoothing algorithms can be applied.
If the algorithm only uses the surface normals and the view direction, it is sufficient to simply smooth the surface normals.
Geometry-based approaches, however, require to smooth the mesh completely.
Operating only on scalar values, an algorithm which smoothes the scalar field around a certain region may be applied, too.\\ \\
\textbf{Frame coherence:} 
The application of feature line approaches or in general for non-photorealistic rendering makes it crucial to provide methods that are frame-coherent.
This means, during the interaction the user should not be distracted by features that pop out or disappear suddenly.
A consistent and continuous depiction of features should be provided in consecutive frames of animation.\\ \\
\textbf{Filtering:} 
Feature line algorithms may generate lines on salient regions as well as lines that result from small local irregularities, which may not be necessary to convey the surface shape or even annoying and distracting.
Filtering of feature lines to set apart relevant lines from distracting ones is a crucial part of a feature line generation.
User-defined thresholds may control the rate of tolerance for line generation.
Some algorithms use an underlying scalar field for thresholding.
Lines are only drawn if the corresponding scalar value exceeds the user-defined threshold.
Other methods integrate along a feature line, determine the value, and decide to draw the whole line instead of filtering some parts.
We will also mention the filtering method of each presented feature line generation method.

%% file: feature_lines.tex
%
%
\section{Feature Lines}\label{feature_lines}
%
%
Line drawings were used extensively for medical visualization tasks, such as displaying tissue boundaries in volume data~\cite{Burns2005,Treavett2000}, vascular structures~\cite{Ritter2006}, neck anatomy~\cite{Krueger2005} and brain data~\cite{Jainek2008,Svetachov2010}.
Furthermore, some higher order feature lines were qualitatively evaluated on medical surface data~\cite{Lawonn2013BVM}.
The importance of feature lines in medical visualization is discussed in~\cite{Preim2014}.
Feature line methods can be divided into \emph{image-based} and \emph{object-based} methods.
Image-based approaches are not in the focus of this survey.
These methods are based on an image as input.
All calculations are performed on the image with the pixels containing, for instance, an RGB or grey value.
Usually, the image is convolved with different kernels to obtain the feature lines.
The resulting feature lines are represented by pixels in the image space.
These lines are mostly not frame-coherent.
Comprehensive overviews of different feature line methods in image space are given by~\cite{Muthukrishnan2011,Nadernejad2008,Senthilkumaran2009}. 
This section presents selected object-based feature line methods.
We will explain the methods and limitations.
Further information on line drawings can be found in ~\cite{Preim2014,Rusinkiewicz2008}.
%
%
\subsection{Contours}\label{subsec:contour}
%
%
We refer to a silhouette as a depiction of the outline of an object as this is the original definition by \'Etienne de Silhouette.
The contour is defined as the loci of points where the normal vector and the view vector are mutually perpendicular:
\begin{wrapfigure}{r}{0.3\textwidth}\label{fig:brain_co}
 \centering
 \includegraphics[width=0.3\textwidth]{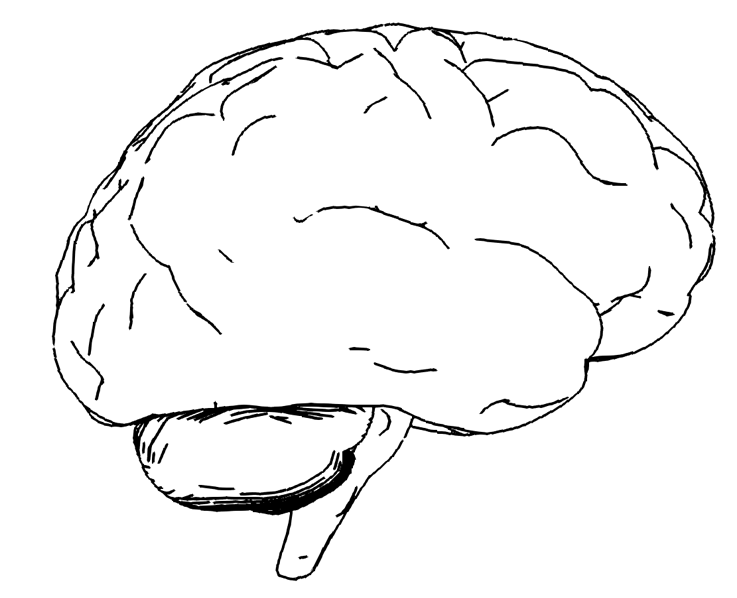}
\caption{The brain model with contours.}
\end{wrapfigure}
\begin{align*}
\boxed{\langle\nn,\vv\rangle=0,}
\end{align*}
where $\nn$ is the normal vector and $\vv$ is the view vector which points towards the camera.
For the discrete case, we highlight edges as a contour whenever the sign of the dot product of the view vector with the normals of the incident triangle normals changes.
The contour yields a first impression of the surface mesh.
On the other hand, it is not sufficient to depict the surface well.
The contour is not appropriate to gain a spatial impression of the object.
Furthermore, it cannot depict salient regions, for instance strong edges.\\
\textbf{Summary:}
In the first place, the contour is necessary for gaining a first impression on the shape of the object.
Unfortunately, spatial cues, as for instance strong edges, are not depicted.
%
%
\subsection{Crease Lines}\label{subsec:creases_lines}
%
%
Crease lines are a set of edges where incident triangles change strongly.
The dihedral angle, i.e., the angle of the normals of the corresponding incident triangles, along the edges is calculated.
The edge belongs to a crease line if the dihedral angle exceeds a user-defined threshold $\tau$.
As the change of the normals is an indicator of the magnitude of the curvature, one can state that all points contribute to a feature line if the underlying absolute value of the maximum curvature exceeds a threshold:
\begin{align*}
\boxed{\kappa_i\ge\tau~\text{ or }~\langle\nn_i,\nn_j\rangle\ge\tau',}
\end{align*}
\begin{wrapfigure}{r}{0.3\textwidth}\label{fig:brain_cr}
 \centering
 \includegraphics[width=0.3\textwidth]{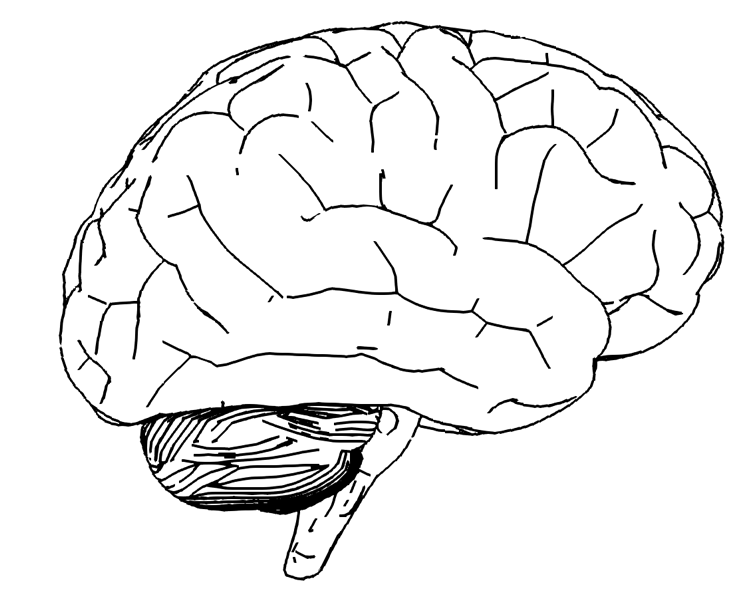}
\caption{The brain model with crease lines and contours.}
\end{wrapfigure}
for adjacent triangles with corresponding normals $\nn_i,\nn_j$.
Afterwards, all adjacent vertices which fulfill the property are connected.
These feature lines need to be computed only once, since they are not view-dependent.
Furthermore, these lines are only drawn along edges.\\
\textbf{Summary:}
Crease lines display edges where the dihedral angle is large.
Strong edges are appropriately depicted, but if the object has small features, this method is not able to depict only important edges.
This is caused by the local determination of the dihedral angle without concerning a neighborhood.
Even smoothing the surface mesh would not deliver proper line drawings.
Furthermore, this method is only able to detect features on edges.
%
%
\subsection{Ridges and Valleys}\label{subsec:ridges_and_valleys}
%
%
\begin{wrapfigure}{r}{0.3\textwidth}\label{fig:brain_rv}
 \centering
 \includegraphics[width=0.3\textwidth]{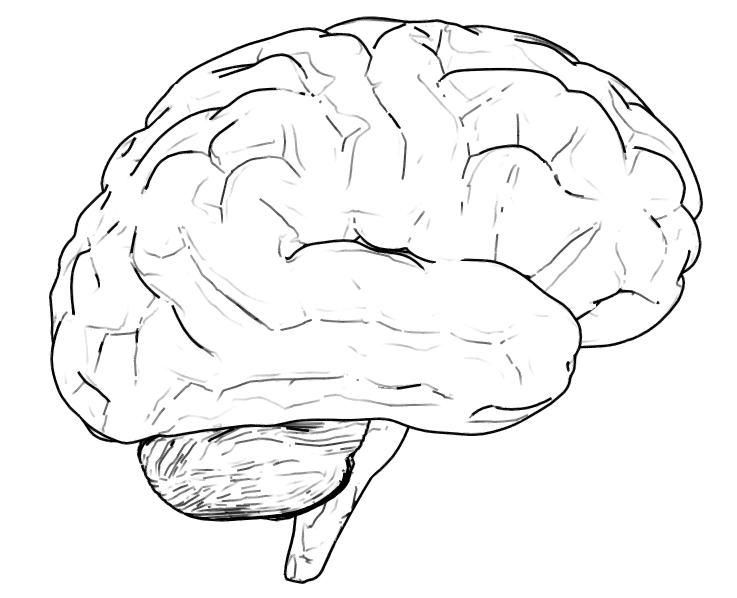}
\caption{The brain model with ridges and valleys, and contours.}
\end{wrapfigure}
\emph{Ridges and valleys} were proposed by Interrante et al.~\cite{Interrante1995} and adapted to triangulated surface meshes by Ohtake et al.~\cite{Ohtake2004}.
These feature lines are curvature-based and not view-dependent.
The computation is based on the principle curvature $\kappa_1$ as well as the associated principle curvature direction $\kk_1$ with $\vert\kappa_1\vert\ge\vert\kappa_2\vert$. 
Formally, ridges and valleys are defined as the loci of points at which the principle curvatures assume an extremum in the principle
direction:
\begin{align*}
\boxed{D_{\kk_1}\kappa_1=0.}
\end{align*}
According to two constraints, the sets of points are called
\begin{align}\label{ridge_valley_condition}
D_{\kk_1}D_{\kk_1}\kappa_1\begin{cases}
  <0,  & \text{and}~\,\kappa_1>0\text{: ridges}\\
  >0,  & \text{and}~\,\kappa_1<0\text{: valleys.}
\end{cases}
\end{align}
To determine the ridge and valley lines, we first need to compute the principle curvatures and their associated principle curvature directions, recall Section~\ref{subsec:discrete_curvature}.
Afterwards, we determine the gradient of $\kappa_1$ for each vertex, see Section~\ref{subsec:discrete_covariant_derivative}.
Finally, we compute the dot product of the gradient and the associated principle curvature direction $\kk_1$.
This yields the scalar value of $D_{\kk_1}\kappa_1$ for each vertex.
Next, we distinguish between ridges and valleys and determine $D_{\kk_1}D_{\kk_1}\kappa_1$ for each vertex.
Here, we need again the gradient of each vertex with the value $D_{\kk_1}\kappa_1$ and determine the dot product of the result with $\kk_1$.
Hence, we gain two scalar values per vertex: $D_{\kk_1}\kappa_1$ and $D_{\kk_1}D_{\kk_1}\kappa_1$.
Afterwards, we assess the zero-crossing of the first scalar value, recall Section~\ref{subsec:isolines}.
We connect the zero crossings in every triangle for which one condition of Equation~\ref{ridge_valley_condition} holds.
The filtering of the lines is again performed by employing an user-defined threshold.
The integral along each ridge and valley line is determined according to the underlying curvature.
If the magnitude of the integral exceeds the threshold for ridges or valleys, the line is drawn.\\
\textbf{Summary:}
The calculation is solely based on the curvature and therefore view-independent.
This method is able to detect small features.
The filtering depends on the underlying curvature and the length of the curve.
Therefore, a long line with small curvature has also the chance to be drawn as a small line with high curvature.
This strategy emphasizes also long feature lines.
Ridges and valley lines are very susceptible to noise, since this method is of 3rd order.
Therefore, small discontinuities on the surface mesh lead to erroneous derivatives and this error propagates for each further derivative.
A crucial task for this method is to guarantee a smoothed mesh to obtain reasonable results. 
From an artist's point of view, some features may be more highlighted than others from different points of view.
This is caused by the different perception of an object and by various light positions.
For this task, the ridge and valley lines are not appropriate due to the restriction of view-independent results.
%
%
\subsection{Suggestive Contours}\label{subsec:suggestive_contours}
%
%
\begin{wrapfigure}{r}{0.3\textwidth}\label{fig:brain_sc}
 \centering
 \includegraphics[width=0.3\textwidth]{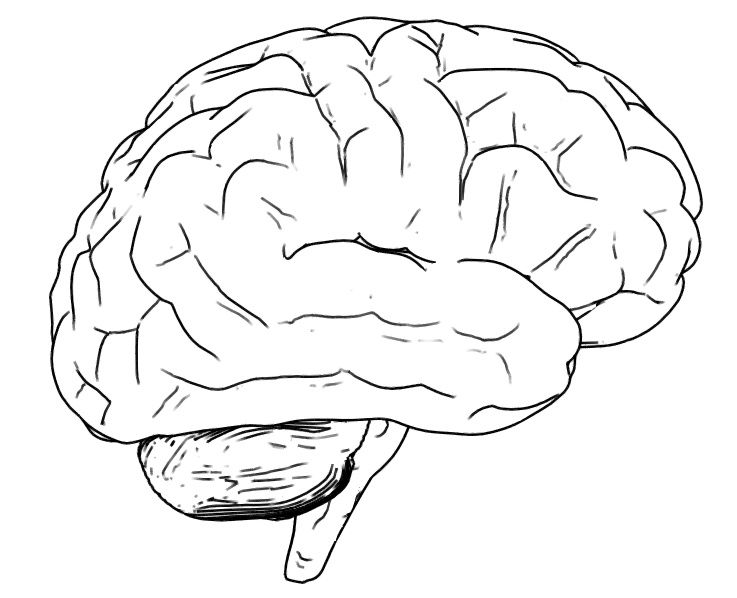}
\caption{The brain model with suggestive contours and contours.}
\end{wrapfigure}
\emph{Suggestive contours} are view-dependent feature lines introduced by DeCarlo et al.~\cite{DeCarlo2003}.
They extend the definition of the contour.
These lines are defined as the set of minima of $\langle\nn,\vv\rangle$ in the direction of $\ww$, where $\nn$ is the surface normal, $\vv$ is the view vector which points towards the camera, and $\ww=(\id-\nn\nn^T)\vv$ is the projection of the view vector on the tangent plane.
Formally:
\begin{align*}
\boxed{D_\ww\,\langle\nn,\vv\rangle = 0~\text{ and }~D_\ww D_\ww\,\langle\nn,\vv\rangle >0.}
\end{align*}
Another equivalent definition of the suggestive contours is given by the radial curvature $\kappa_r$.
It is defined as the curvature in direction of $\ww$.
As seen in Equation~\ref{curvature_direction}, this curvature can be determined by knowing the principle curvature directions as well as the corresponding curvatures. 
Therefore, the definition of the suggestive contours is equivalent to the set of points at which the radial curvature $\kappa_r$ is equal 0 and the directional derivative of $\kappa_r$ in direction $\ww$ is positive:
\begin{align*}
\boxed{\kappa_r = 0~\text{ and }~D_\ww\kappa_r >0.}
\end{align*}
The filtering strategy is to apply a small threshold to eliminate suggestive contour points where the radial curvature in direction of the projected view vector is very low.
Additionally, a hysteresis threshold is applied to increase granularity.\\
\textbf{Summary:}
Suggestive contours extend the normal definition of the contour.
This method depicts zero crossing of the diffuse light in view direction.
This can be seen as inflection points on the surface.
This method is of 2nd order only and thus less susceptible to noise.
Unfortunately, suggestive contours are not able to depict some sorts of sharp edges, which are in fact noticeable features.	
For instance, a rounded cube has no suggestive contours.
%
%
\subsection{Apparent Ridges}\label{subsec:apparent_ridges}
%
%
\begin{wrapfigure}{r}{0.3\textwidth}\label{fig:brain_ar}
 \centering
 \includegraphics[width=0.3\textwidth]{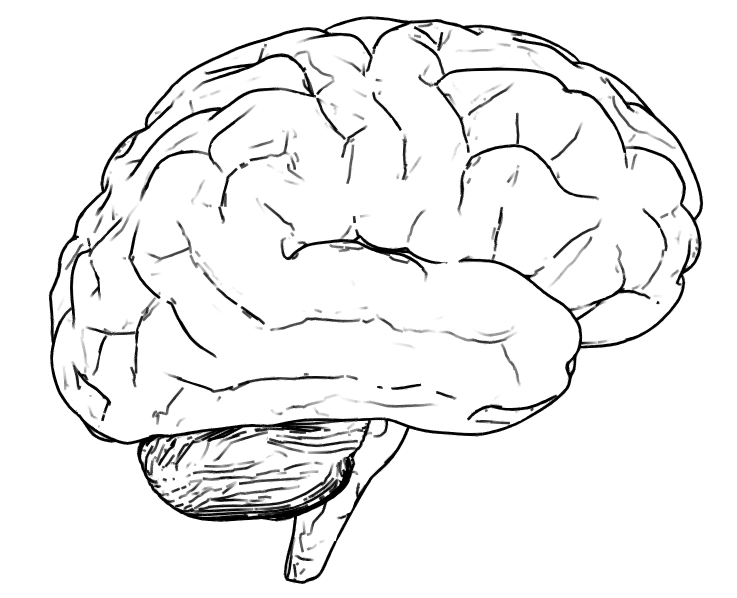}
\caption{The brain model with apparent ridges.}
\end{wrapfigure}
\emph{Apparent ridges} were proposed by Judd et al.~\cite{Judd2007}.
These feature lines extend the definition of ridges by a view-dependent curvature term.
Therefore, a projection operator $P$ is used to map the vertices on a screen plane $V$.
The orthonormal basis of the screen plane is given by $(\vv_1,\vv_2)$.
Assume we have a parametrized surface $f\colon I\subset\R^2\rightarrow\R^3$.
Then the projection of $f$ onto $V$ is given by:
\begin{align*}
P(\xx)=\begin{pmatrix}
\langle \vv_1,f(\xx)\rangle\\ \langle\vv_2,f(\xx)\rangle
\end{pmatrix}.
\end{align*}
The Jacobian $J_P$ of $P$ can be expressed as:
\begin{align*}
J_P=\begin{pmatrix}
\langle \vv_1, \frac{\partial f}{\partial x_1}\rangle &\langle \vv_1, \frac{\partial f}{\partial x_2}\rangle\\
\langle \vv_2, \frac{\partial f}{\partial x_1}\rangle &\langle \vv_2, \frac{\partial f}{\partial x_2}\rangle
\end{pmatrix}.
\end{align*}
In the discrete case with surface meshes, the Jacobian can be expressed by a basis for the tangent plane $(\ee_1,\ee_2)$:
\begin{align*}
J_P=\begin{pmatrix}
\langle \vv_1, \ee_1\rangle &\langle \vv_1, \ee_2\rangle\\
\langle \vv_2, \ee_1\rangle &\langle \vv_2, \ee_2\rangle
\end{pmatrix}.
\end{align*}
If a point $\pp'$ on the screen plane is not a contour point, there exists a small neighborhood where the inverse of $P$ exists.
Normal vectors $\nn'$ at a point $\pp'$ on the screen plane are defined as $\nn'(\pp')\coloneqq \nn(P^{-1}(\pp'))$.
The main idea is to build a view-dependent shape operator $S'$ at a point $\pp'$ on the screen as
\begin{align*}
S'(\ww')=D_{\ww'}\nn'
\end{align*}
where $\ww'$ is a vector in the screen plane.
The view-dependent shape operator is therefore defined as:
\begin{align*}
S'=S\,J_P^{-1}.
\end{align*} 
Here, the basis of the tangent space expressing $S$ and $J_P$ must be the same.
In contrast to the shape operator, the view-dependent shape operator is not a self-adjoint operator, recall Section~\ref{subsec:weingarten}.
Therefore, it is not guaranteed that $S'$ has two eigenvalues, but it has a maximum singular value $\kappa_1'$:
\begin{align*}
\kappa_1'=\max_{\Vert \ww\Vert=1}\Vert S'(\ww')\Vert.
\end{align*}
This is equivalent to find the maximum eigenvalue of $S'^TS'$ and to take the square root.
The corresponding singular eigenvector $\ttb'$ is called the maximum view-dependent principle direction.
The rest of the method is similar to the ridge and valley methods.
Formally, apparent ridges are defined as the loci of points at which the view-dependent principle curvature assumes an extremum in the view-dependent principle
direction:
\begin{align*}
\boxed{D_{\ttb'}\kappa_1'=0~\text{ and }~D_{\ttb'}D_{\ttb'}\kappa_1' <0.}
\end{align*}
The sign of $\kappa'$ is always positive.
To distinguish between ridge lines and valley lines, we may compare the sign of the object-space curvature:
\begin{align*}
\kappa_1\begin{cases}
  <0, &\text{ridges}\\
  >0, &\text{valleys.}
\end{cases}
\end{align*}
The calculation of the directional derivative is different from the other methods.
This calculation is performed with finite differences.
Therefore, we transform the singular eigenvector $\ttb'$ to object space $\ttb$ using the corresponding basis of the associated vertex $i$.
Furthermore, we need the opposite edges of the vertex and determine two points $\ww_1$, $\ww_2$ on the edges such that $\ttb$ and the edges are orthogonal and $\ww_1$, $\ww_2$ are the dropped perpendiculars of $\ttb$ to the corresponding edges.
The directional derivatives are determined by averaging the finite differences of the curvatures between $\pp_i$ and $\ww_1$, $\ww_2$.
The curvature of $\ww_1$, $\ww_2$ is assessed by linear interpolation of the endpoints of the associated edge.
Having the principle view-dependent curvature direction $\ttb'$, we need to make it consistent over the mesh because it is not well-defined.
Therefore, $\ttb'$ is flipped in opposite direction whenever it does not point to the direction where the view-dependent curvature is increasing.
The zero-crossings are determined by checking if the principle view-dependent curvature directions of the vertices along an edge point are in the same direction.
Only in this case there is no zero-crossing.
Pointing in different directions means that the enclosing angle is greater than 90 degrees.
The zero crossing is determined by interpolating the values of the derivatives.
To locate only maxima, a perpendicular is dropped from each vertex to the zero crossing line.
If the perpendiculars of the vertices of an edge make an acute angle with their principle view-dependent curvature directions, the zero crossing is a maximum.
Otherwise, the zero crossing is a minimum.
To eliminate unimportant lines, a threshold based on the view-dependent curvature is used.\\
\textbf{Summary:}
Apparent ridges incorporate the advantages of the ridges and valley lines as well as the view dependency.
They extend the ridge and valley definition by introducing view-dependent curvatures.
This method is able to depict salient regions as sharp edges.
Unfortunately, the 3rd order computation leads to low frame rates and to visual clutter if the surface mesh is not sufficiently smoothed.
%
%
\subsection{Photic Extremum Lines}\label{subsec:photic_extremum_lines}
%
%
\begin{wrapfigure}{r}{0.3\textwidth}\label{fig:brain_pel}
 \centering
 \includegraphics[width=0.3\textwidth]{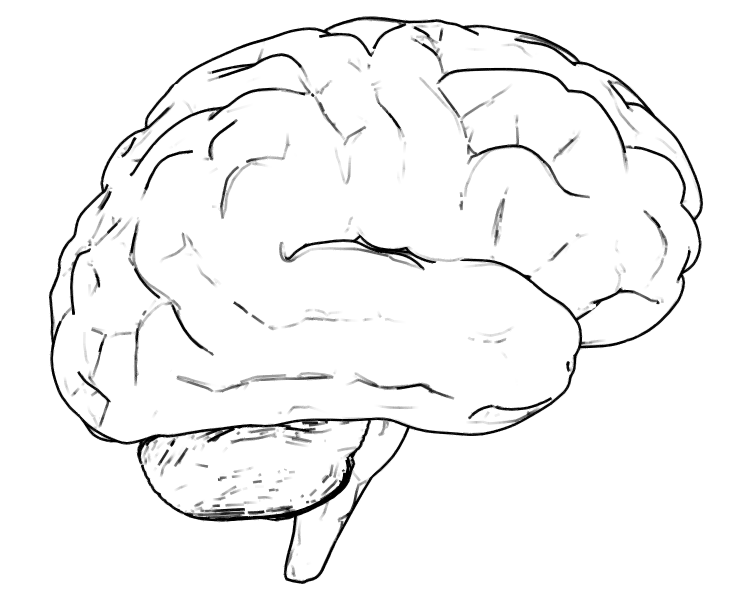}
\caption{The brain model with photic extremum lines.}
\end{wrapfigure}
\emph{Photic extremum lines} (PELs) were introduced by Xi et al.~\cite{Xi2007}.
These feature lines depict regions of the surface mesh with significant variations of illuminations.
This method is based on the magnitude of the light gradient.
Formally, these lines are defined as the set of points where the variation of illumination along its gradient direction reaches a local maximum:
\begin{align*}
\boxed{D_\ww\Vert\nabla f \Vert=0 ~\text{ and }~D_\ww D_\ww\Vert\nabla f\Vert<0,}
\end{align*}
with $\ww=\tfrac{\nabla f}{\Vert\nabla f\Vert}$.
Normally, $f$ is used as the headlight illumination: $f\coloneqq \langle\nn,\vv\rangle$ with $\nn$ as the normal vector and $\vv$ as the view-vector.
PELs have more degrees of freedom to influence the result by adding more light sources.
Thus, the scalar value of $f$ changes by adding the light values of the vertices by other lights.
Noisy photic extremum lines are filtered by a threshold which is based on the integral of single connected lines.
The strength $T$ of a line with points $\xx_0,\ldots,\xx_n$ is determined by:
\begin{align*}
T=\int \Vert \nabla f\Vert=\sum_{i=0}^{n-1}\frac{\Vert \nabla f(\xx_i)\Vert+\Vert \nabla f(\xx_{i+1})\Vert}{2}\Vert \xx_i-\xx_{i+1}\Vert.
\end{align*}
If $T$ is less than a user-defined threshold, the line is canceled out.\\
\textbf{Summary:}
Photic extremum lines are strongly inspired by edge detection in image processing and by human perception of a change in luminance.
It uses the variation of illumination.
The result may be improved by adding lights.
Beside the filtering strategy to integrate over the lines and accumulate the magnitude of the gradient, the noise can also be reduced by adding a spotlight that directs to certain regions.
Nevertheless, smoothing is necessary to gain reasonable results.
Here, the smoothing of the normal is sufficient as the computation is mainly based on the normals.
However, the computation has high performance costs.
The original work was improved by Zhang et al.~\cite{Zhang2010} to significantly increase the runtime.
\\
\\

%
%
\subsection{Demarcating Curves}\label{subsec:demarcating_curves}
%
%
\begin{wrapfigure}{r}{0.3\textwidth}\label{fig:brain_dem}
 \centering
 \includegraphics[width=0.3\textwidth]{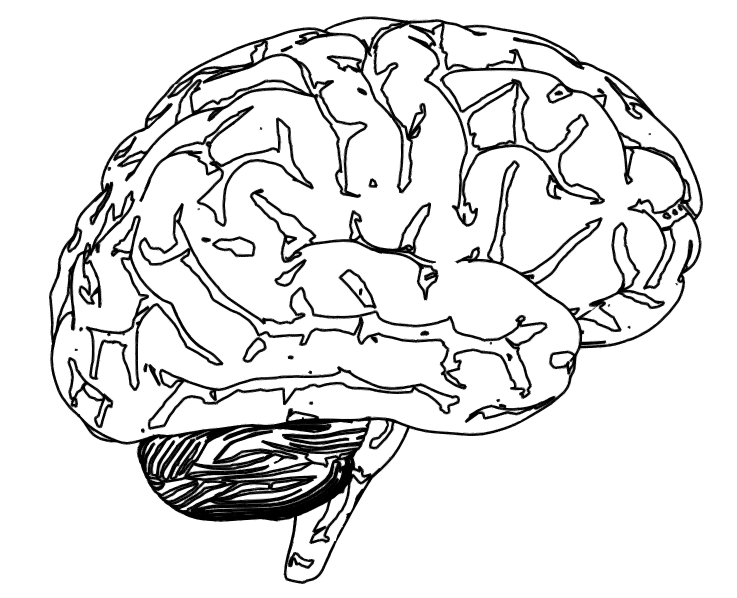}
\caption{The brain model with demarcating curves and contours.}
\end{wrapfigure}
\emph{Demarcating curves} were proposed by Kolomenkin et al.~\cite{Kolomenkin2008}.
These feature lines are defined as the transition of a ridge to a valley line.
To determine these lines, the derivative of the shape operator has to be calculated, recall Equation~\ref{Eq:C}:
\begin{align*}
C=\begin{pmatrix}D_\vv S & D_\ww S \end{pmatrix}.
\end{align*}
The demarcating curves are defined as the set of points where the curvature derivative is maximal:
\begin{align*}
\boxed{\langle \ww, S \ww\rangle=0~\text{ with }~\ww=\arg\max_{\Vert \vv\Vert=1}D_\vv\kappa.}
\end{align*}
The values for $\ww$ can be analytically found as the roots of a third order polynom.
This is obtained by setting $\vv=\binom{\sin(\theta)}{\cos(\theta)}$ and combining this with Equation~\ref{Eq:D_ukappa}.
A user-defined threshold eliminates demarcating curves, if it exceeds the value of $D_\ww\kappa$.\\
\textbf{Summary:}
Demarcating curves are view-independent feature lines displaying regions where the change of the curvature is maximal.
Therefore, higher-order derivatives are used.
A $2\times 2\times 2$ rank-3 tensor is determined.
This method can be used to illustrate bumps by surrounding curves.
The advantage of the method is to enhance small features.
Especially when combined with shading, this approach has its strength in illustrating archaeology objects where specific details are important, e.g., old scripts.
For this application, view-dependent illustration techniques are not recommended because details need to be displayed for every camera position.
Contrary, due to higher-order derivatives, the method is sensitive to noise and is not well suited for illustrative visualization.
%
%
\subsection{Laplacian Lines}\label{subsec:laplacian_lines}
%
%
\emph{Laplacian lines} were proposed by Zhang et al.~\cite{Zhang2011}.
The introduction of these lines was inspired by the Laplacian-of-Gaussian (LoG) edge detector in image processing and aims at a similar effect for surface meshes.
The idea of the LoG method is to determine the Laplacian of the Gaussian function and to use this kernel as a convolution kernel for the image.
Laplacian lines calculate the Laplacian of an illumination function $f$ and determine the zero crossing as feature lines.
To remove noisy lines, the lines are only drawn if the magnitude of the illumination gradient exceeds a user-defined threshold $\tau$:
\begin{align*}
\boxed{\Delta f =0~\text{ and }~\Vert\nabla f\Vert\geq\tau,}
\end{align*}
where $\Delta$ is the discrete Laplace-Beltrami operator on the surface mesh and $f$ is the illumination with $f\coloneqq \langle\nn,\vv\rangle$.
Here, the discrete Laplace-Beltrami operator with the Belkin weights is used, as introduced in Section~\ref{subsec:discrete_laplace_operator}.
The advantage of this method is the simplified representation of the Laplacian of the illumination:
\begin{align*}
\Delta f(\pp) &=\Delta\langle\nn,\vv\rangle\\
&=\langle \Delta\nn,\vv\rangle.
\end{align*}
Here, $\Delta\nn$ is the vector Laplace operator in the Euclidean space.
\begin{wrapfigure}{r}{0.3\textwidth}\label{fig:brain_ll}
 \centering
 \includegraphics[width=0.3\textwidth]{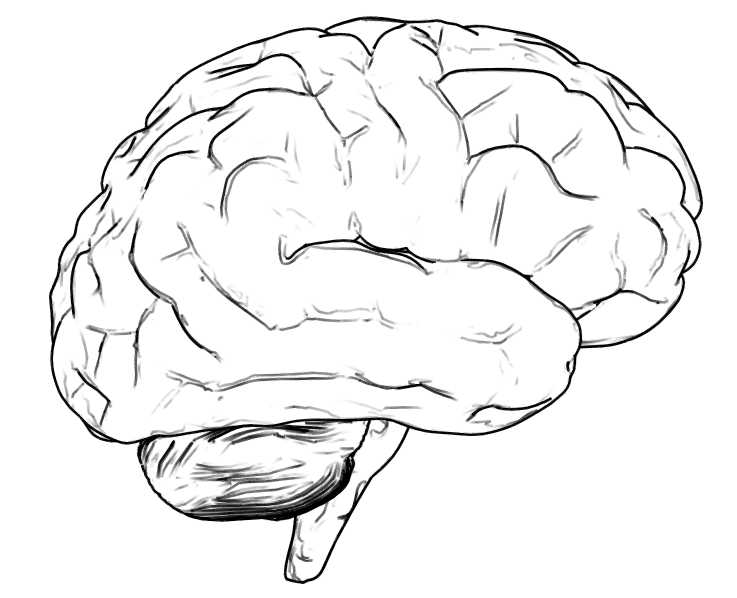}
\caption{The brain model with Laplacian lines.}
\end{wrapfigure}

This is just a composite of the Laplacian of the different components.
Thus, the algorithm consists of a preprocessing step to calculate the Laplace-Beltrami operator with the Belkin weights of the components of the normal $\Delta\nn$.
During runtime, the algorithm detects the zero crossings of $\langle \Delta\nn,\vv\rangle$ and checks if the magnitude of $\Vert \nabla f\Vert$ exceeds the user-defined threshold.\\
\textbf{Summary:}
The Laplacian lines are strongly inspired by edge detection algorithms in image processing.
This method is based on the Laplacian-of-Gaussian.
Basically, the method searches for zero crossings in the Laplacian of the illumination.
The computational effort can be simplified by a preprocessing step.
Thus, interactive frame rates for geometric models of moderate size are possible during the interaction.
Similar to other higher order methods, this approach also assumes well smoothed surface normals.
The Belkin weights for the Laplace-Beltrami operator have a smoothing effect for the Laplacian line generation.
This method illustrates sharp edges well, but is not suitable for round corners.

%% file: discussion.tex
%
%
\section{Discussion and Comparison}\label{discussion}
\begin{figure*}[tb]
 \centering
  \subfigure[Ridges and Valleys, Apparent Ridges]{
 \includegraphics[width=0.33\textwidth]{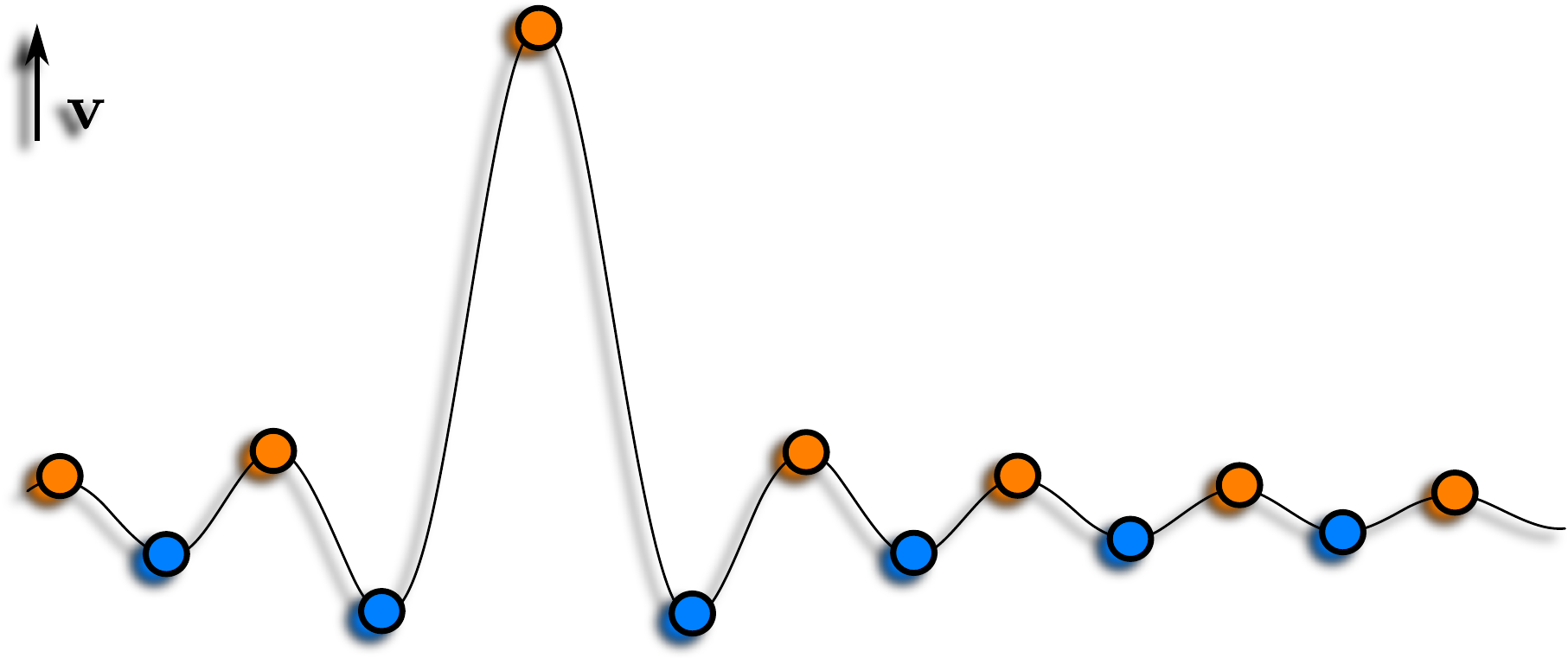}
  \label{fig:fe_rva}
  }~
  \subfigure[Suggestive\,Contours, Demarcating Curves]{
 \includegraphics[width=0.33\textwidth]{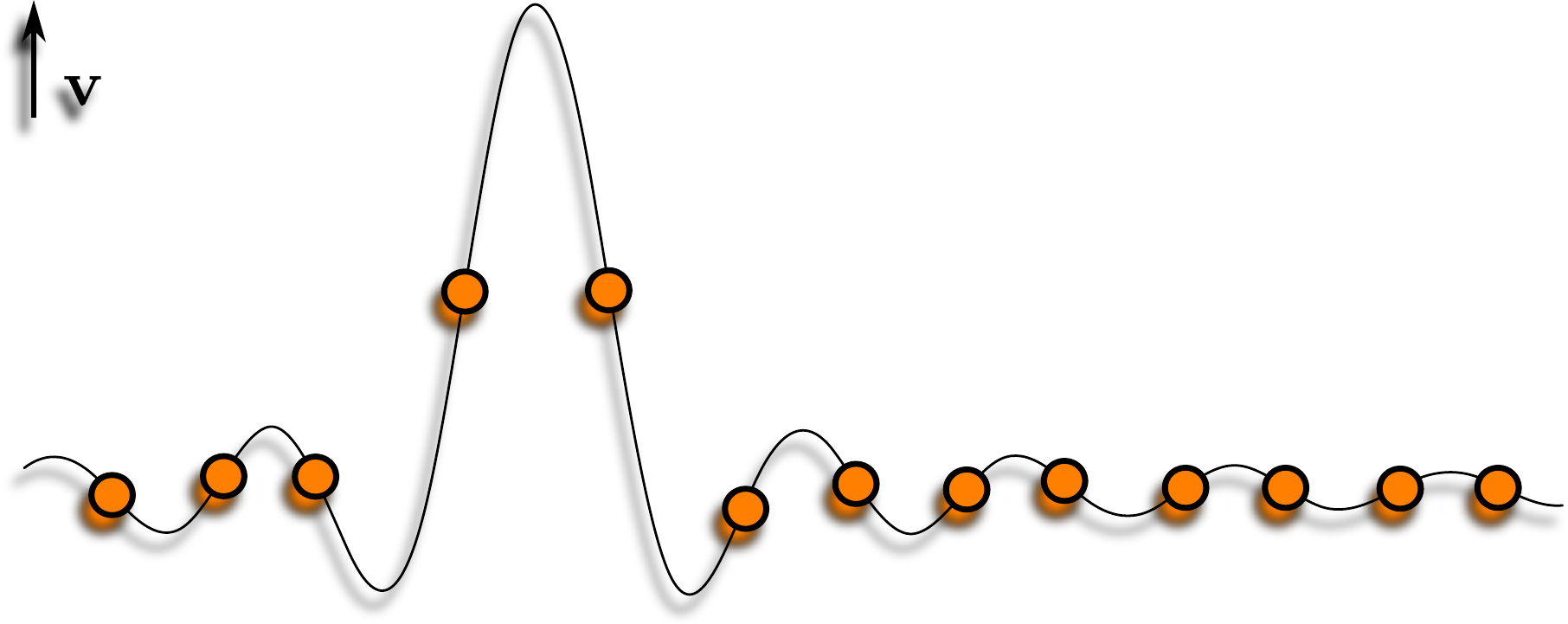}
 \label{fig:fe_sc}
 }~
 \subfigure[Photic\,Extremum Lines, Laplacian Lines]{
 \includegraphics[width=0.33\textwidth]{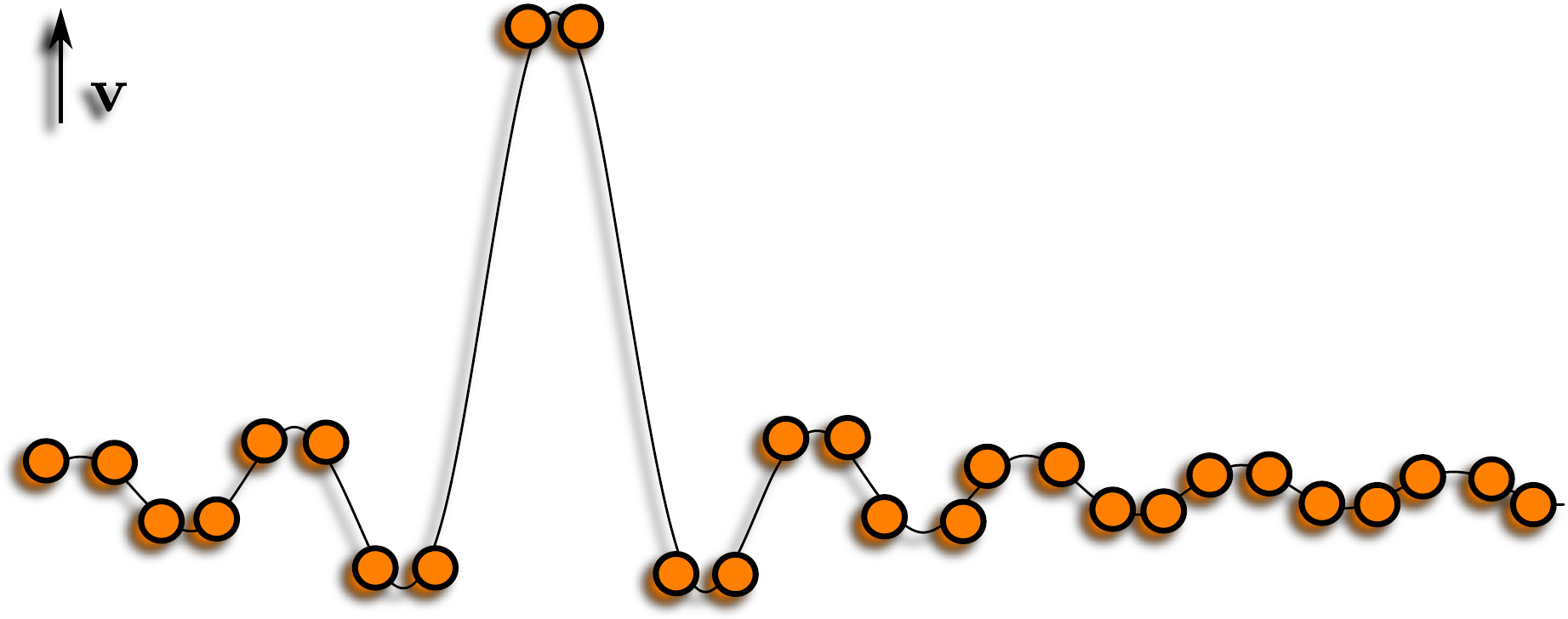}
 \label{fig:fe_pelll}
}
 \caption{Drawing of an analytic function with illustrated feature line positions. In \subref{fig:fe_rva} the ridges are denoted in orange and the valleys are illustrated in cyan. For this function with fixed view direction, the apparent ridges coincide with ridge and valley lines. In \subref{fig:fe_sc} the suggestive contours and the demarcating curves are the same. In \subref{fig:fe_pelll} the photic extremum lines and the Laplacian lines coincide.}
 \label{fig:feature_explain}
\end{figure*}

%
%
\begin{wraptable}{r}{0.45\textwidth}\vspace{-1cm}
\begin{tabular}{ l| c| c }
  Name 					& Order & View-dep. \\ \hline
  Contours 				& 1 & yes \\
  Crease Lines 			& 1 & no \\
  Ridges $\&$ Valleys 	& 3 & no \\
  Suggestive Contours 	& 2 & yes \\
  Apparent Ridges 		& 3 & yes \\
  Photic Extremum Lines	& 3 & yes \\
  Demarcating Curves	& 3 & no  \\
  Laplacian Lines		& 3 & yes \\
\end{tabular}
\caption{List of different feature line methods with derivative order and view dependency.}
\label{tab:feature_line_list_tab}
\end{wraptable}
This section deals with general properties of the different feature line methods.
We discuss the different approaches to derive first recommendations which method may be used for which kind of geometry.
First, we list all feature line methods in Table~\ref{tab:feature_line_list_tab} and name different properties and the order of the corresponding method.
Furthermore, in Figure~\ref{fig:feature_explain} the higher-order feature lines are illustrated on an analytic function.\\ \\ \\ \\
\begin{table*}[th]
 \begin{tabular}{ l| c| c|c| c|c|c   }
  Name 					& Sharp Edges 	& Round Edges 	& Bumps (s.w.)	& Bumps (top)	& Contour 	& Deformation \\  
  \hline
  Contour 				& \no 			& \no  			& \no				& \no			& \yes 		& \yes	\\	
  Crease Lines			& \yes 			& \no  			& \no				& \no			& \no		& \yes	\\	
  Ridges $\&$ Valleys 	& \yes 			& \yes			& \no				& \yes			& \no		& \no	\\	
  Suggestive Contours 	& \no			& \no			& \yes				& \yes			& \no		& \yes	\\	
  Apparent Ridges 		& \yes			& \yes 			& \yes				& \yes			& \yes		& \no	\\	
  Photic Extremum Lines	& \yes			& \no 			& \no				& \yes			& \yes		& \yes	\\	
  Demarcating Curves	& \no			& \no			& \no				& \yes			& \no		& \no	\\	
  Laplacian Lines		& \yes 			& \no 			& \no				& \yes			& \yes		& \no	\\	
\end{tabular}
\caption{List of supported feature by the methods. The different features are illustrated in Figure~\ref{fig:features_properties}.}
\label{tab:feature_line_support}
\end{table*}
%
%
%
%
The benefit of feature lines is motivated by the visual perception.
In~\cite{Marr1976} it is stated that the first stage of the assessment of the shape is done by extracting features, such as contours.
These characteristics help to understand the shape.
The illustration of shapes with feature lines cannot be seen as an alternative to shading.
It is rather an additional concept.
Kolomenkin et al.~\cite{Kolomenkin2008} showed that their demarcating lines support the shading and can extract text from archaeology objects.
However, for examining structures where the whole object inherits important information, feature lines should not be used solely.
For data where the scene can be divided into focus and context objects, feature lines can be applied to the context objects.
Furthermore, feature lines can also be used to enhance focus with additional shading.

Depending on the underlying model, we may recommend different techniques.
Most of the feature lines are able to depict the contour, but this depends strongly on the bending of the surface at the contour.
Especially apparent ridges and photic extremum lines are able to draw contour lines, but in our experiments we noticed that activating the contour enhances the visual impression because some parts of the contour were missing.
If the surface model is an assembly with sharp edges we recommend to use ridge and valley lines or apparent ridges.
These features often appear in medical models models like implants or prostheses. 
For simple models with only a few sharp edges, crease lines may be appropriate as well.
If the models have a lot of round edges the answer for the right feature line method is a matter of taste.
These features appear in models like vascular surfaces or organs.
For scenarios where it is important to illustrate details for every camera position, we recommend ridges and valleys as well as demarcating curves.
From an artistic point of view, suggestive contours, photic extremum lines, and Laplacian lines should be chosen.
The reason for this suggestion is that especially for a rounded cube the photic extremum lines and Laplacian lines generate double lines around the feature to denote the rounded edge.
If the user wants to visualize the line along the edge, the ridge and valley lines or the apparent ridges should be used.
For this case, the crease line approach is not useful because it depicts only edges with specific greater value of the dihedral angle.
Therefore, too many lines may be generated.
If the surface has many crevices, we recommend the suggestive contours.
They illustrate the inflection points of valleys.
Round corners are often represented in many organic structures like livers or bones, see Figure~\ref{fig:models} for a femur model or a skull model.

Table~\ref{tab:feature_line_support} lists all possible features and Figure~\ref{fig:features_properties} shows the different features.
Please note that the assessment of the suitability of a method -- marked in the table -- necessarily is a subjective assessment by the authors and two artists.
For instance, regarding the property whether the methods are able to detect round edges, we mean if it detects the specific round feature.
As already mentioned, it does not reflect the ability to enhance the round edge from an illustrative or artist point of view.
This concerns the ability to depict bumps.
In agreement with artist, the bump shown from a sideway (s.w.) perspective would be illustrated such that it depicts the smooth transition from the ground to the dent.
The drawing of the surrounding circle of the bump is not desirable as it conveys a sharp transition from the bump to the ground.
For bumps shown from the top perspective it is sufficient if a round circle is drawn.
Bumps can occur as polyps or blebs on a cerebral aneurysm.
Especially blebs are important anatomical features to be detected because they are an indicator for rupture.
Blebs can also occur as polyps in CT colonography.

We also listed the property \emph{deformation} in the table.
This characteristic means if the corresponding method is able to illustrate the features of deformable surfaces, e.g., animated objects, in real-time.
As an example, Oeltze et al.~\cite{Oeltze2008} analyzed myocardial perfusion data.
The focus lies on the examination of the infarction scarf on the left ventricle.
In this paper, the left ventricle is illustrated as context information.
Using the time-dependent data, it would also be possible to illustrate the context information with some feature line methods during the animation.
\newcommand{\lineWPic}{0.12\textwidth}
\begin{figure*}[tb]
 \centering
  \subfigure{
 \includegraphics[width=\lineWPic]{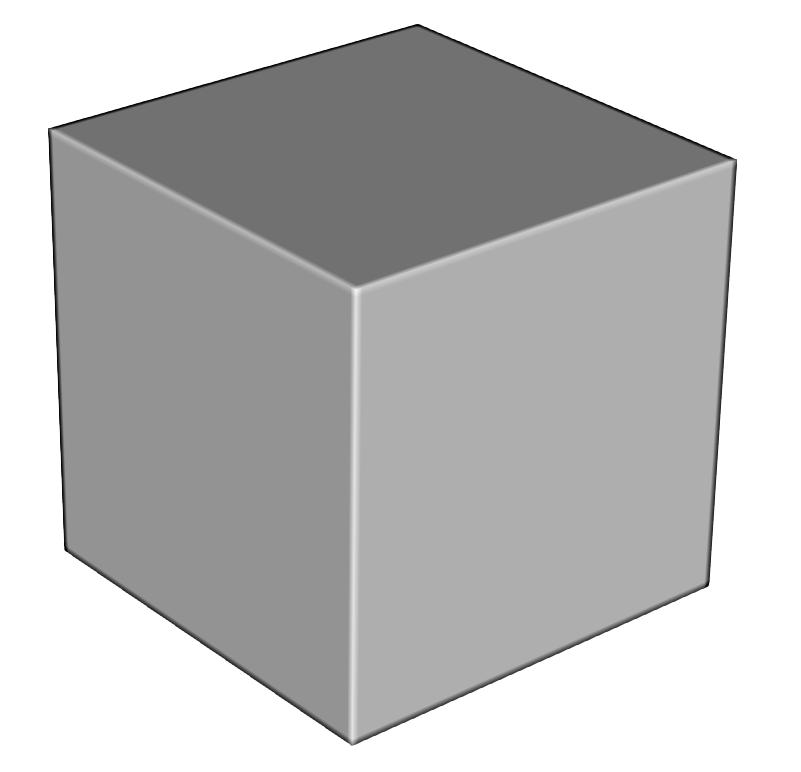}
  \label{fig:cmp_cube_sha}
  }~
  \subfigure{
 \includegraphics[width=\lineWPic]{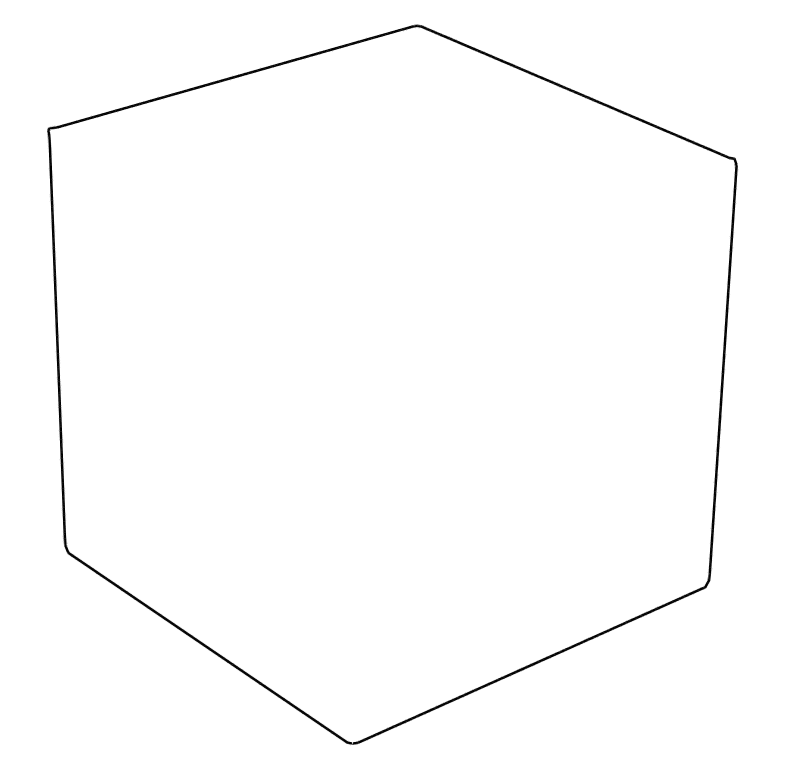}
  \label{fig:cmp_cube_sha}
  }~
  \subfigure{
 \includegraphics[width=\lineWPic]{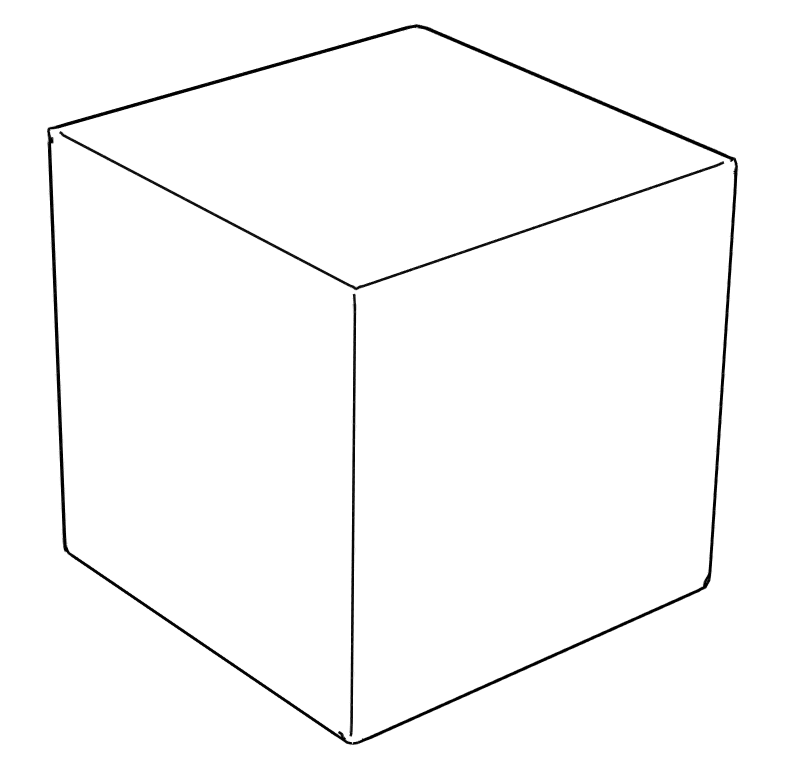}
  \label{fig:cmp_cube_sha}
  }~
  \subfigure{
 \includegraphics[width=\lineWPic]{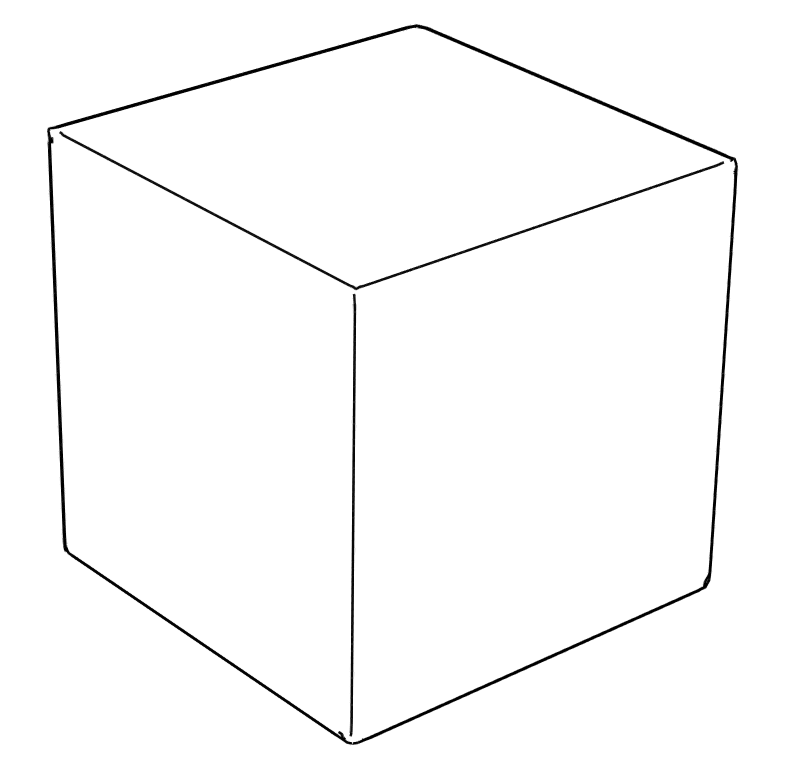}
  \label{fig:cmp_cube_sha}
  }~
  \subfigure{
 \includegraphics[width=\lineWPic]{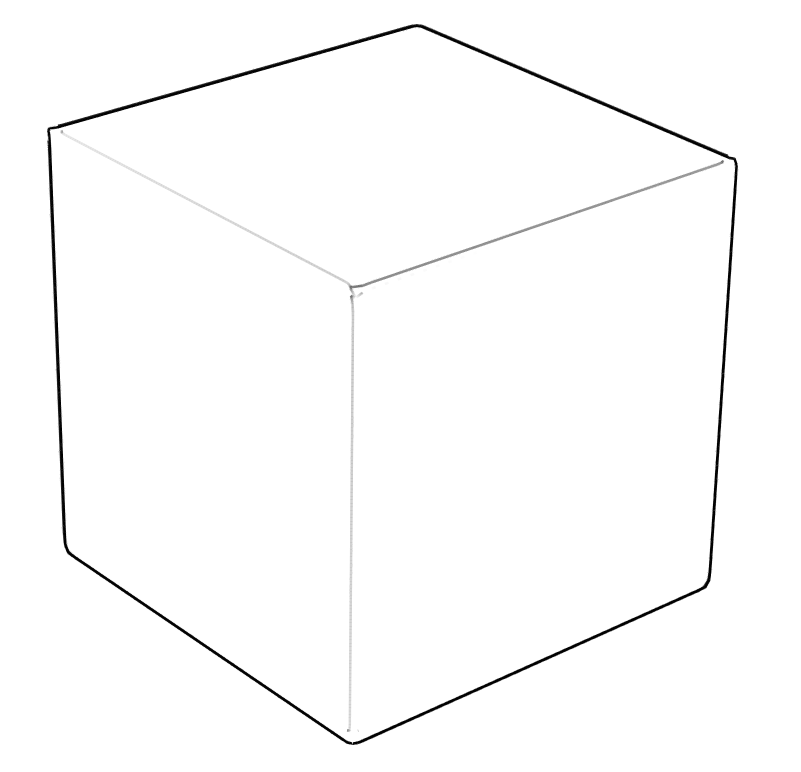}
  \label{fig:cmp_cube_sha}
  }~
  \subfigure{
 \includegraphics[width=\lineWPic]{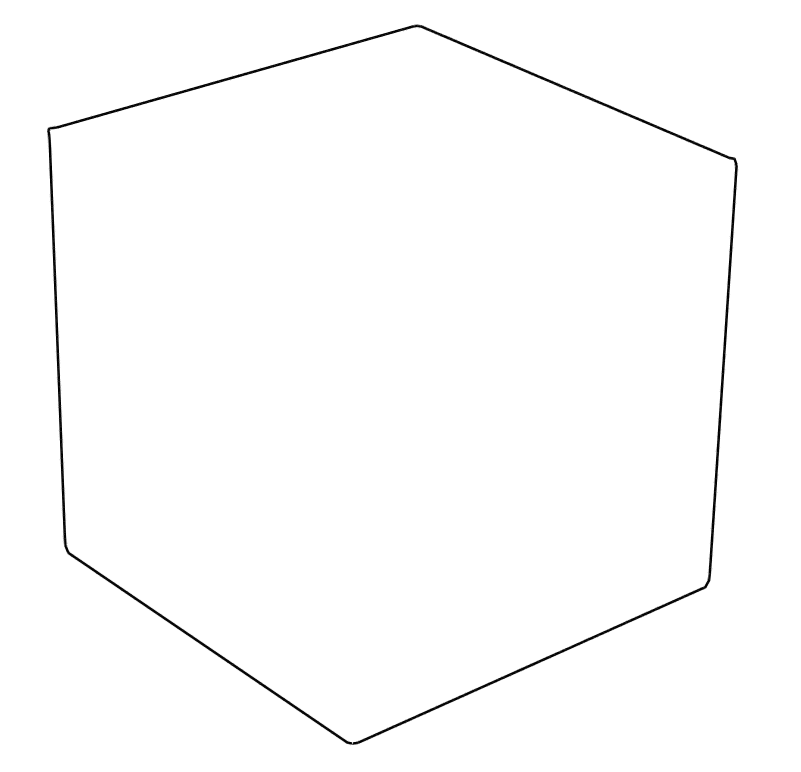}
  \label{fig:cmp_cube_sha}
  }~
  \subfigure{
 \includegraphics[width=\lineWPic]{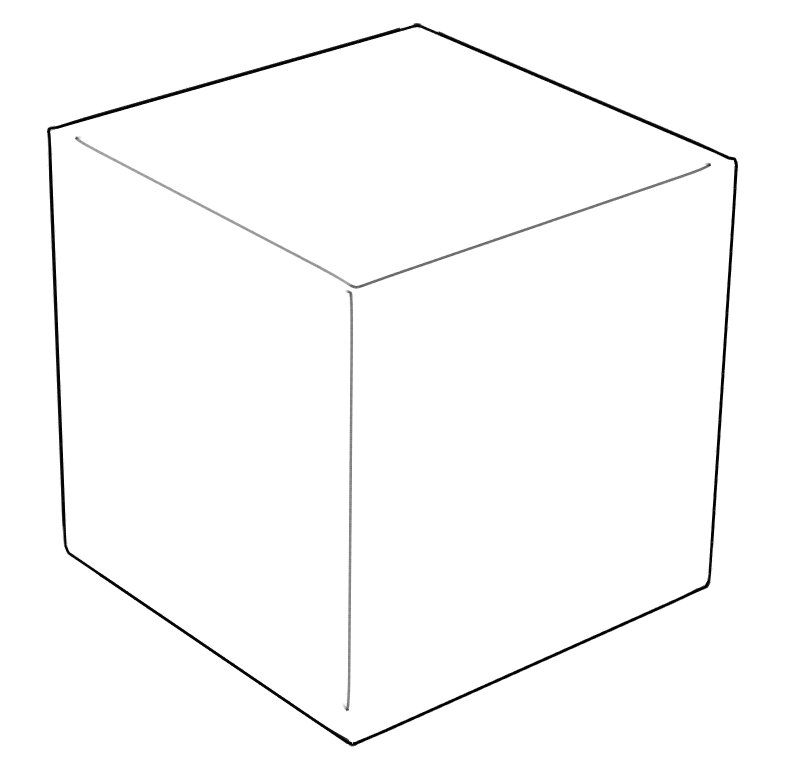}
  \label{fig:cmp_cube_sha}
  }
    \\
 \subfigure{
 \includegraphics[width=\lineWPic]{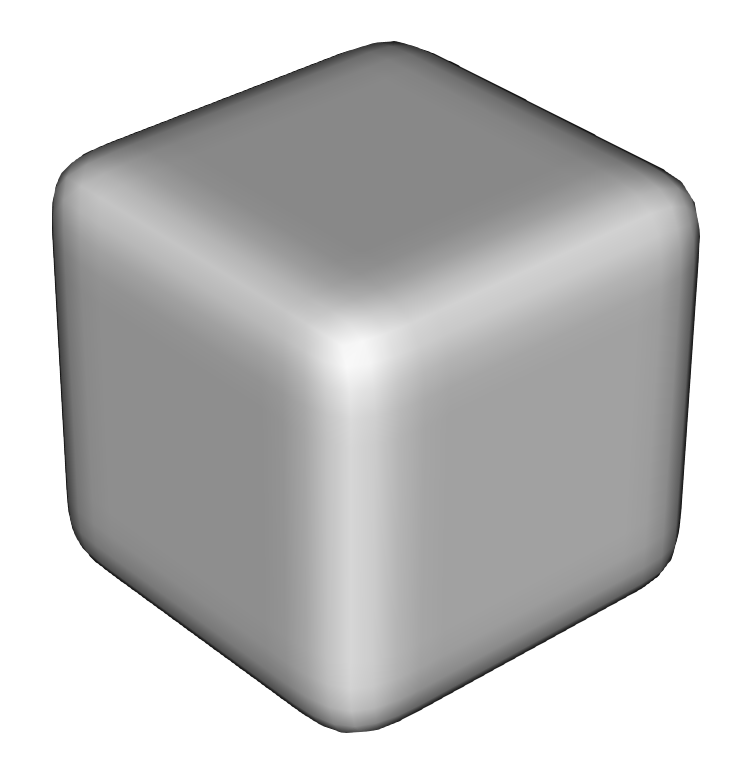}
  \label{fig:cmp_cube_sha}
  }~
  \subfigure{
 \includegraphics[width=\lineWPic]{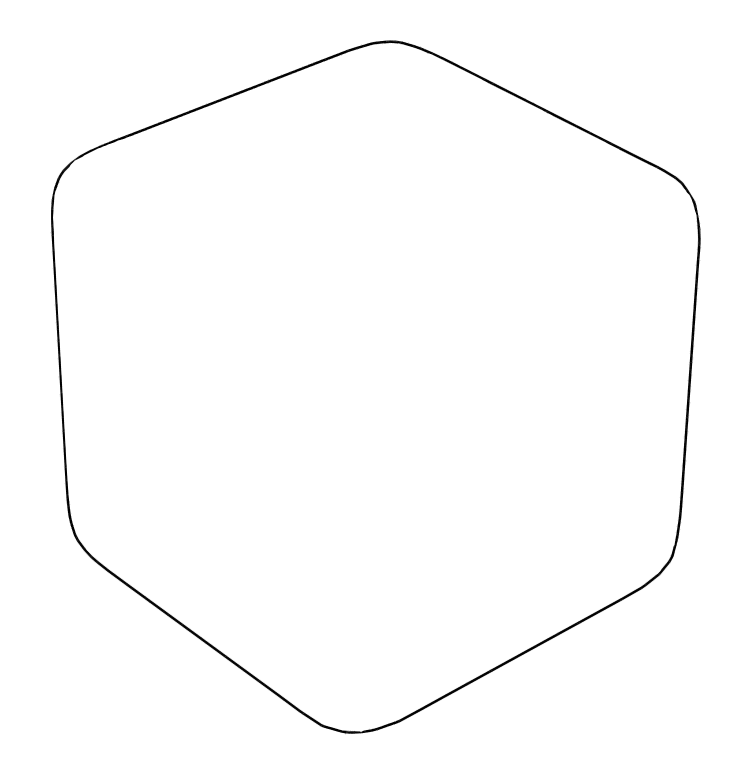}
  \label{fig:cmp_cube_sha}
  }~
  \subfigure{
 \includegraphics[width=\lineWPic]{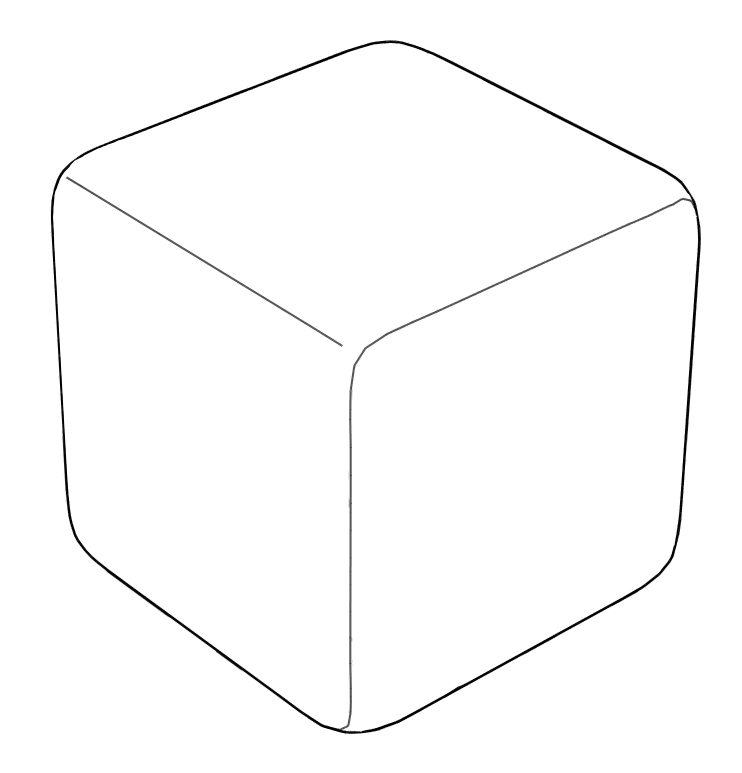}
  \label{fig:cmp_cube_sha}
  }~
  \subfigure{
 \includegraphics[width=\lineWPic]{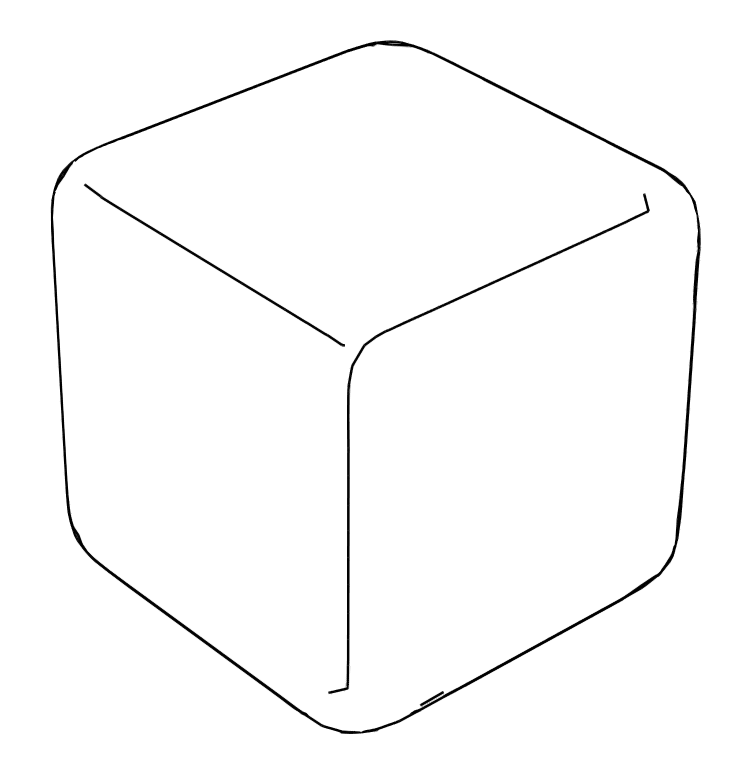}
  \label{fig:cmp_cube_sha}
  }~
  \subfigure{
 \includegraphics[width=\lineWPic]{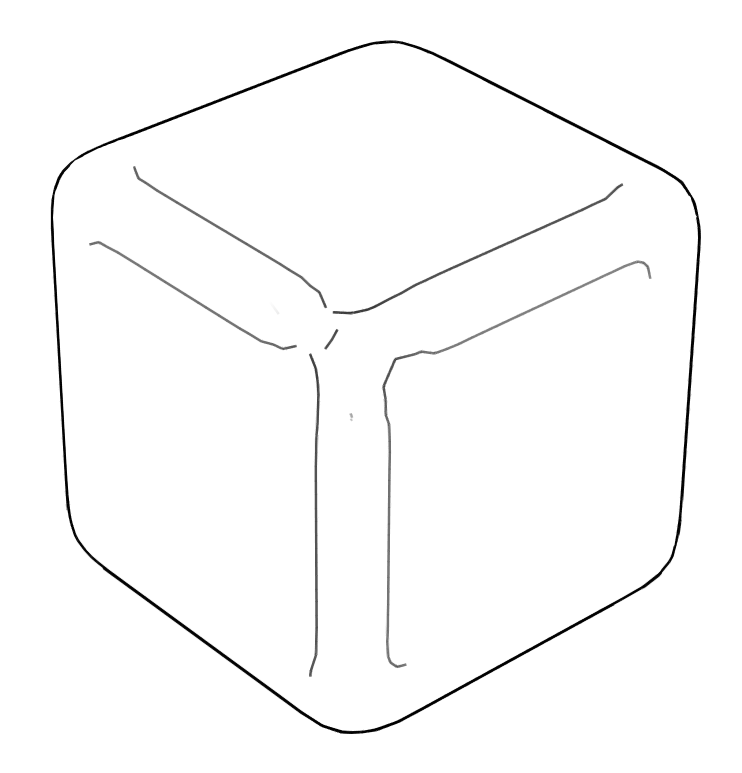}
  \label{fig:cmp_cube_sha}
  }~
   \subfigure{
 \includegraphics[width=\lineWPic]{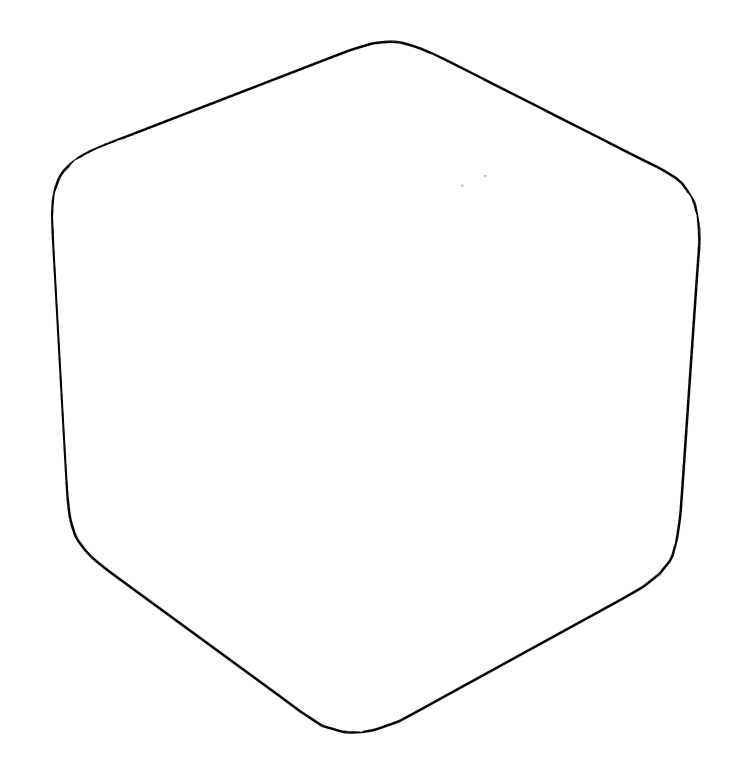}
  \label{fig:cmp_cube_sha}
  }~
  \subfigure{
 \includegraphics[width=\lineWPic]{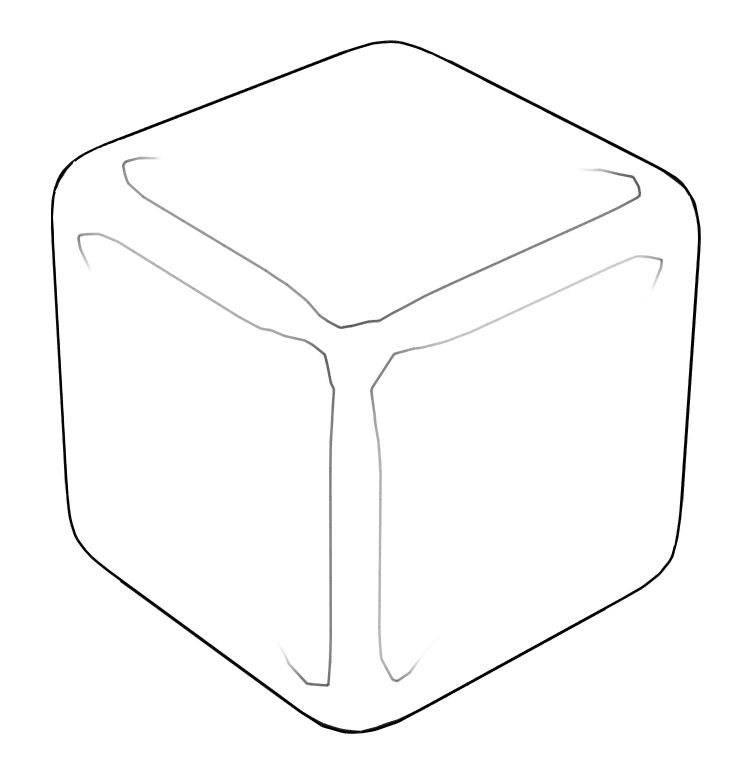}
  \label{fig:cmp_cube_sha}
  }
   \\
   \subfigure{
 \includegraphics[width=\lineWPic]{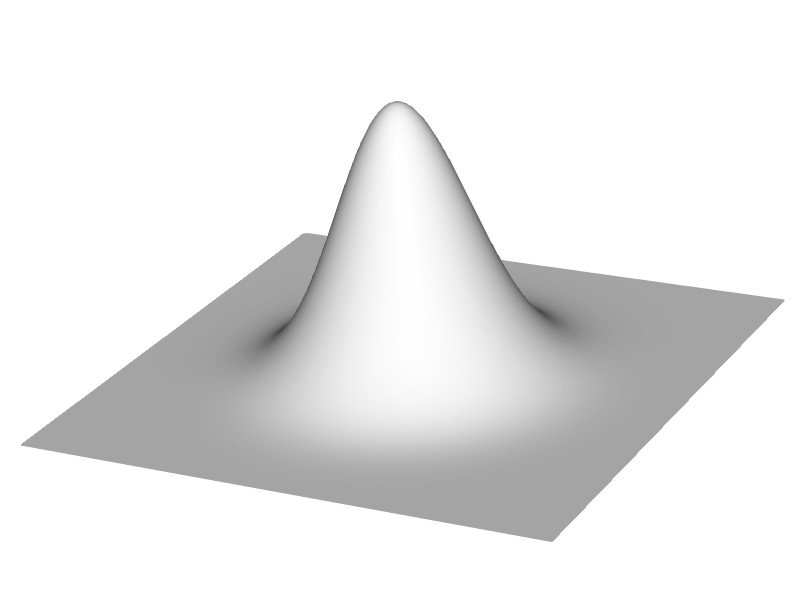}
  \label{fig:cmp_cube_sha}
  }~
  \subfigure{
 \includegraphics[width=\lineWPic]{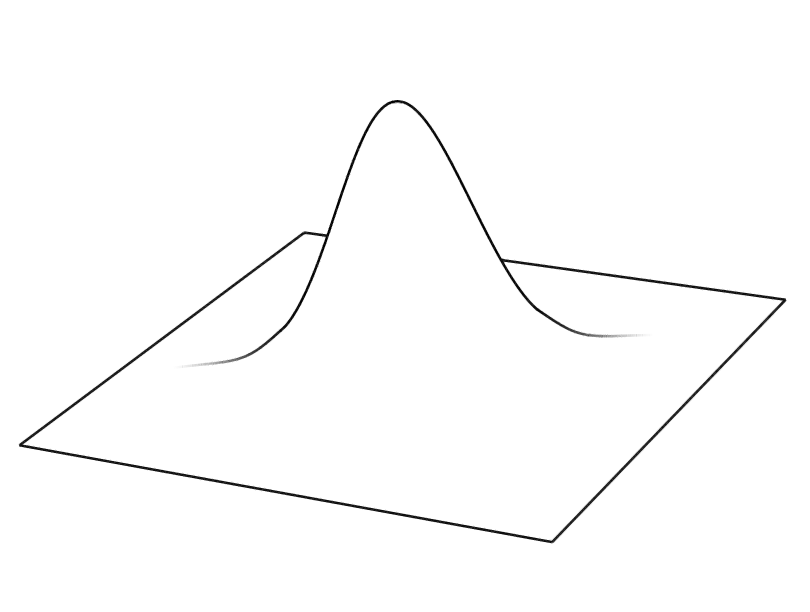}
  \label{fig:cmp_cube_sha}
  }~
  \subfigure{
 \includegraphics[width=\lineWPic]{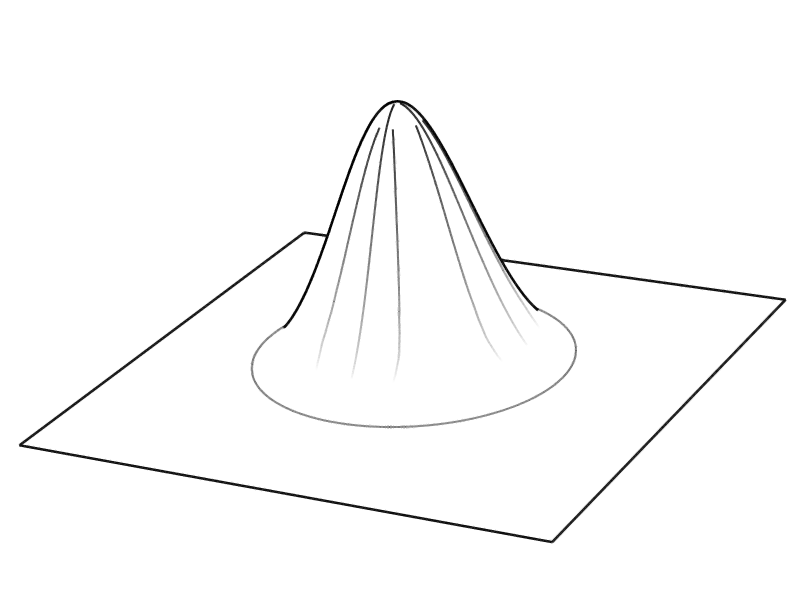}
  \label{fig:cmp_cube_sha}
  }~
  \subfigure{
 \includegraphics[width=\lineWPic]{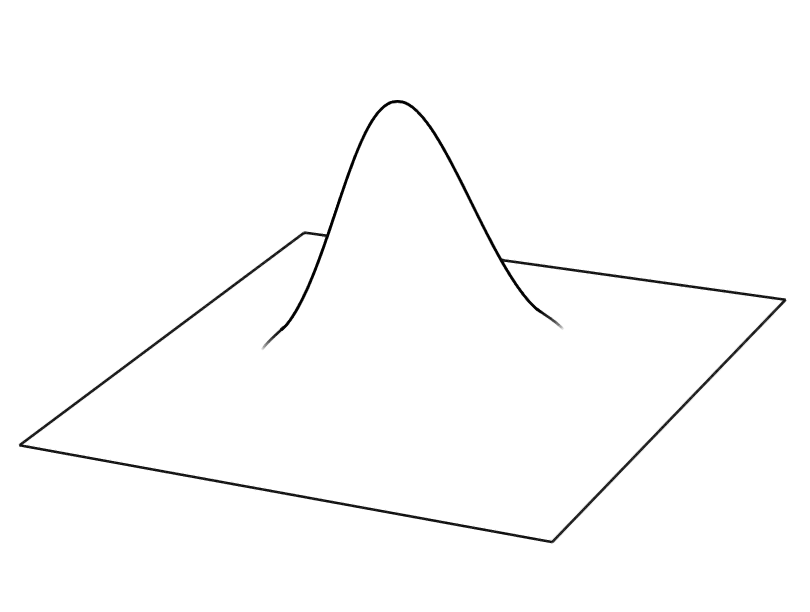}
  \label{fig:cmp_cube_sha}
  }~
  \subfigure{
 \includegraphics[width=\lineWPic]{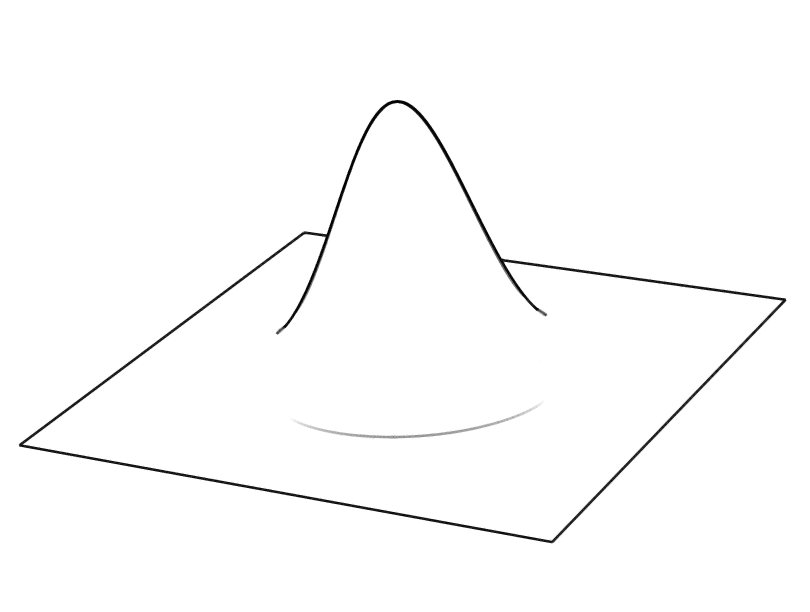}
  \label{fig:cmp_cube_sha}
  }~
   \subfigure{
 \includegraphics[width=\lineWPic]{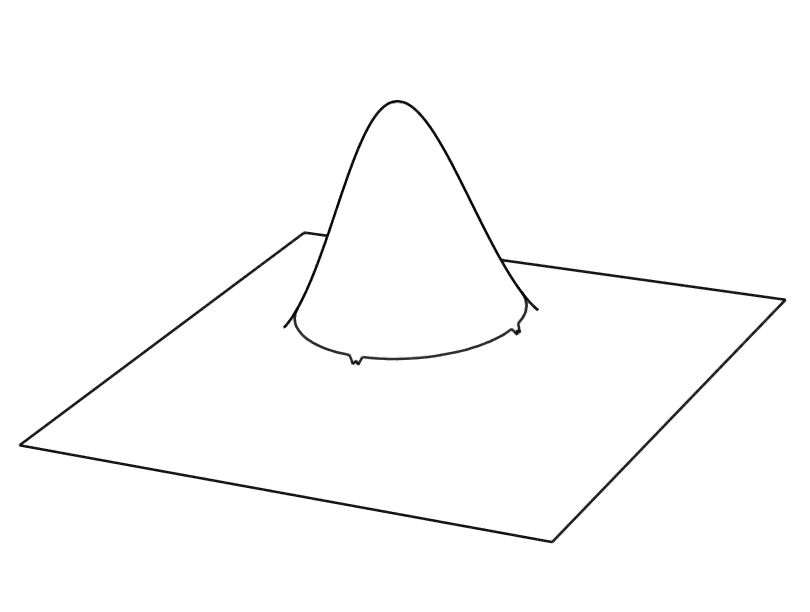}
  \label{fig:cmp_cube_sha}
  }~
  \subfigure{
 \includegraphics[width=\lineWPic]{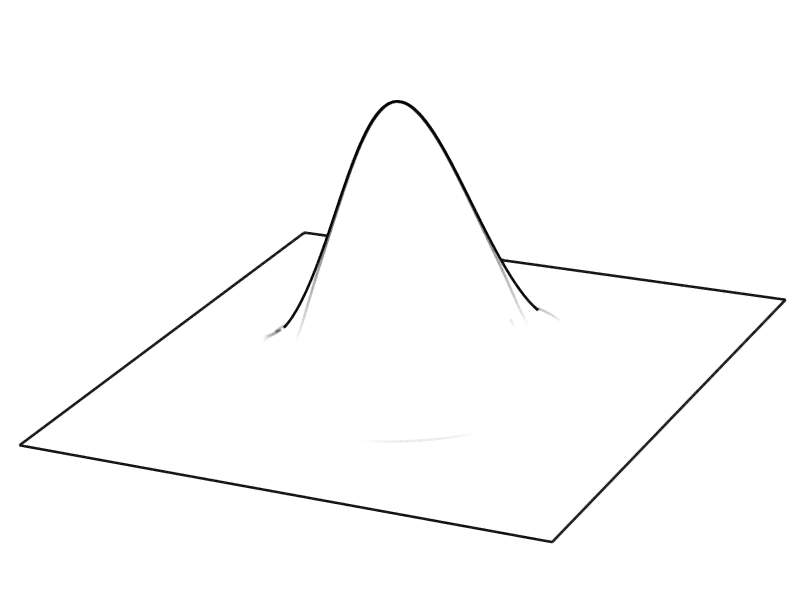}
  \label{fig:cmp_cube_sha}
  }
   \\
  \subfigure{
 \includegraphics[width=\lineWPic]{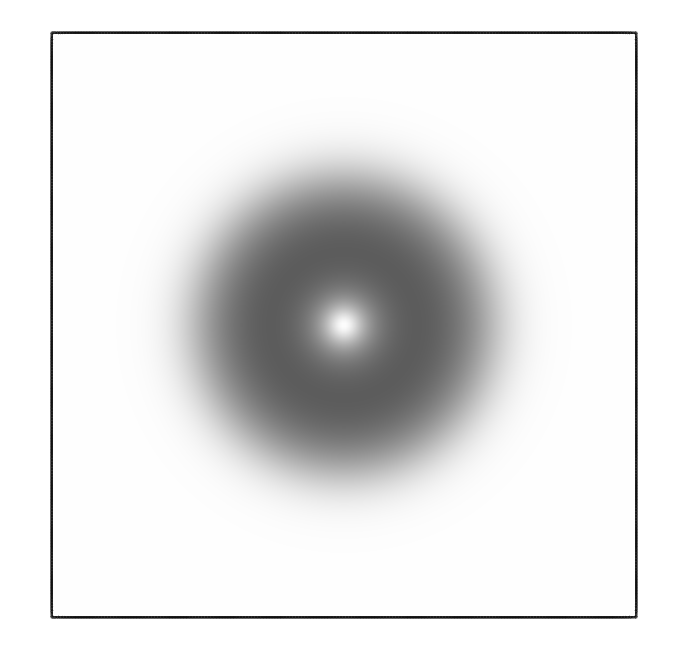}
  \label{fig:cmp_cube_sha}
  }~
  \subfigure{
 \includegraphics[width=\lineWPic]{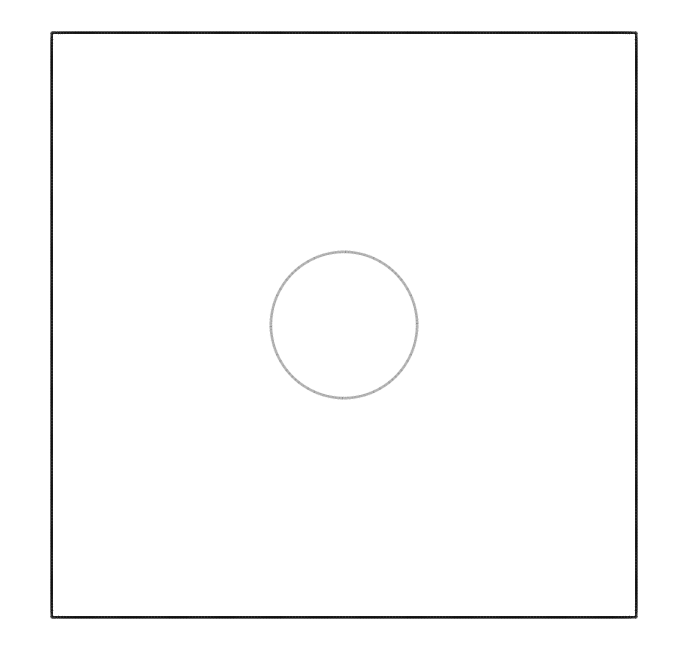}
  \label{fig:cmp_cube_sha}
  }~
  \subfigure{
 \includegraphics[width=\lineWPic]{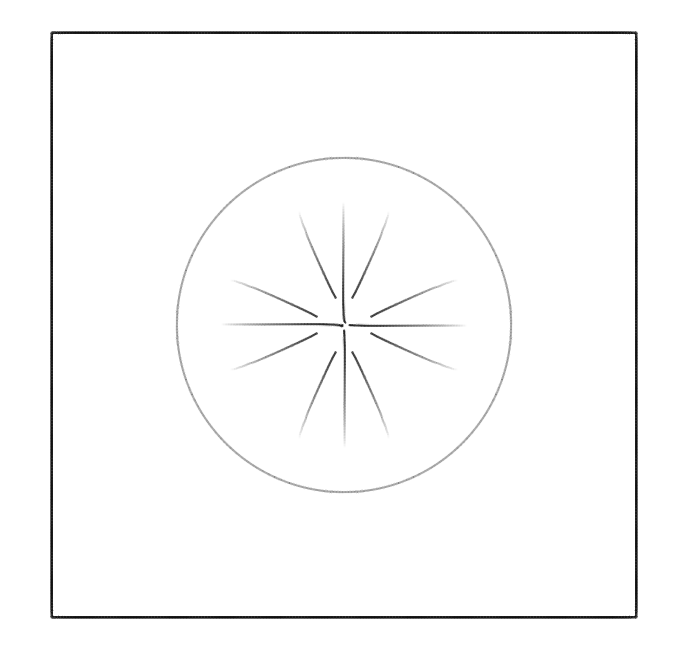}
  \label{fig:cmp_cube_sha}
  }~
  \subfigure{
 \includegraphics[width=\lineWPic]{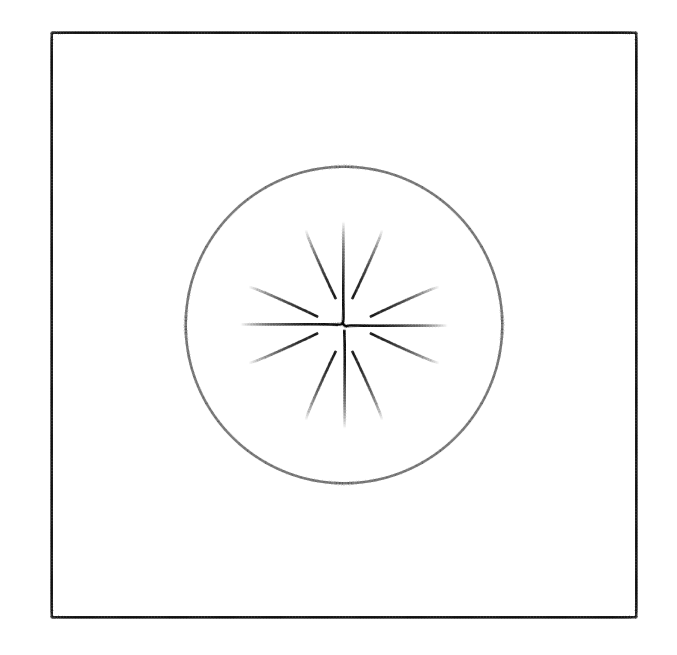}
  \label{fig:cmp_cube_sha}
  }~
  \subfigure{
 \includegraphics[width=\lineWPic]{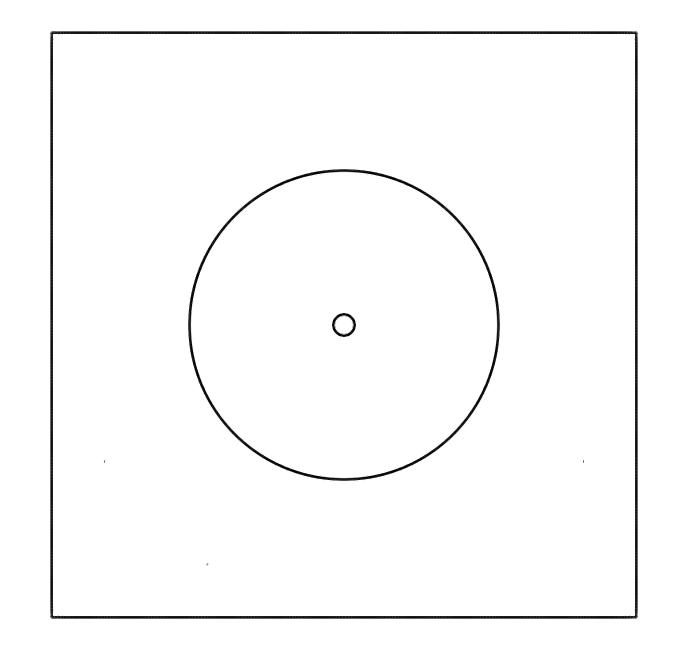}
  \label{fig:cmp_cube_sha}
  }~
   \subfigure{
 \includegraphics[width=\lineWPic]{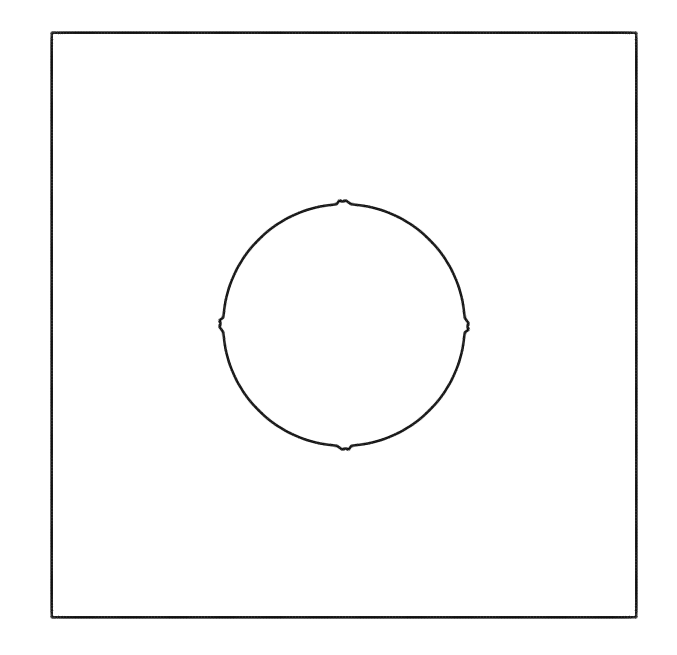}
  \label{fig:cmp_cube_sha}
  }~
  \subfigure{
 \includegraphics[width=\lineWPic]{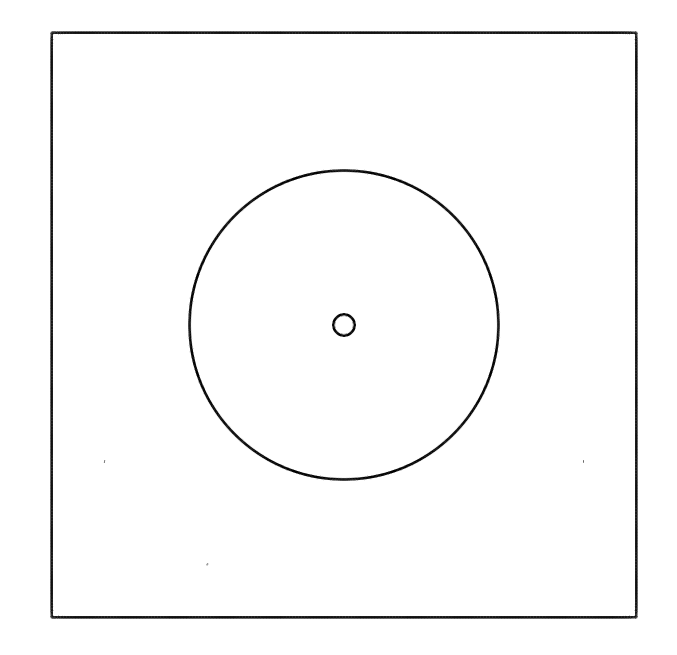}
  \label{fig:cmp_cube_sha}
  }
   \\
  \newcommand{\lineW}{0.04325}
   \begin{tabular}{c c c c c c c} 			\hline
			\hspace{\lineW\textwidth} SH \hspace{\lineW\textwidth} &  
			\hspace{\lineW\textwidth} SC \hspace{\lineW\textwidth} & 
			\hspace{\lineW\textwidth} RV \hspace{\lineW\textwidth} & 
			\hspace{\lineW\textwidth} AR \hspace{\lineW\textwidth} & 
			\hspace{\lineW\textwidth} PEL \hspace{\lineW\textwidth} & 
			\hspace{\lineW\textwidth} DC \hspace{\lineW\textwidth} &
			\hspace{\lineW\textwidth} LL \hspace{\lineW\textwidth}  \\
		\end{tabular}\\
 \caption{Different surface features are illustrated with shading (SH) and in different high-order feature line methods: suggestive contours (SC), ridge and valley (RV), apparent ridges (AR), photic extremum lines (PEL), demarcating curves (DC), and Laplacian lines (LL).}
 \label{fig:features_properties}
\end{figure*}
\begin{figure*}[tb]
 \centering
  \subfigure{
 \includegraphics[width=\lineWPic]{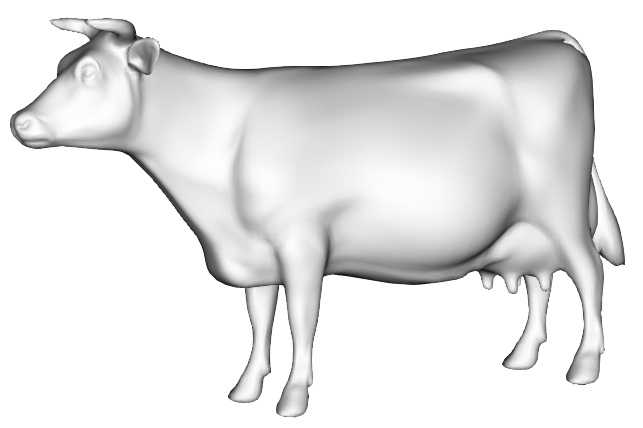}
  \label{fig:cmp_cube_sha}
  }~
  \subfigure{
 \includegraphics[width=\lineWPic]{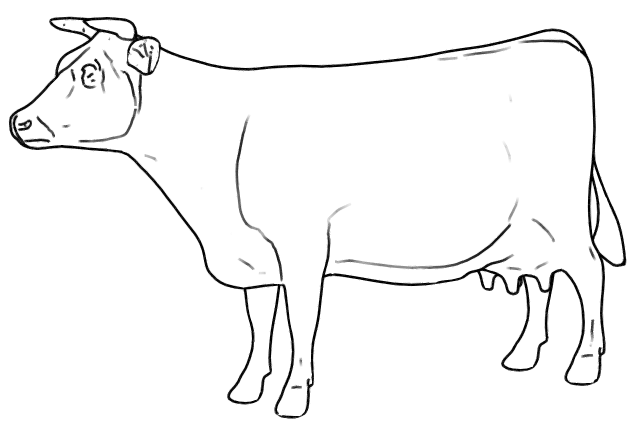}
  \label{fig:cmp_cube_sha}
  }~
  \subfigure{
 \includegraphics[width=\lineWPic]{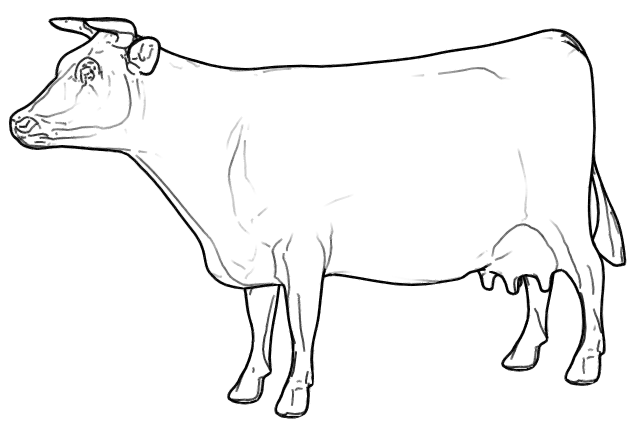}
  \label{fig:cmp_cube_sha}
  }~
  \subfigure{
 \includegraphics[width=\lineWPic]{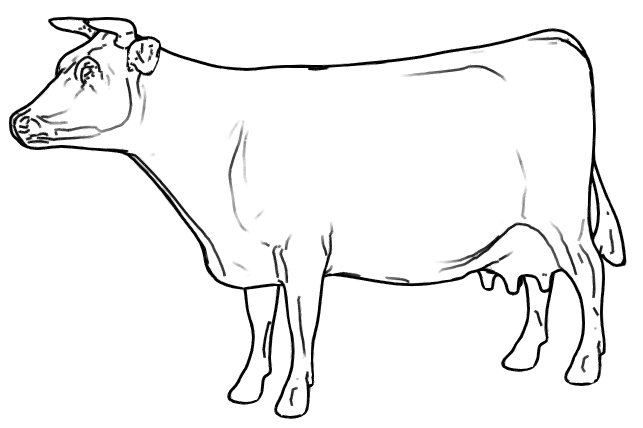}
  \label{fig:cmp_cube_sha}
  }~
  \subfigure{
 \includegraphics[width=\lineWPic]{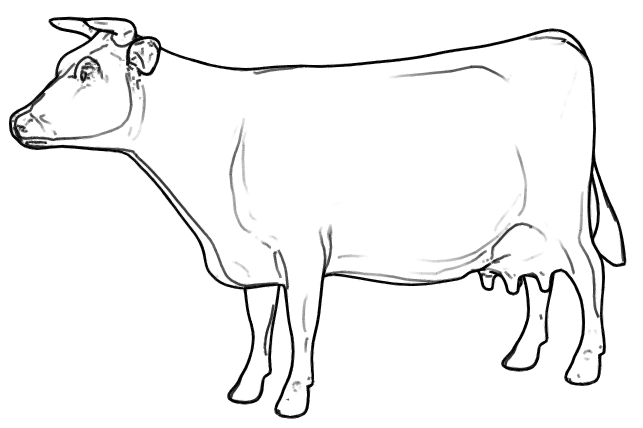}
  \label{fig:cmp_cube_sha}
  }~
  \subfigure{
 \includegraphics[width=\lineWPic]{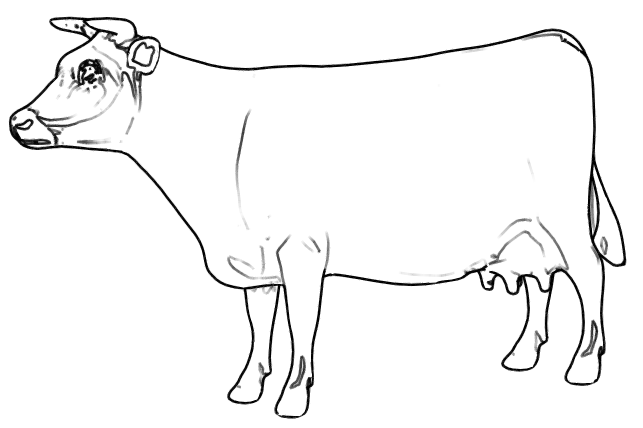}
  \label{fig:cmp_cube_sha}
  }~
  \subfigure{
 \includegraphics[width=\lineWPic]{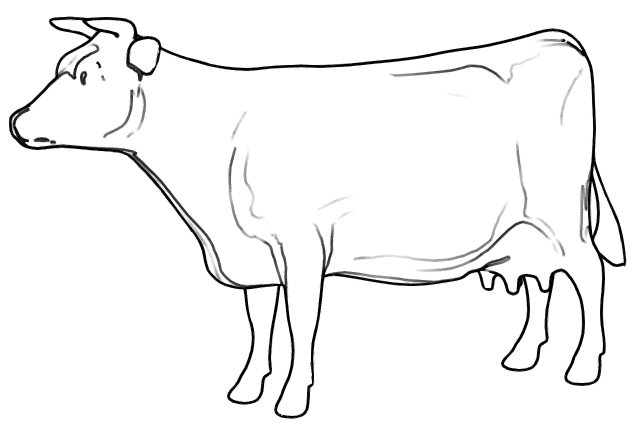}
  \label{fig:cmp_cube_sha}
  }\\
   \subfigure{
   \includegraphics[width=\lineWPic]{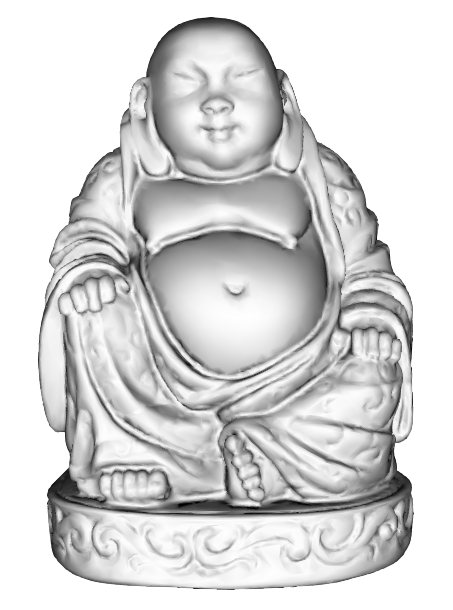}
  \label{fig:cmp_cube_sha}
  }~
  \subfigure{
 \includegraphics[width=\lineWPic]{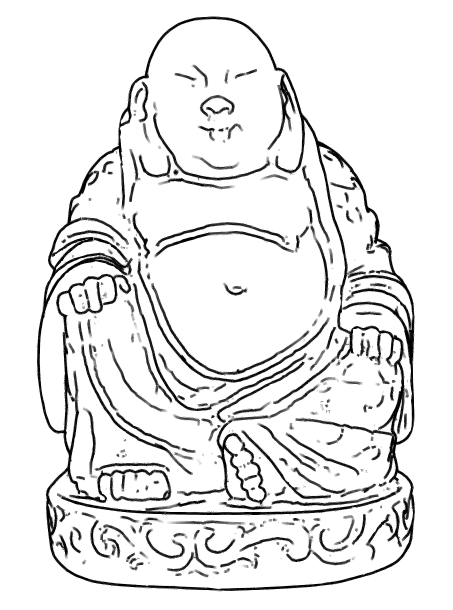}
  \label{fig:cmp_cube_sha}
  }~
  \subfigure{
 \includegraphics[width=\lineWPic]{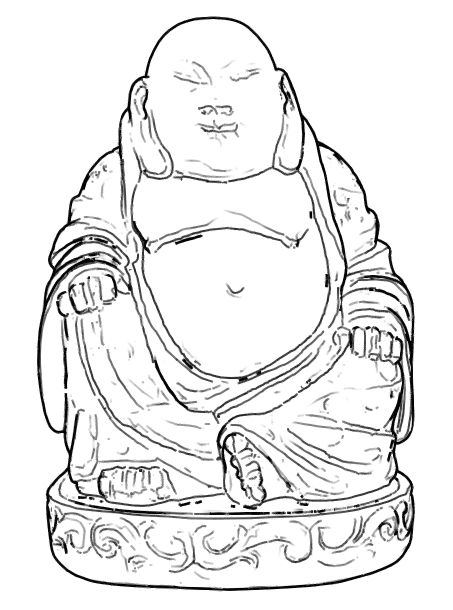}
  \label{fig:cmp_cube_sha}
  }~
  \subfigure{
 \includegraphics[width=\lineWPic]{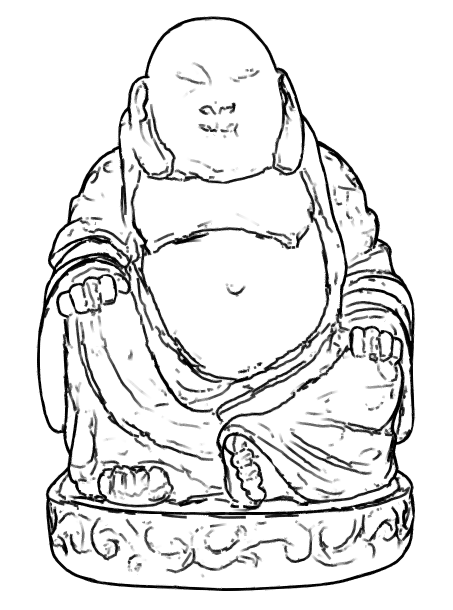}
  \label{fig:cmp_cube_sha}
  }~
  \subfigure{
 \includegraphics[width=\lineWPic]{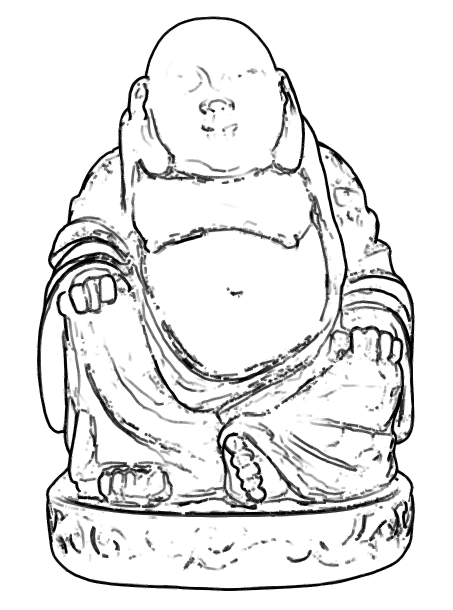}
  \label{fig:cmp_cube_sha}
  }~
  \subfigure{
 \includegraphics[width=\lineWPic]{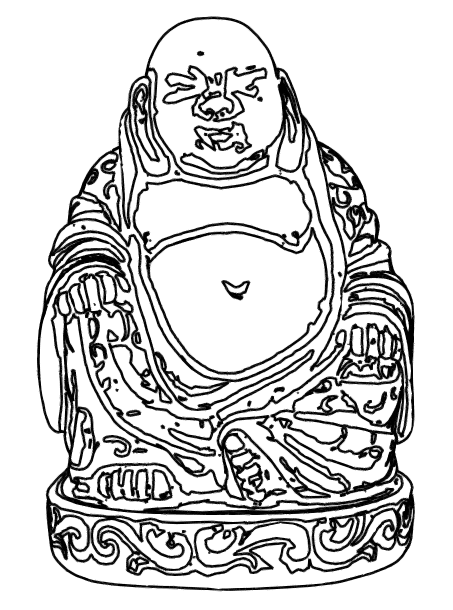}
  \label{fig:cmp_cube_sha}
  }~
  \subfigure{
 \includegraphics[width=\lineWPic]{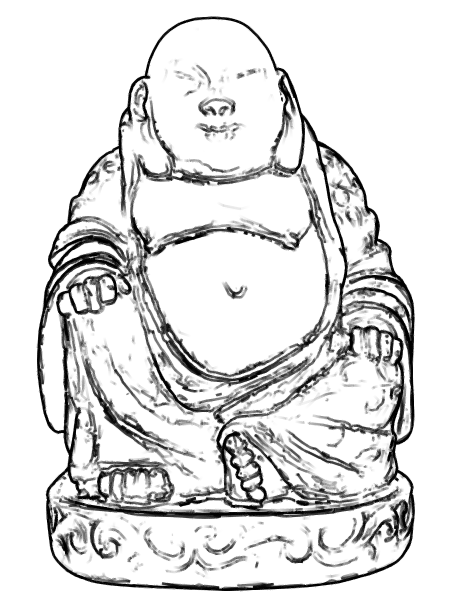}
  \label{fig:cmp_cube_sha}
  }\\
  \subfigure{
   \includegraphics[width=\lineWPic]{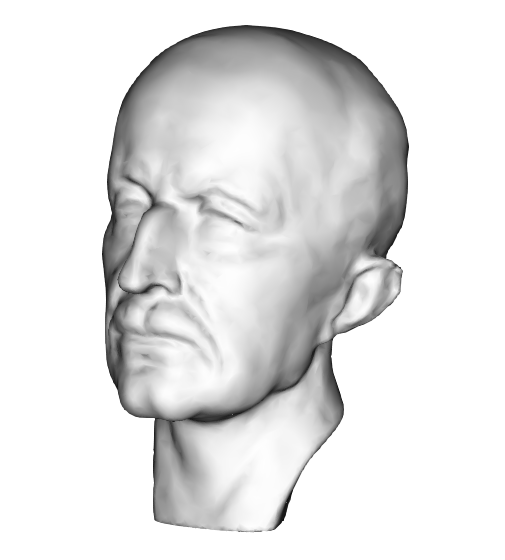}
  \label{fig:cmp_cube_sha}
  }~
  \subfigure{
 \includegraphics[width=\lineWPic]{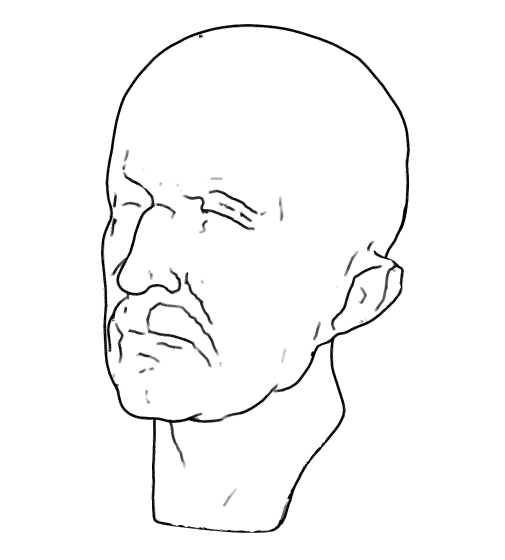}
  \label{fig:cmp_cube_sha}
  }~
  \subfigure{
 \includegraphics[width=\lineWPic]{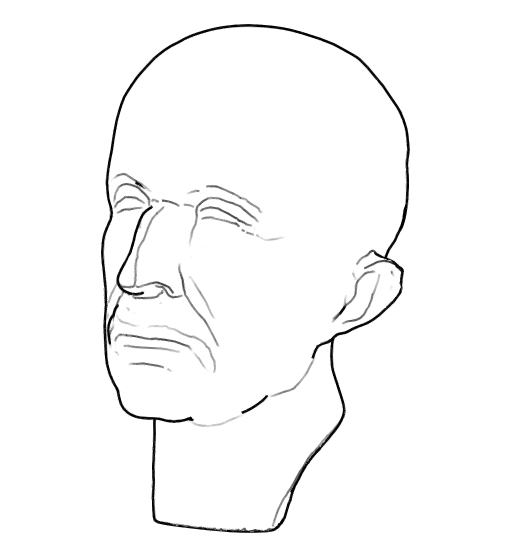}
  \label{fig:cmp_cube_sha}
  }~
  \subfigure{
 \includegraphics[width=\lineWPic]{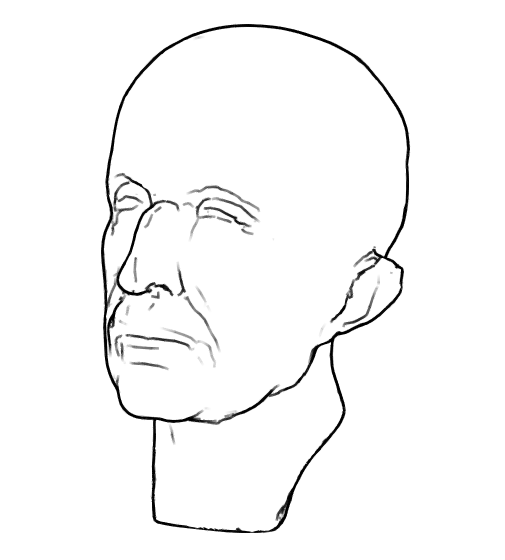}
  \label{fig:cmp_cube_sha}
  }~
  \subfigure{
 \includegraphics[width=\lineWPic]{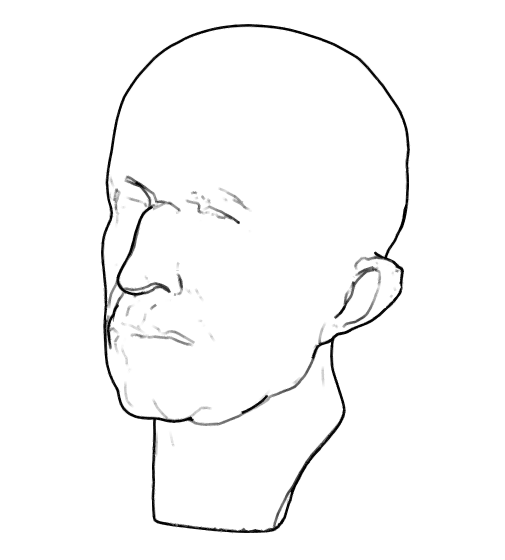}
  \label{fig:cmp_cube_sha}
  }~
  \subfigure{
 \includegraphics[width=\lineWPic]{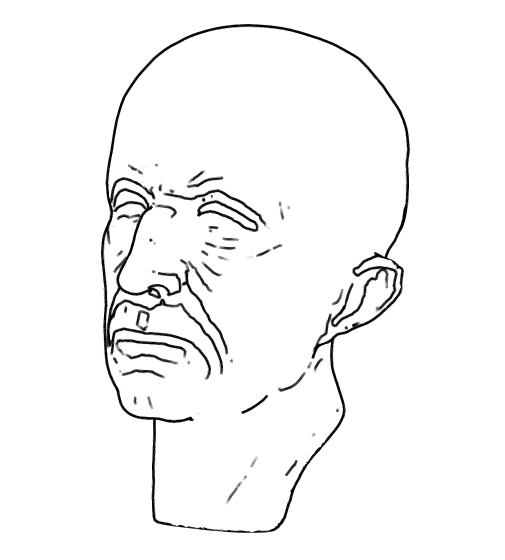}
  \label{fig:cmp_cube_sha}
  }~
  \subfigure{
 \includegraphics[width=\lineWPic]{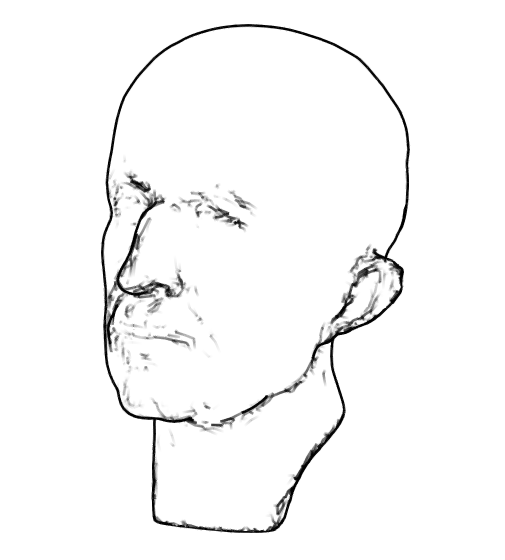}
  \label{fig:cmp_cube_sha}
  }\\
  \subfigure{
   \includegraphics[width=\lineWPic]{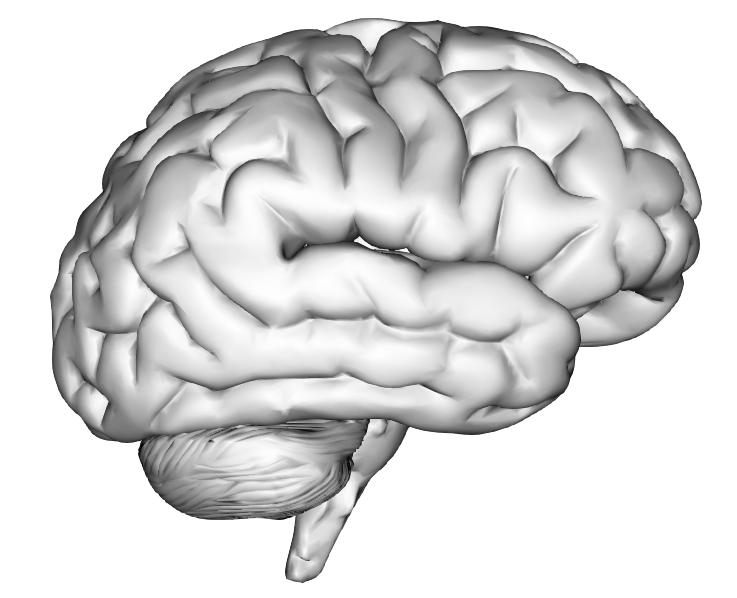}
  \label{fig:cmp_cube_sha}
  }~
  \subfigure{
 \includegraphics[width=\lineWPic]{compare//brain_sc}
  \label{fig:cmp_cube_sha}
  }~
  \subfigure{
 \includegraphics[width=\lineWPic]{compare//brain_rv}
  \label{fig:cmp_cube_sha}
  }~
  \subfigure{
 \includegraphics[width=\lineWPic]{compare//brain_ar}
  \label{fig:cmp_cube_sha}
  }~
  \subfigure{
 \includegraphics[width=\lineWPic]{compare//brain_pel}
  \label{fig:cmp_cube_sha}
  }~
  \subfigure{
 \includegraphics[width=\lineWPic]{compare//brain_dem}
  \label{fig:cmp_cube_sha}
  }~
  \subfigure{
 \includegraphics[width=\lineWPic]{compare//brain_ll}
  \label{fig:cmp_cube_sha}
  }\\
  \subfigure{
   \includegraphics[width=\lineWPic]{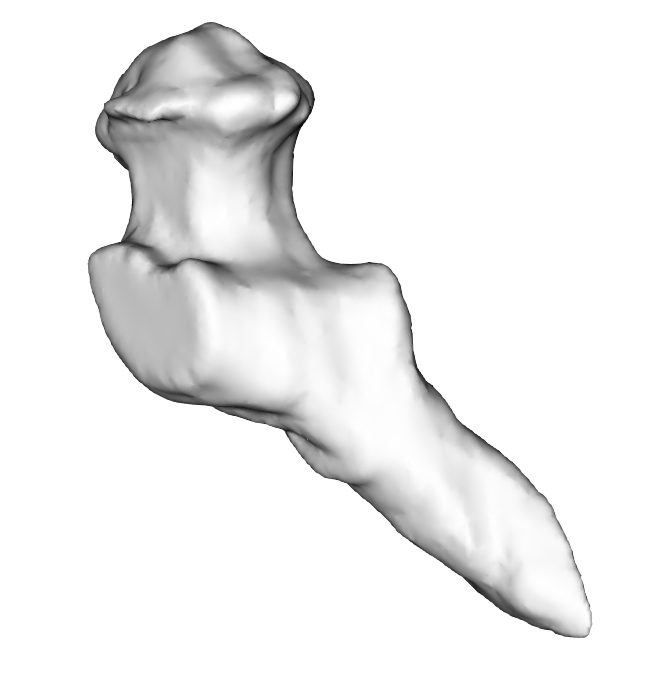}
  \label{fig:cmp_cube_sha}
  }~
  \subfigure{
 \includegraphics[width=\lineWPic]{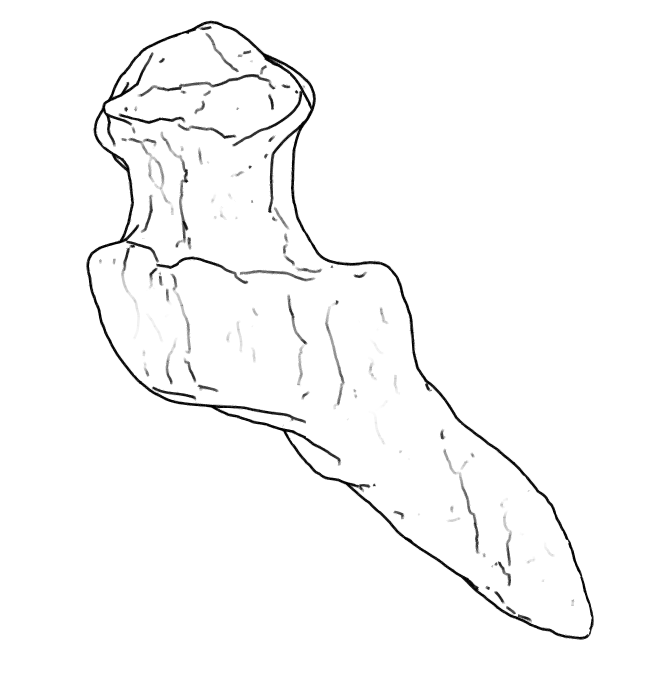}
  \label{fig:cmp_cube_sha}
  }~
  \subfigure{
 \includegraphics[width=\lineWPic]{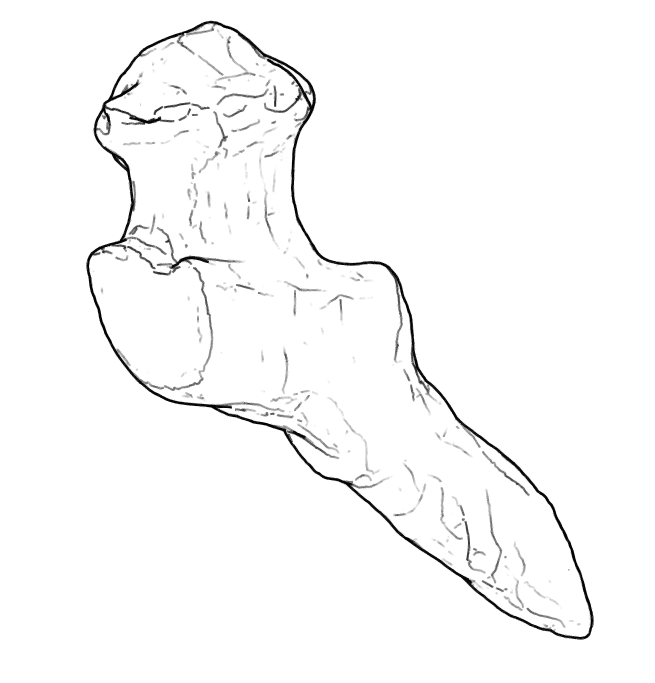}
  \label{fig:cmp_cube_sha}
  }~
  \subfigure{
 \includegraphics[width=\lineWPic]{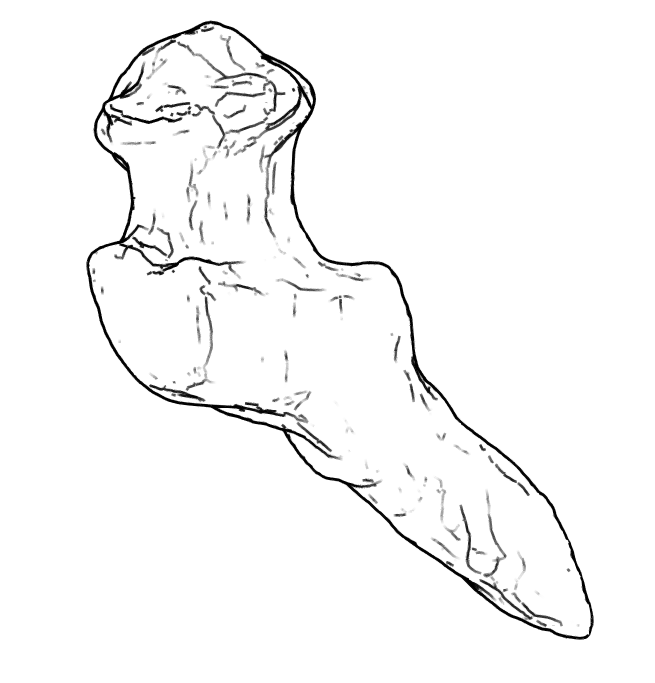}
  \label{fig:cmp_cube_sha}
  }~
  \subfigure{
 \includegraphics[width=\lineWPic]{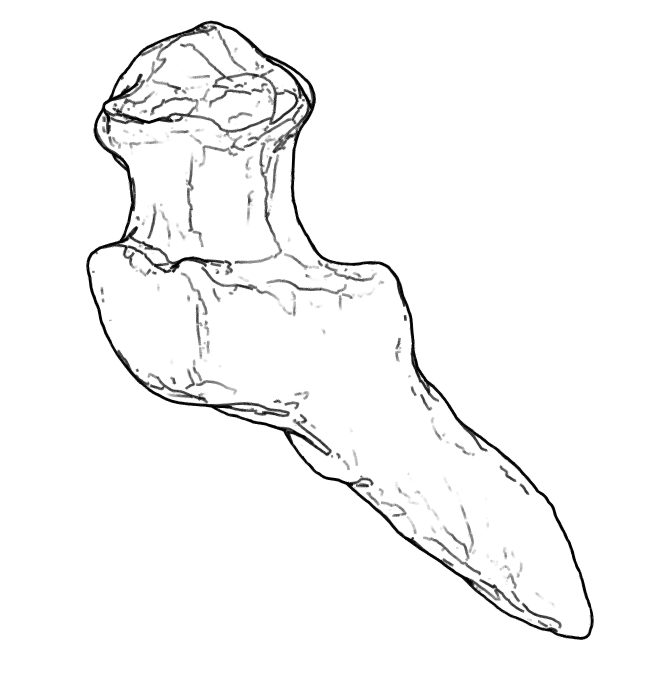}
  \label{fig:cmp_cube_sha}
  }~
  \subfigure{
 \includegraphics[width=\lineWPic]{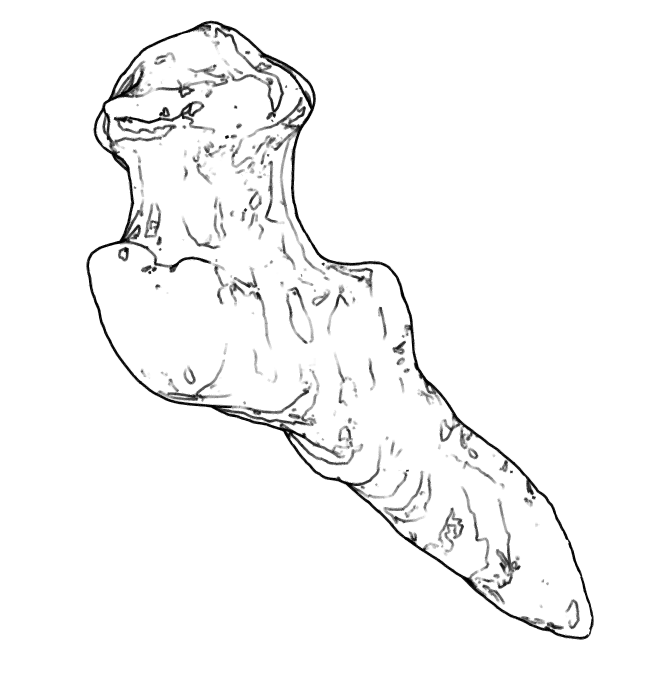}
  \label{fig:cmp_cube_sha}
  }~
  \subfigure{
 \includegraphics[width=\lineWPic]{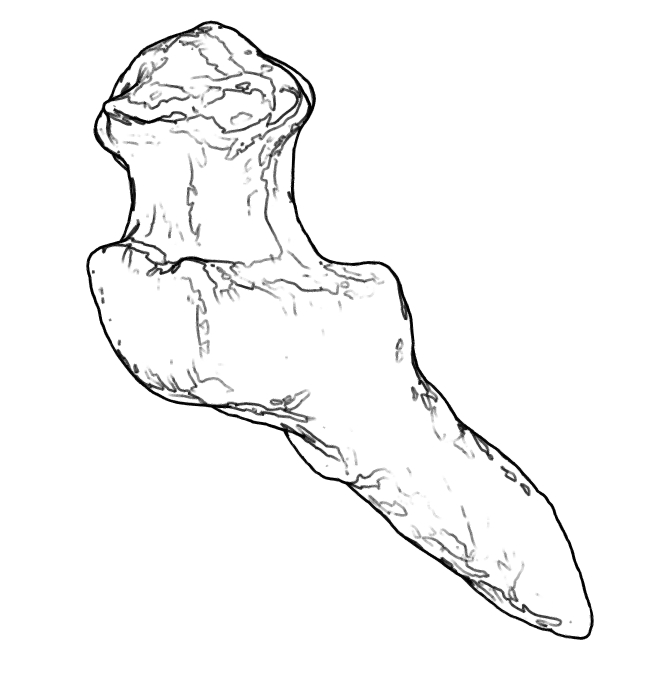}
  \label{fig:cmp_cube_sha}
  }\\
  \subfigure{
   \includegraphics[width=\lineWPic]{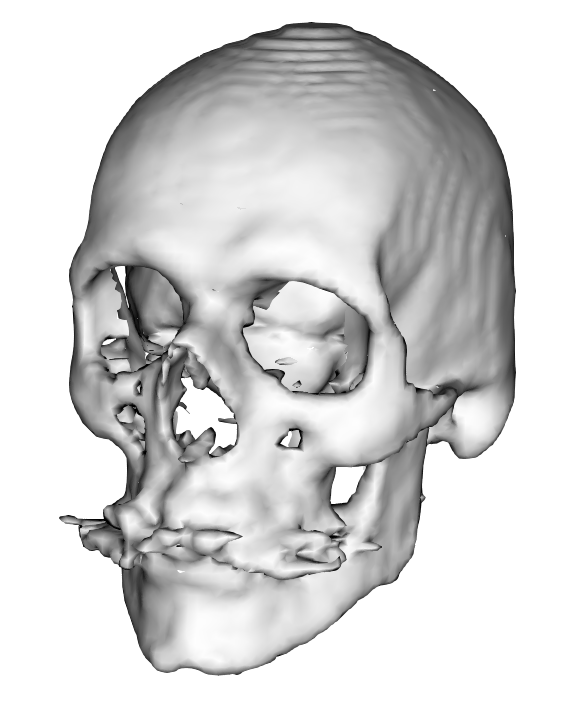}
  \label{fig:cmp_cube_sha}
  }~
  \subfigure{
 \includegraphics[width=\lineWPic]{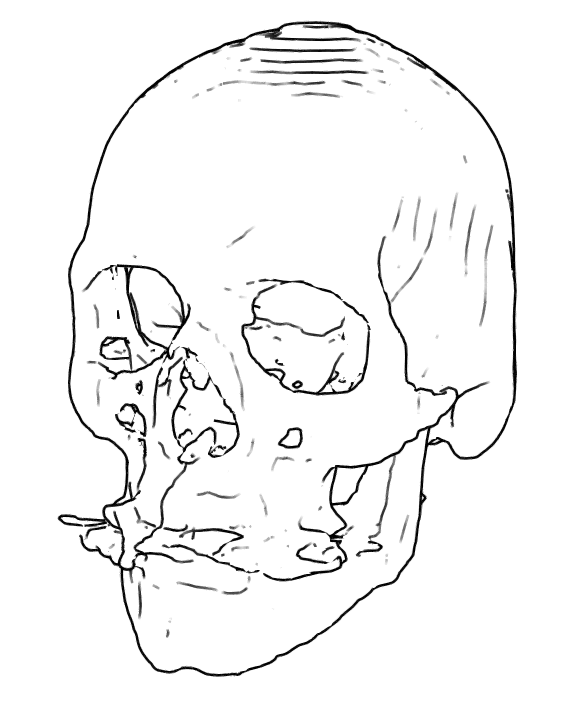}
  \label{fig:cmp_cube_sha}
  }~
  \subfigure{
 \includegraphics[width=\lineWPic]{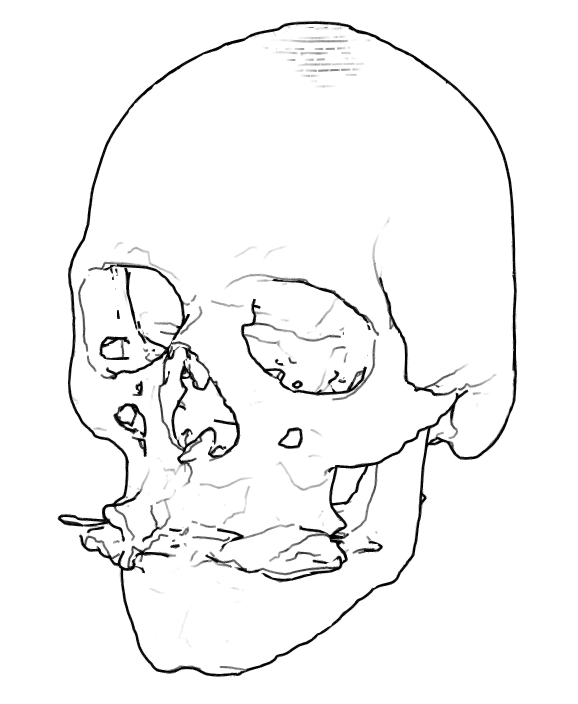}
  \label{fig:cmp_cube_sha}
  }~
  \subfigure{
 \includegraphics[width=\lineWPic]{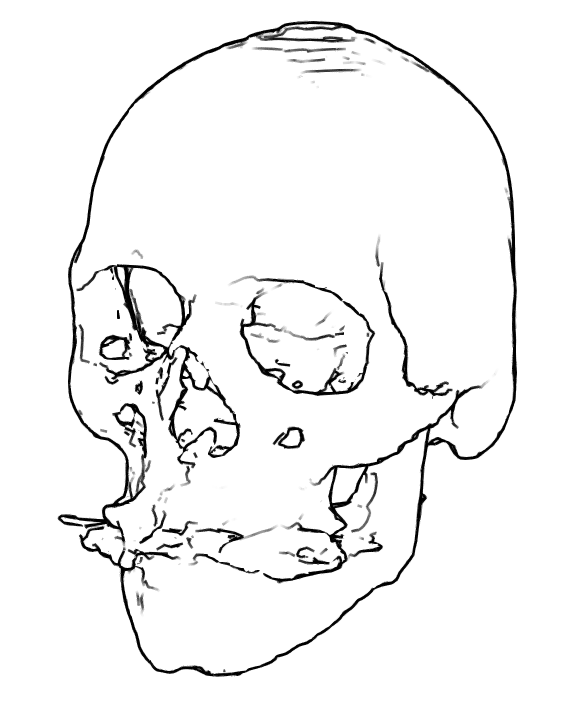}
  \label{fig:cmp_cube_sha}
  }~
  \subfigure{
 \includegraphics[width=\lineWPic]{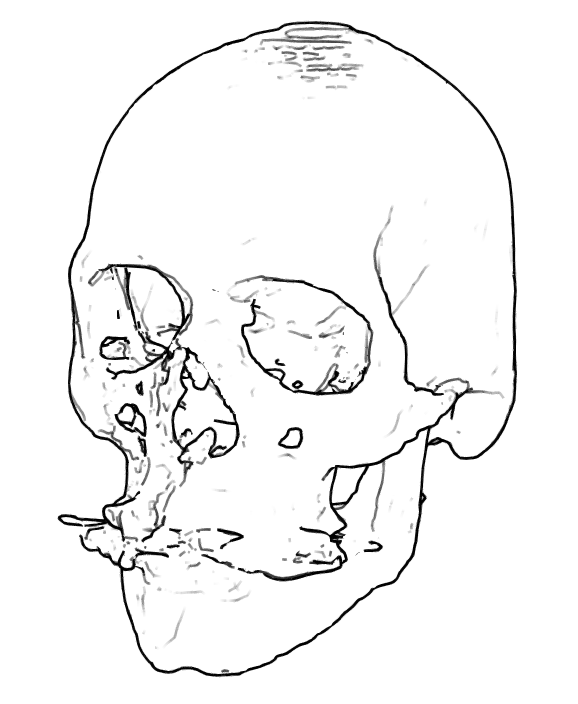}
  \label{fig:cmp_cube_sha}
  }~
  \subfigure{
 \includegraphics[width=\lineWPic]{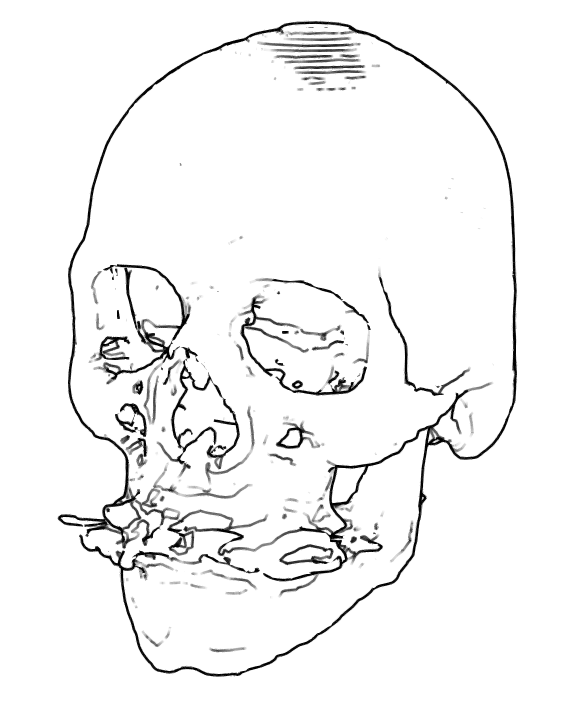}
  \label{fig:cmp_cube_sha}
  }~
  \subfigure{
 \includegraphics[width=\lineWPic]{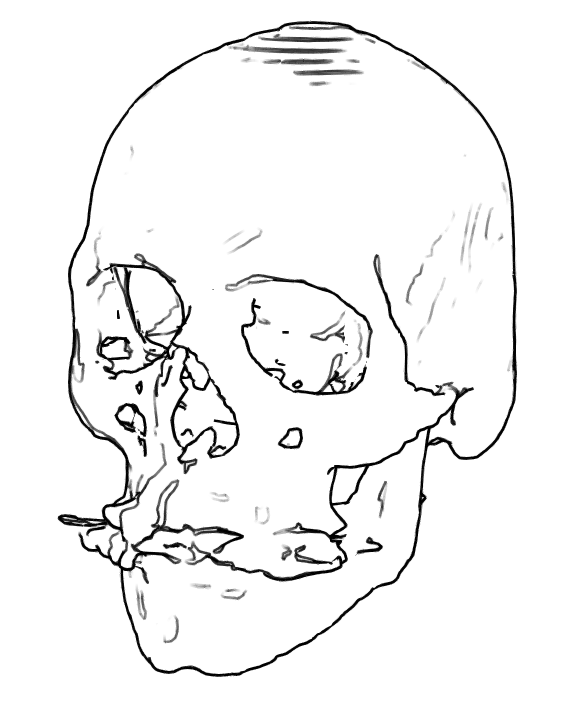}
  \label{fig:cmp_cube_sha}
  }
    \newcommand{\lineW}{0.04325}
   \begin{tabular}{c c c c c c c} 			\hline
			\hspace{\lineW\textwidth} SH \hspace{\lineW\textwidth} &  
			\hspace{\lineW\textwidth} SC \hspace{\lineW\textwidth} & 
			\hspace{\lineW\textwidth} RV \hspace{\lineW\textwidth} & 
			\hspace{\lineW\textwidth} AR \hspace{\lineW\textwidth} & 
			\hspace{\lineW\textwidth} PEL \hspace{\lineW\textwidth} & 
			\hspace{\lineW\textwidth} DC \hspace{\lineW\textwidth} &
			\hspace{\lineW\textwidth} LL \hspace{\lineW\textwidth}  \\
		\end{tabular}\\
 \caption{Selected models depicted in shading and higher-order feature lines.}
 \label{fig:models}
\end{figure*}

For example suggestive contours have two definitions of how to assess the feature lines.
One is curvature-based and the other is light-based.
With the second definition, no preprocessing is needed to assess the curvature and the principle curvature directions.
This is in contrast to ridges and valleys and apparent ridges.
Therefore, these algorithms are not able to compute the feature lines during the deformation. 
Photic extremum lines are also able to compute the feature lines during runtime because of the light and view dependency.
The Laplacian lines need to precompute the Laplacian of the normals.
Hence, this method is not suited for deformations.

Furthermore, Figure~\ref{fig:models} shows some exemplary models illustrated with higher order feature lines.
Three typical models in the discrete differential geometry field (cow, Buddha, Max Planck) as well as three models from the medical image data (brain, femur, skull) are presented.

In summary, current feature lines are not suitable for the depiction of anatomical structures directly derived from medical image data because the underlying surfaces are too noisy.
Advanced smoothing algorithms are necessary to reduce artifacts, but preserve important anatomical structures. 
For the depiction of a sparse representation of the model in a context-aware manner, the feature line methods can be used.

%
%
\subsection{Medical Application}
%
%
\begin{figure*}[h!]
 \centering
  \subfigure[]{
 \includegraphics[width=0.5\textwidth]{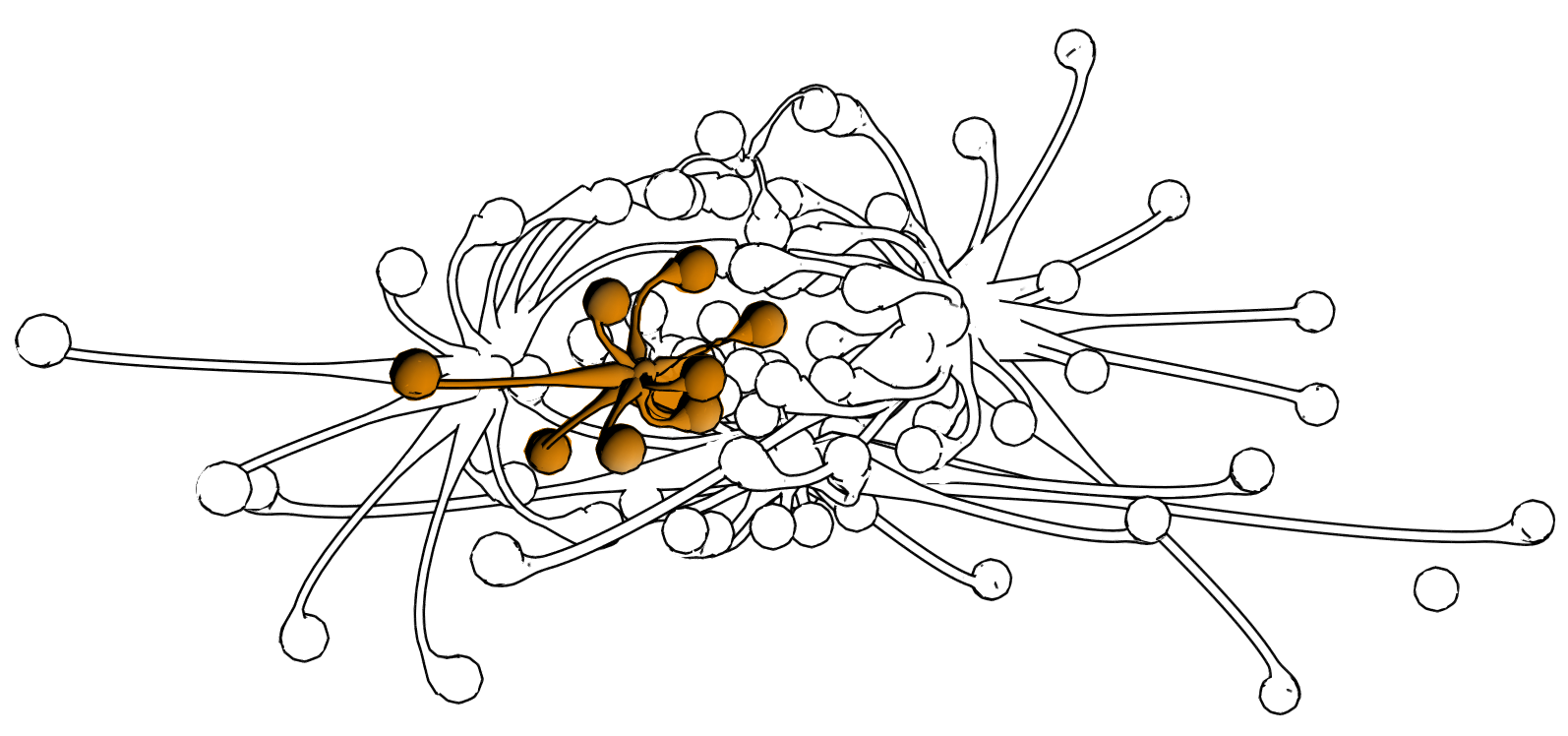}
  \label{fig:ma_clustervis}
  }~
  \subfigure[]{
 \includegraphics[width=0.25\textwidth]{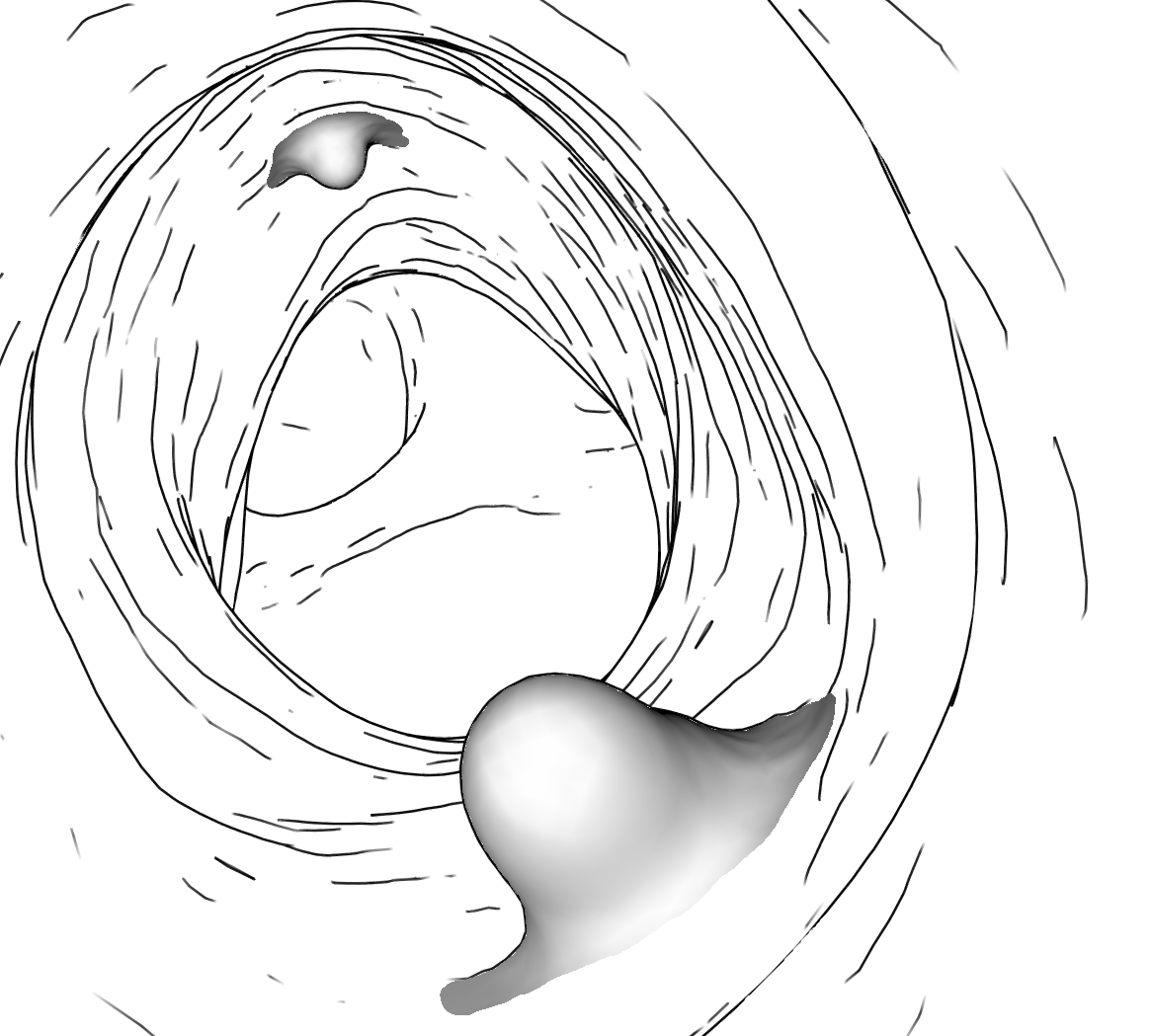}
 \label{fig:ma_polyp}
 }~
 \subfigure[]{
 \includegraphics[width=0.25\textwidth]{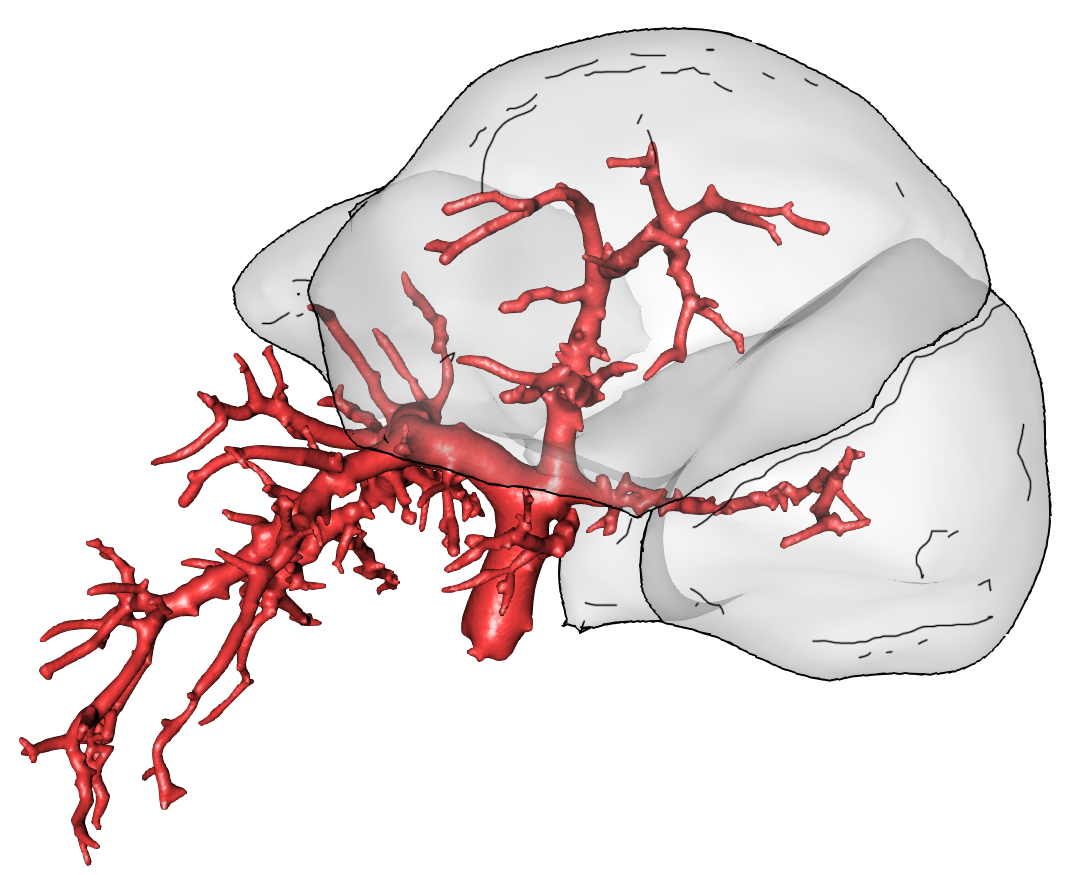}
 \label{fig:ma_liver}
}
 \caption{Different medical application fields where feature lines can be used to illustrate surrounding objects.}
 \label{fig:medical_application}
\end{figure*}
As stated, feature line methods can be seen as an illustrative visualization method that can enhance shading or as an alternative in the focus-and-context visualization.
In this section, we list different application fields where illustrative visualization is useful for effectively depicting medical data.
At the end, we list possible fields  where feature lines can be used to encode context information.

Fischer et al.~\cite{Fischer2005} proposed to use illustrative visualizing tools to depict structures of hidden surfaces.
The rendering style is tailored for understanding spatial relationships and for visualizing hidden objects.
Born et al.~\cite{Born2010} used illustrative techniques to depict stream surfaces.
Their techniques are very useful for the visualizing of complex flow structures.
In the area of brain data, Jainek et al.~\cite{Jainek2008} suggested to use a hybrid visualization method to illustrate mesh and volume rendering.
Their approach is efficient for the exploration in clinical research.
Chu et al.~\cite{Chu2008} proposed a guideline of various rendering techniques.
They combined, e.g., isophote-based line hatching and silhouette drawing, for illustrative vascular visualization.
Different rendering techniques for medical applications were presented by Tietjen et al.~\cite{Tietjen2005}.
An example for illustrative visualization for liver surgery can be found in~\cite{Hansen2010}.

Gla{\ss}er et al.~\cite{Glasser2014} presented an approach to visualize 3D cluster results. 
Here, the medical researcher can analyze the whole 3D scene with different cluster results where he can also select interesting objects.
The surrounding objects become context information.
Thus, we propose to illustrate them with feature lines.
In this case, we used the contour because the objects does not inherit much features.
Figure~\ref{fig:ma_clustervis} illustrates the main object with unselected objects illustrated with feature lines.

In the field of endoscopic views, the identification of polyps is necessary.
Once the polyps are detected, they can be illustrated in such a way that the endoscopic views are used for context information.
In Figure~\ref{fig:ma_polyp}, we used suggestive contours for the vessel and diffuse shading for the polyps.

In Figure~\ref{fig:ma_liver}, we visualized the portal vein and three liver segments.
The portal vein is illustrated in diffuse shading in red.
The liver segments are visualized in diffuse shading with transparency and photic extremum lines.

%% file: conclusion.tex
%
%
\section{Conclusion}\label{conclusion}
%
%
We have summarized the most common feature line methods for object space-based presentations of 3D meshes as they are frequently used in medicine and molecular biology.
The presentation of the different methods was also covered by two basic sections.
We did not only list the most common feature line methods and their calculation in the discrete space, but also provided the mathematical background to explain the calculation from the differential geometry point of view.
Our goal was to present an extensive list of feature lines on the one hand, and to equip the reader with  basic knowledge of differential geometry on the other hand.
The graduated student may be able to follow the different methods and to implement every feature line algorithm based on our explanations in the field of discrete differential geometry.
Therefore, our survey and tutorial may also be used by students who are new in the field of illustrative rendering.
Furthermore, this survey may also be used as a starting point for the development of new feature line methods.
The potential of advanced and recently introduced feature line techniques is currently not exploited in the display of medical surface models.
The careful application of these methods and perception-based evaluations are left open for future work.